\documentclass[journal]{IEEEtran}
\usepackage{times}
\usepackage{epsfig}
\usepackage{graphicx}
\usepackage{amsmath}
\usepackage{amssymb}
\usepackage[utf8]{inputenc}
\usepackage{mathtools}
\usepackage{xcolor}
\usepackage{multirow,color}
\definecolor{snm}{rgb}{0,0,0}
\newcommand{\snm}[1]{\textcolor{snm}{#1}}
\usepackage[noadjust]{cite}

\ifCLASSINFOpdf
\else
\fi
\begin{document}

%%%%%%%%% TITLE
% \title{Self-Supervised Learning of Inter-Label Geometric Relationships For Segmenting Gleason Graded Histopathology Images}

\title{Learning of Inter-Label Geometric Relationships Using Self-Supervised Learning: Application To Gleason Grade Segmentation}

\author{Dwarikanath Mahapatra

\thanks{D. Mahapatra is with the Inception Institute of Artificial Intelligence, Abu Dhabi, United Arab Emirates  (e-mail: dwarikanath.mahapatra@inceptioniai.org).}
}

\maketitle
%\thispagestyle{empty}

%%%%%%%%% ABSTRACT
\begin{abstract}
Segmentation of Prostate Cancer (PCa) tissues from Gleason graded histopathology images is vital for accurate diagnosis. Although deep learning (DL) based segmentation methods achieve state-of-the-art accuracy, they rely on large datasets with manual annotations. We propose a method to synthesize for PCa histopathology images by learning the geometrical relationship between different disease labels using self-supervised learning. We use a weakly supervised segmentation approach that uses Gleason score to segment the diseased regions and the resulting segmentation map is used to train a Shape Restoration Network (ShaRe-Net) to predict missing mask segments in a self-supervised manner. Using DenseUNet as the backbone generator architecture  we incorporate latent variable sampling to inject diversity in the image generation process and thus improve robustness. 
   Experiments on multiple histopathology datasets demonstrate the superiority of our method over competing image synthesis methods for segmentation tasks. Ablation studies show the benefits of integrating geometry and diversity in generating high-quality images, and our self-supervised approach with limited class-labeled data achieves similar performance as fully supervised learning. 
\end{abstract}


%%%%%%%%% BODY TEXT

\section{Introduction}
\label{sec:intro}

Prostate Cancer (PCa) is the sixth most common and second deadliest cancer among men worldwide. The most accurate PCa detection and staging is obtained from the analysis of stained biopsy tissue images. Each tissue region is assigned a Gleason grade between $1$ and $5$ \snm{(corresponding to the observed cellular patterns)} and the final score is the sum of the most prominent and second most prominent patterns. %,. 
%Figure~\ref{fig:disImage} shows examples of Gleason graded images.
%
% 
Gleason scoring is subjective due to  high level of heterogeneity % in the cellular and glandular patterns associated with each grade, 
leading to significant inter-observer and intra-observer variability %with reported Gleason score disagreements varying between $30\%$ to $53\%$ 
\cite{GGL6,GGL8,GGL14,MonusacTMI,Mahapatra_Thesis,KuanarVC,MahapatraTMI2021,JuJbhi2020,Frontiers2020,Mahapatra_PR2020,ZGe_MTA2019,Behzad_PR2020}.
Computer-aided methods can potentially improve consistency, speed, accuracy, and reproducibility of  diagnosis. Automated methods, especially state-of-the-art deep learning (DL) methods, require large image datasets for training. Owing to scarcity of such datasets, most approaches apply image augmentation to increase dataset size \snm{for network training}.

Traditional augmentations such as image rotations or deformations do not fully represent the underlying data distribution of the training set, \snm{do not add qualitatively novel information,} and are sensitive to parameter choices. 
Recent data augmentation methods of \cite{han2018gan, nielsen2019gan,Mahapatra_CVIU19,Mahapatra_CVIU2019,Mahapatra_CMIG2019,Mahapatra_LME_PR2017,Zilly_CMIG_2016,Mahapatra_SSLAL_CD_CMPB,Mahapatra_SSLAL_Pro_JMI,Mahapatra_LME_CVIU,LiTMI_2015,MahapatraJDI_Cardiac_FSL,Mahapatra_JSTSP2014,MahapatraTIP_RF2014} have used generative adversarial networks (GANs), \cite{goodfellow2014generative,MahapatraTBME_Pro2014,MahapatraTMI_CD2013,MahapatraJDICD2013,MahapatraJDIMutCont2013,MahapatraJDIGCSP2013,MahapatraJDIJSGR2013,MahapatraTrack_Book,MahapatraJDISkull2012,MahapatraTIP2012,MahapatraTBME2011}, and show success for medical image classification. However, \emph{they have limited relevance for segmentation} since they do not model geometric relation between different organs. %,   
Hence there is a need for augmentation methods that consider the geometric relation between different anatomical regions (labels) and generate realistic images for diseased and healthy cases.

\begin{figure*}[t]
 \centering
\begin{tabular}{cccccc}
\includegraphics[height=3cm, width=2.5cm]{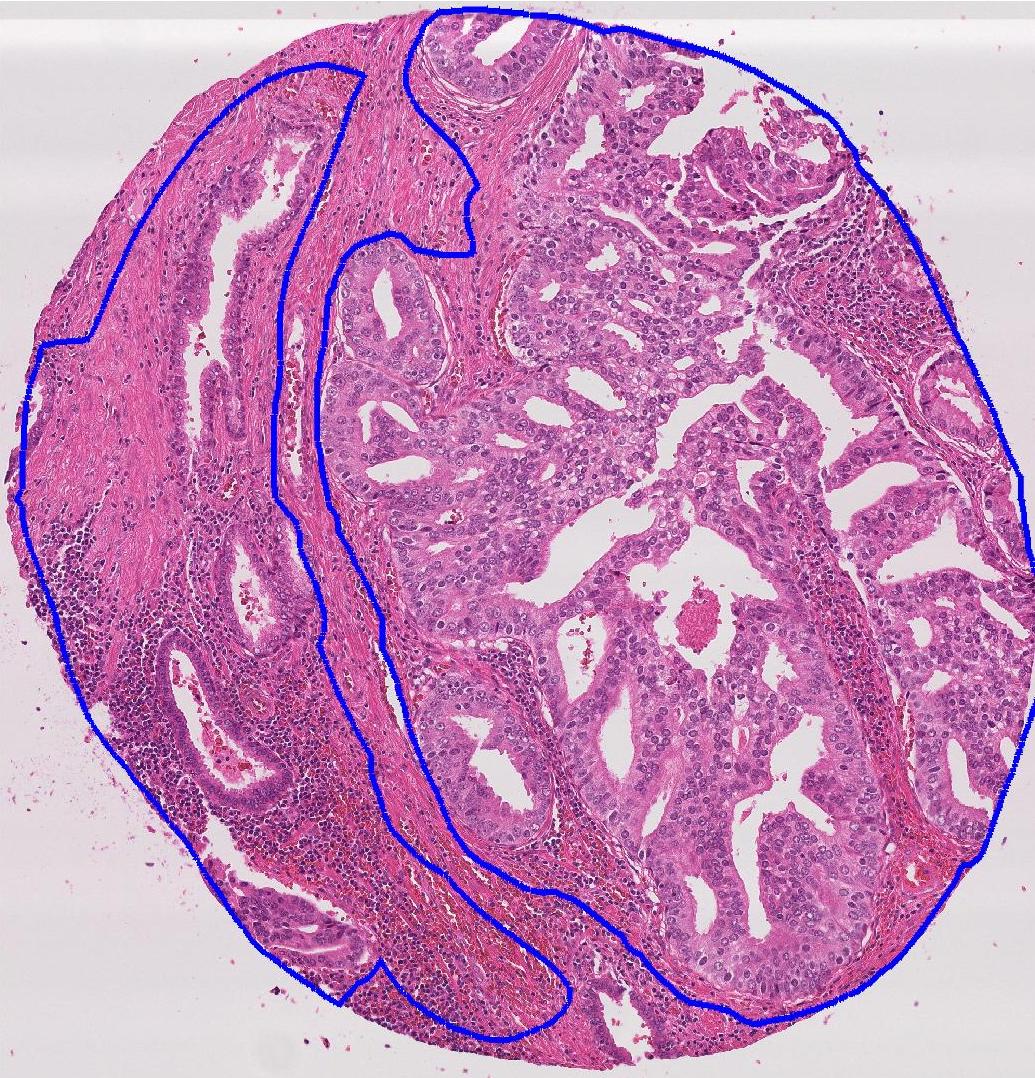} &
\includegraphics[height=3cm, width=2.5cm]{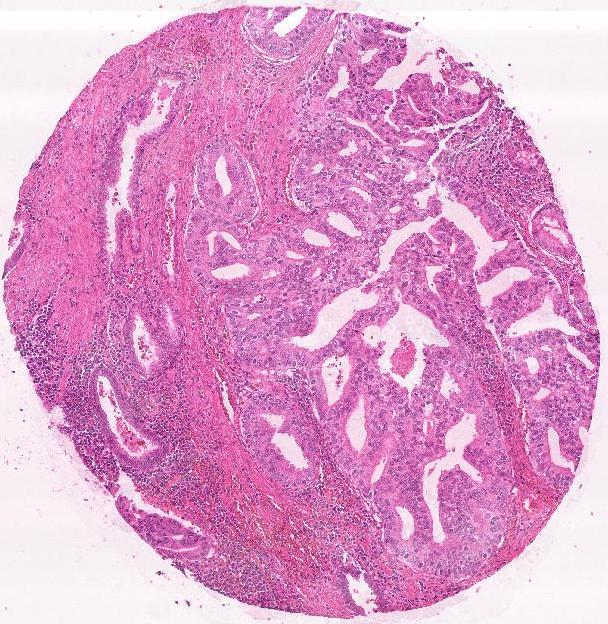} &
\includegraphics[height=3cm, width=2.5cm]{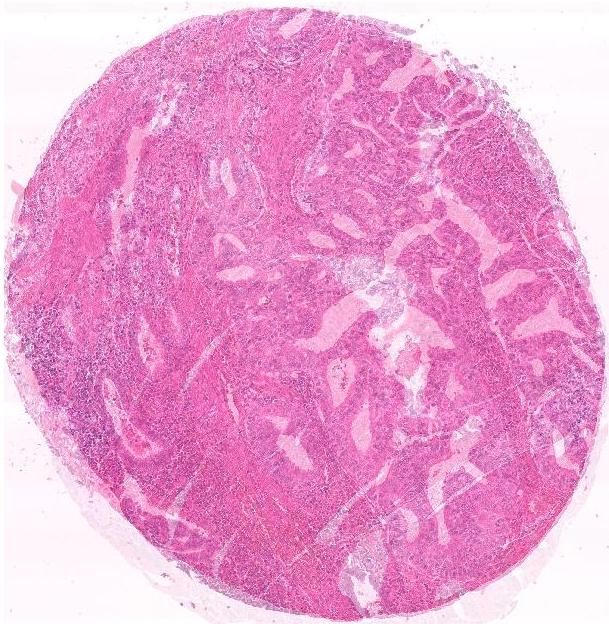} &
\includegraphics[height=3cm, width=2.5cm]{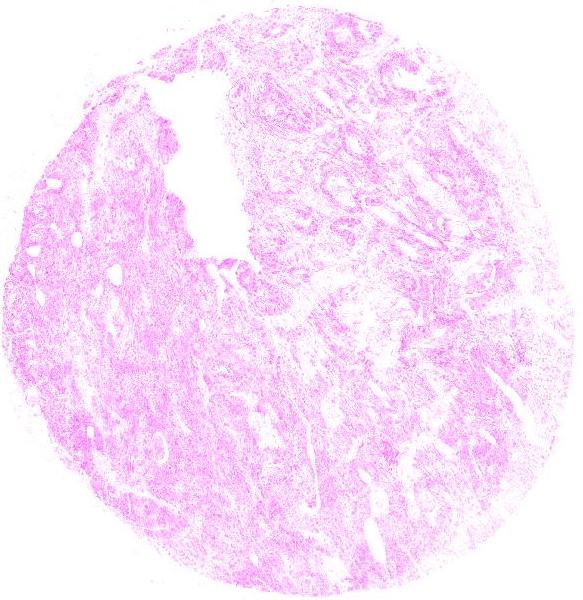} &
\includegraphics[height=3cm, width=2.5cm]{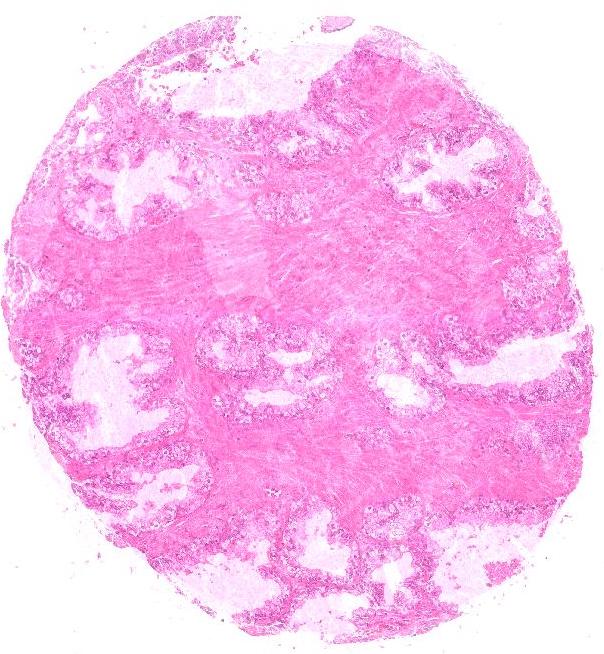} &
\includegraphics[height=3cm, width=2.5cm]{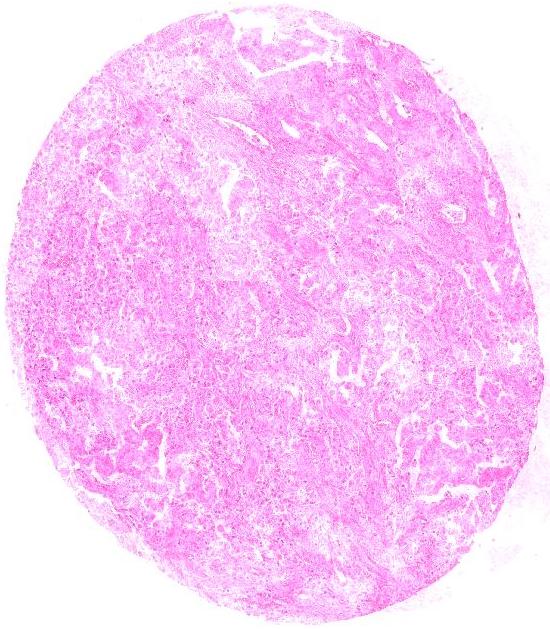}\\
(a) & (b) & (c) & (d) & (e) & (f)\\
\end{tabular}
\caption{(a) Base image with Gleason graded regions shown as blue outline; Example of generated images using: (b) Our proposed $GeoGAN$ method; (c) \cite{HGAN}; (d) \cite{Gupta}; (e) $DAGAN$ method by \cite{DAGAN}; (f) $cGAN$ method by \cite{Mahapatra_MICCAI2018}. }
\label{fig:SynImages_comp}
\end{figure*}

In this paper we propose a method to learn inter label geometric relationships using self-supervised learning. The learned model is used in a GAN framework to generate synthetic PCa histopathology images and used for segmentation of Gleason graded regions. 
Figure~\ref{fig:SynImages_comp} shows example cases of synthetic images generated by our method and other competing techniques. A sample image \snm{generated by our approach} is more realistic in appearance compared to the the base image, while other images have artifacts, noise, or distorted regions.  

% \begin{figure}[t]
%  \centering
% \begin{tabular}{cc}
% \includegraphics[height=4cm, width=4cm]{DisExample_1.png} &
% \includegraphics[height=4cm, width=4cm]{DisExample_2.png} \\
% (a) & (b)\\
% \end{tabular}
% \caption{Example of images showing disease pathologies. Images show the labels of the corresponding pathological regions (delineated as blue and green contours). Label $2$ corresponds to ``Benign'', while Labels $3,4,5$ correspond to Gleason grades ``3,4,5''.}
% \label{fig:disImage}
% \end{figure}

\subsection{Our Contribution}

% Since improved data augmentation yields better segmentation performance in a DL system, 
We propose to generate more informative images by considering the intrinsic relationships between shape and geometry of anatomical structures. We present a Geometry-Aware Shape Generative Adversarial Network (GeoGAN) that learns to generate plausible images of different Gleason graded regions while preserving learned relationships between geometry and shape. 
Since annotating medical images is a time-consuming task, it is challenging to obtain manually annotated segmentation masks to model the geometrical relation between different image labels. 
To overcome this challenge, we propose to use a weakly supervised segmentation approach to generate labeled segmentation maps which are used to learn the geometric relationship between different pathological regions.

In an earlier version of our method, \cite{Mahapatra_CVPR2020}, we propose a GAN based approach for generating optical coherence tomography (OCT) images of retinal diseases. Apart from applying the method to histopathology images, our current work has the following additional novelties compared to \cite{Mahapatra_CVPR2020}:
\begin{enumerate}
    \item \textbf{Weakly Supervised Segmentation:} In \cite{Mahapatra_CVPR2020} we used manual segmentation maps to learn inter-label geometrical relationships. However, that restricts the method to those datasets having manually annotated segmented maps, which are time-consuming to obtain. We introduce a novel weakly supervised semantic segmentation (WSSS) method that generates segmentation maps of histopathology images using the knowledge of disease grades. 
    \snm{Different from \cite{yamamoto}: 1) we use semi-supervised learning and label refinement in addition to clustering of autoencoder features; 2) we apply our approach to semantic segmentation while \cite{yamamoto} use their method only for classification.
}    
    \item \textbf{Self-Supervised Learning:} We introduce a self-supervised learning based method to learn the inter-label geometrical relationships. In \cite{Mahapatra_CVPR2020} we model this knowledge using a class conditional classifier. However, our experiments show that using the conditional classifier on the weakly supervised segmentation output affects the method's robustness. \snm{Possible reasons are inherent errors} in the segmentation output. Hence we use a \snm{pre-text task based} self-supervised learning approach in this work to introduce greater robustness.
    
    \item \textbf{Uncertainty sampling:} Previously in \cite{Mahapatra_CVPR2020} we used a UNet based generator network for generating samples, and the uncertainty sampling was accordingly formulated. In our current work, we use Dense UNet \cite{DenseUNet}, whose dense connections across previous layers facilitate more significant interaction between different levels of image representation.
    
\end{enumerate}

\section{Related Work}
\label{sec:prior}

\paragraph{Histopathology Image Analysis For Gleason Grading }

% Machine learning, especially deep learning methods have
% achieved promising results on general histopathology image
% classification. 
Introduction of whole slide image (WSI) scanners has opened up the opportunity for computer-aided diagnosis (CAD) to aid pathologists and reduce inter-observer variability \cite{Lucas12,MahapatraEURASIP2010,MahapatraTh2012,MahapatraRegBook,PandeyiMIMIC2021,SrivastavaFAIR2021,Mahapatra_DART21b,Mahapatra_DART21a,LieMiccai21,TongDART20,Mahapatra_MICCAI20,Behzad_MICCAI20}.
To overcome issue of high dimensional WSI commonly used methods \cite{HG19,HG41,Mahapatra_CVPR2020,Kuanar_ICIP19,Bozorgtabar_ICCV19,Xing_MICCAI19,Mahapatra_ISBI19,MahapatraAL_MICCAI18,Mahapatra_MLMI18,Sedai_OMIA18,Sedai_MLMI18,MahapatraGAN_ISBI18} apply patch-level image classification or use sliding windows.
Existing work on prostate biopsies use convolution neural networks \cite{Lucas18,Toro,Sedai_MICCAI17,Mahapatra_MICCAI17,Roy_ISBI17,Roy_DICTA16,Tennakoon_OMIA16,Sedai_OMIA16,Mahapatra_OMIA16,Mahapatra_MLMI16,Sedai_EMBC16,Mahapatra_EMBC16}, semantic segmentation \cite{Lucas19,Mahapatra_MLMI15_Optic,Mahapatra_MLMI15_Prostate,Mahapatra_OMIA15,MahapatraISBI15_Optic,MahapatraISBI15_JSGR,MahapatraISBI15_CD,KuangAMM14,Mahapatra_ABD2014,Schuffler_ABD2014,Schuffler_ABD2014_2}, feature-engineering \cite{GGL30,MahapatraISBI_CD2014,MahapatraMICCAI_CD2013,Schuffler_ABD2013,MahapatraProISBI13,MahapatraRVISBI13,MahapatraWssISBI13,MahapatraCDFssISBI13,MahapatraCDSPIE13,MahapatraABD12,MahapatraMLMI12,MahapatraSTACOM12}, and \snm{multiple instance learning} for binary classification of clinical specimens \cite{GGL26,VosEMBC,MahapatraGRSPIE12,MahapatraMiccaiIAHBD11,MahapatraMiccai11,MahapatraMiccai10,MahapatraICIP10,MahapatraICDIP10a,MahapatraICDIP10b,MahapatraMiccai08,MahapatraISBI08,MahapatraICME08,MahapatraICBME08_Retrieve,MahapatraICBME08_Sal,MahapatraSPIE08,MahapatraICIT06}. %,  and deep learning \cite{GGL27}.
% Kallen et al. \cite{Lucas20} differentiate between Gleason grade (GG) 3 and GG 5, yielding an accuracy of $81\%$ in homogeneous GG regions of interest. %
%
Nagpal et al.\cite{NagpalGGL}  developed a DL system  to perform Gleason scoring and quantification of prostatectomy specimens including fine-grained measures of tumor grading.

%
% Del toro
% Manually annotating tumors in whole slide images is time-consuming and not scalable to the large
% number of slides produced daily in hospitals. Inaccurate manual annotations  lead to missed isolated tumor cells \cite{Lucas18} and imprecise borders of the segmented tumors \cite{Toro19}. Toro et al. \cite{Toro}  propose an approach
% that automatically generates patches from WSIs in regions–of–interest and trains a CNN model to classify Gleason grades. % using a large data set of non–manually segmented tissue slides.
% from Google AI paper
 
% Previous computation approaches
% for Gleason grading used feature-engineering
% \cite{GGL30}, \snm{multiple instance learning} for binary classification of clinical specimens \cite{GGL26},  and
% deep learning \cite{GGL27}. 
% %
% Consistency of Gleason scoring has been shown
% to improve its prognostic utility \cite{GGL9}. 
% Nagpal et al.\cite{NagpalGGL}  developed a DL system  to perform Gleason scoring and quantification of prostatectomy specimens including fine-grained
% measures of tumor grading.

\paragraph{Data Augmentation (DA)}

Conventional augmentation approaches (such as rotation, scaling, etc.) do not add data diversity and are sensitive to parameter values \cite{Zhao19_25,Covi19_Ar,JBHISyn_Ar,Kuanar_AR2,TMI2021_Ar,Kuanar_AR1,Lie_AR2,Lie_AR,Salad_AR,Stain_AR,DART2020_Ar,CVPR2020_Ar,sZoom_Ar,CVIU_Ar,AMD_OCT}, variation in image resolution, appearance, and quality \cite{Zhao19_45,GANReg2_Ar,GANReg1_Ar,PGAN_Ar,Haze_Ar,Xr_Ar,RegGan_Ar,ISR_Ar,LME_Ar,Misc,Health_p,Pat2,Pat3,Pat4,Pat5}.
%
%
%  Recent DL based methods trained with synthetic images outperform those trained with standard DA over classification and segmentation tasks. 
Generative adversarial networks (GANs) have been very effective for data augmentation in few shot learning systems \cite{DAGAN}, chest xray images \cite{Mahapatra_MICCAI2018}, cervical histopathological images \cite{HGAN}, and to generate virtual histopathology images \cite{Gupta}. Zhao et al. \cite{Zhao_CVPR2019} proposed a learning-based registration method to register images to an atlas, use the corresponding deformation field to deform a segmentation mask and obtain new data. %.
 %
 %
%
%
% GANs have also been used for generating synthetic retinal images \cite{ZhaoMIA2018} and brain magnetic resonance images (MRI) \cite{han2018gan,ShinSASHIMI2018}, facial expression analysis \cite{Behzad_PR2020}, for super resolution \cite{SRGAN,MahapatraMICCAI_ISR}, %image registration \cite{Mahapatra_PR2020,Mahapatra_ISBI19,Mahapatra_MLMI18} 
% and generating higher strength MRI from their low strength acquisition counterparts \cite{GAN_MI_Rev}. 
GANs have also been used for generating synthetic retinal images in \cite{ZhaoMIA2018} and brain magnetic resonance images (MRI) in \cite{han2018gan,ShinSASHIMI2018}, 
and generating higher strength MRI from their low strength acquisition counterparts \cite{GAN_MI_Rev}. 
% 
 %Generated images have implicit variations in intensity distribution. 
 However, there is no explicit attempt to model attributes such as shape variations that are important to capture different conditions across a population. 
 \cite{Zhao19_51,Pat6,Pat7,Pat8,Pat9,Pat10,Pat11,Pat12,Pat13,Pat14,Pat15,Pat16,Pat17,Pat18} augmented medical images with simulated anatomical changes but demonstrate inconsistent performance based on transformation functions and parameters.% settings.
 
%  
%------------------------------------

% \subsection{Image Generation Using Uncertainty}
\paragraph{Image Generation Using Uncertainty}
 
 Uncertainty based image generation has  been realized using approximate Bayesian inference in scene understanding \cite{PhS2} while Lakshminarayanan et al. \cite{PhS5} generate different samples using an ensemble of $M$ networks and  \cite{PhS7} use a single network with $M$ heads for image generation. Sohn et al. \cite{PhS8} use conditional variational autoencoders (cVAE) to model segmentation masks, while in probabilistic UNet \cite{PhS4}, cVAE is combined with UNet \cite{Unet} to generate multiple segmentation masks. Baumgartner et al. \cite{PhiSeg} generate images with greater diversity by injecting randomness at multiple levels. These approaches do not capture complex correlations between different labels

\subsection{\snm{Self-Supervised Learning}}

\snm{
Self-supervised learning methods consist of two distinct approaches: 1) pretext tasks and 2) loss functions used for down-stream tasks. %Solving pre-text tasks learns a useful data representation, although the task itself may not be relevant, while methods focusing on loss functions are independent of the pretext task and evaluate the quality of features learned by self-supervised learning. Contrastive learning approaches such as MoCo \cite{HeMoco} and SimCLR \cite{ChenSimCLR} are popular and give state-of-the-art results for loss function based methods. 
We use the pre-text approach to leverage the learned representation for generating synthetic images. 
}
%
% \snm{
% An appropriate pretext task contributes towards improved accuracy and robustness of the primary task by providing initial knowledge (in the form of network weights) to successfully train the main network. Common pretext tasks include estimating relative position of patches \cite{Doersch,Noroozi}, local context \cite{Pathak}, and colour \cite{Zhang}. 
% %
% Exemplar learning has also been proposed as a self- supervised learning strategy \cite{Dosovitskiy} by classifying each data instance into a unique class. 
% }
%
\snm{
%Due to the unique challenges of medical image analysis, pre-text tasks are important in the final outcome. %F
In one of the first works on self-supervised learning in medical imaging, Jamaludin et al. \cite{Jamaludin} propose a siamese CNN to recognize patients’ MR scans and predict the level of vertebral bodies. 
Other works include surgical video re-colourisation as a pretext task for surgical instrument segmentation \cite{BaiMic19_10}, rotation prediction for lung lobe segmentation and nodule detection \cite{BaiMic19_11}, and disease bounding box localization for cardiac MR image segmentation \cite{BaiMiccai19}. 
Chen et al. \cite{ChenMedIA19} use context restoration as a pre-text for classification, segmentation, and disease localization in medical images. Self supervised learning has also been applied to histopathology images, using domain specific pretext tasks \cite{SelfPath}, semi-supervised histology classification \cite{LuMahmood} and cancer subtyping  using visual dictionaries \cite{MuhammadFuchs}.
}

\section{Method}
\label{sec:method}

\begin{figure}[ht]
 \centering
\begin{tabular}{c}
\includegraphics[height=5.0cm, width=8.8cm]{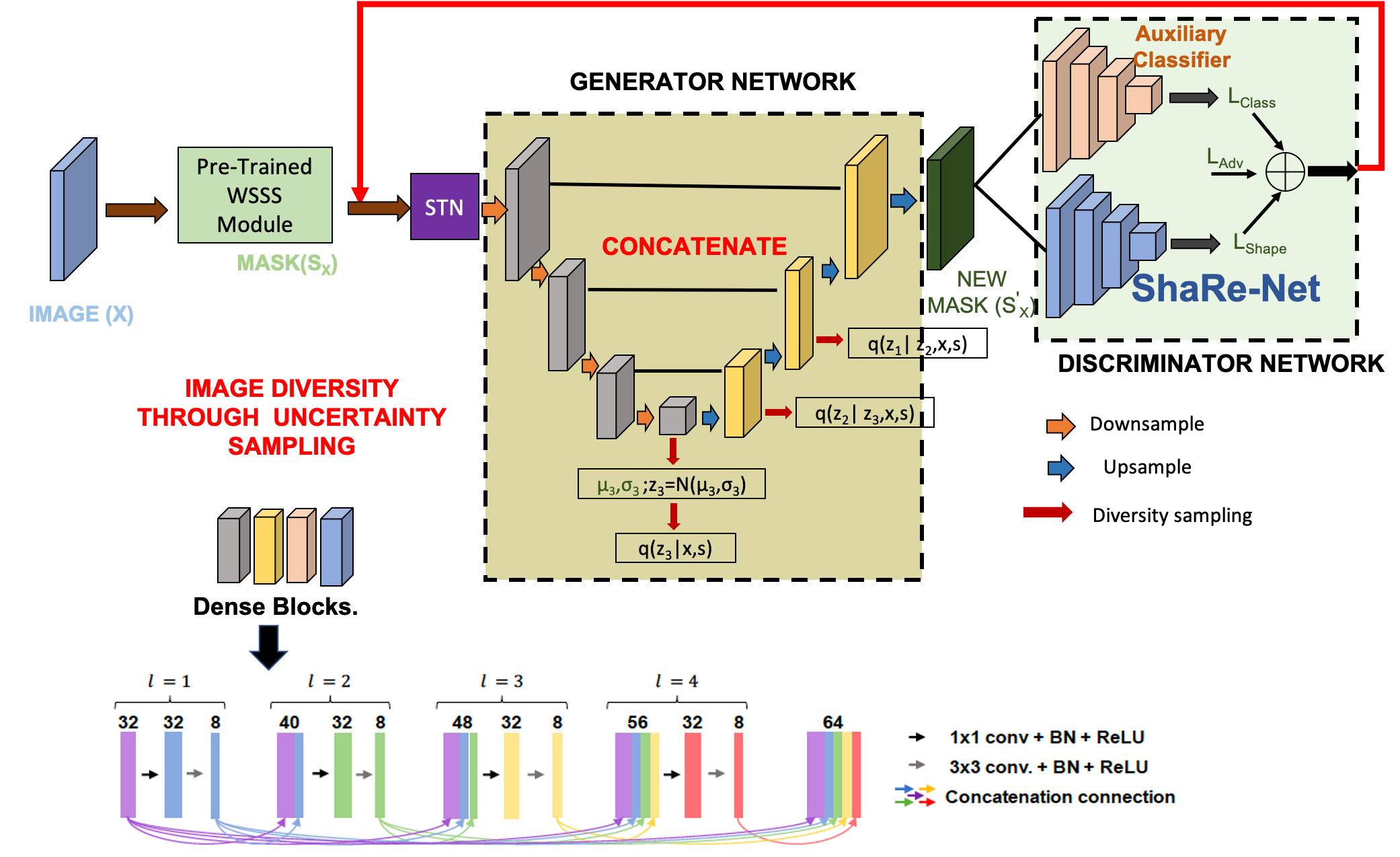}
\end{tabular}
\caption{Overview of the training stage of our method. The images ($X$) are input to the WSSS module which outputs corresponding segmentation masks ($S_X$) $S_X$ is input to a STN whose output is fed to the generator network. The   DenseUNet based Generator network injects diversity at different levels through uncertainty sampling to generate a new mask $S_X^{'}$. $S_X^{'}$ is fed to the discriminator which evaluates its accuracy based on $L_{class}$, $L_{shape}$ and $L_{adv}$. The provided feedback is used for weight updates to obtain the final model. The architecture of each dense block is shown. Feature-maps from previous layers are concatenated together as the input to the following layers.}
\label{fig:workflow}
\end{figure}

\snm{
Our objective is to synthesize informative histopathology patches from a base image patch and train a segmentation model. Before the actual training of the synthetic image generator, the base image is segmented using a weakly supervised semantic segmentation step (the ``Pre-Trained WSSS Module'' in Figure~\ref{fig:workflow}). This gives us a base mask to learn inter-label geometric relationships. 
The first stage of synthetic image generation is a spatial transformation network (STN) that transforms the base mask with different location, scale, and orientation attributes. This initially transformed mask is input to the Dense-UNet based `Generator' that injects diversity, and the discriminator ensures that the final generated mask is representative of the dataset. In addition to the adversarial loss ($L_{Adv}$) we incorporate losses from an auxiliary classifier ($L_{Class}$ to ensure the image has the desired label) and geometric relationship module ($L_{Shape}$ from ShaRe-Net). 
}

%  \begin{figure}[h]
%  \centering
% \begin{tabular}{c}
% \includegraphics[height=3.1cm, width=7.9cm]{TrnTest.png}  \\
% \end{tabular}
% \caption{Depiction of mask generation. The trained generator network is used on validation set base images and masks to generate new images that are used to train a segmentation network (UNet or Dense UNet). The model then segments infected regions from test images.}
% \label{fig:workflow2}
% \end{figure}

\subsection{Weakly Supervised Semantic Segmentation}
\label{met:wss}

To obtain pixel labels from  image labels in a weakly supervised setting, we solve an instance-level classification problem with pixels as instances. %Subsequently, existing fully supervised segmentation methods can be applied.
%
% Our method borrows concepts from sub-category exploration \cite{ChangWssCvpr20} and uses them in a combined Multiple Instance Learning (cMIL) for instance classification \cite{CAMEL}. %However there are significant novelties to deal with our unique problem.
% 
In a weakly supervised setting, the image (bag) labels are known while the individual pixel (instance) labels need to be determined. A `normal' bag indicates that all instances in the bag are labeled normal, while a `diseased' bag indicates that at least one instance is diseased. %\snm{Previous solutions have used class activation maps and multiple instance learning \cite{GGL26,CLAM}.}  

 Gleason grades range from $1-5$ and describe resemblance of a biopsy to healthy tissue (lower score) or abnormal tissue (higher score).  Most cancers score a grade of 3 or higher. %
Since prostate tumors are often made of cancerous cells with different grades, two grades are assigned for each patient.  A primary grade describes the cells making up the largest area of the tumor, and a secondary grade describes the cells of the next largest area.  
A Gleason score of, for example, ``5+4'' indicates that the two most prominent patterns are Gleason grades $5$ and $4$.
If the cancer is almost entirely made up of cells with the same score, the grade for that area is counted twice to calculate the total Gleason Score. 
However, there is also the possibility of the occurrence of other grades (such as $3$), which is not indicated in the bag label. To overcome this obstacle, we combine clustering with bag label information and semi-supervised learning to estimate the instance labels and then use these labels to train a multilabel segmentation network.

\snm{The work by Yamamoto et al. \cite{yamamoto} investigate explainable features for predicting PCa recurrence where they employ autoencoder feature extraction and clustering. Our approach also uses autoencoder feature extraction followed by clustering to get a first estimate of labels of some pixels. However, not all pixel labels can be estimated by this approach for which we use additional steps such as: 1) semi-supervised learning for identifying labels of unlabeled samples; and 2) label refinement step for correcting the error of previous label prediction step. %We also show that these additional steps contribute towards better segmentation performance.
}

  The image pixels in our dataset can have any one of $5$ labels - background, benign, Gleason grades $3-5$. We denote these labels, respectively, as $[1,2,3,4,5]$. The background (label $1$) is identified by simple thresholding, and hence the challenge is to identify the instances of other labels. We rarely observe an image with a single label. Hence it is challenging to obtain examples of so-called normal labels (for example, benign). Therefore, we identify all possible subcategories and use knowledge of bag labels to assign labels to samples from all classes. %' samples. %  take a slightly different approach to to solve our problem. 

\begin{figure}[ht]
 \centering
\begin{tabular}{c}
\includegraphics[height=5cm, width=8.8cm]{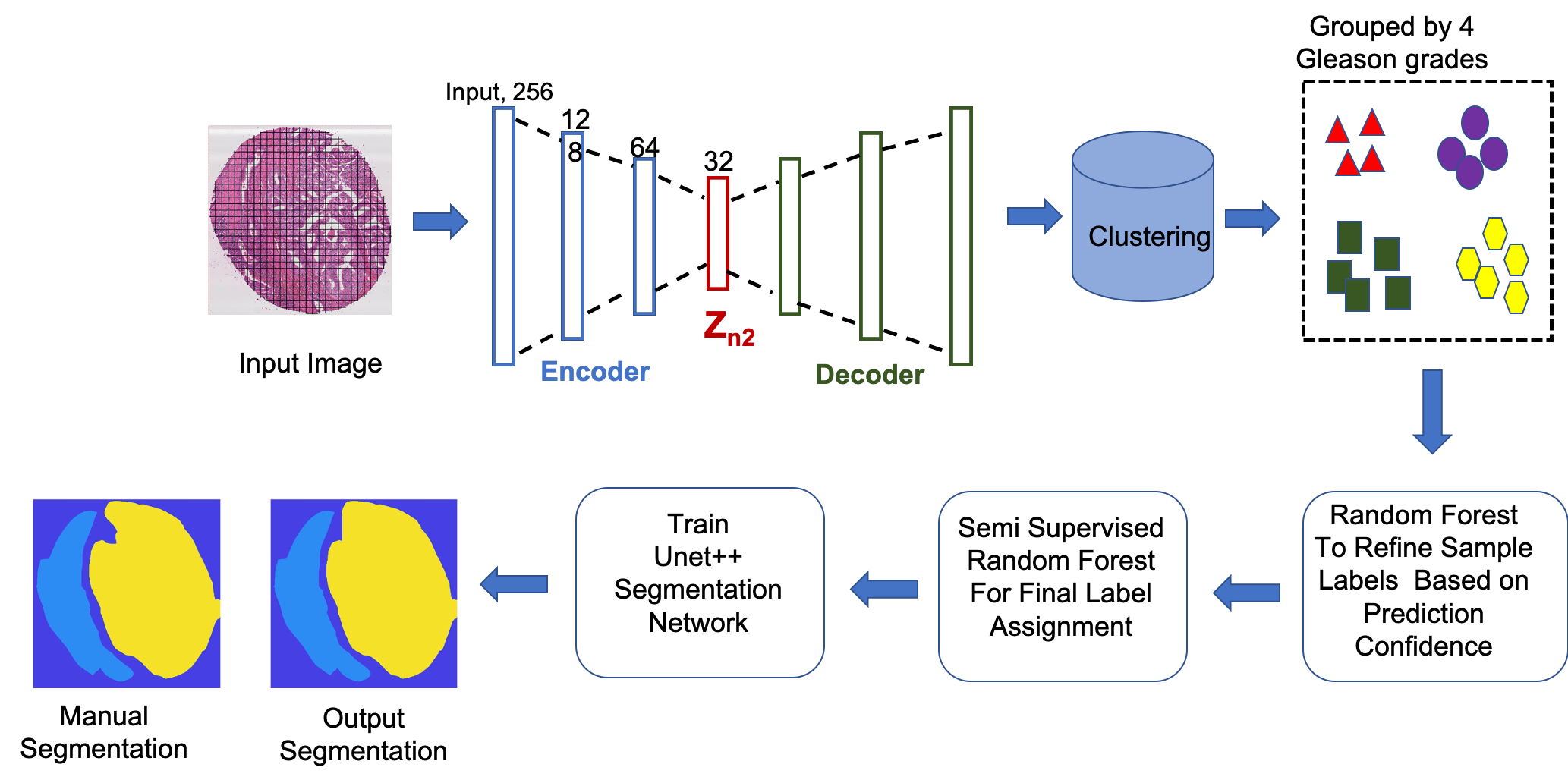}
\end{tabular}
\caption{The workflow of weakly supervised segmentation. The image patch is processed through an autoencoder to extract latent features which are clustered using k-means. The two largest clusters are assigned labels of the Gleason grade and then input to a random forest-based semi-supervised classifier that outputs the unlabeled samples' labels. The labeled samples are used to train a UNet++ network for obtaining the weakly supervised segmentation output.}
\label{fig:WSS}
\end{figure}

\subsubsection{\textbf{k-Means Clustering}}

Most weak segmentation methods' start with class activation maps (CAMs) to get the initial labels. and that requires a sufficiently high number of annotations. To use our method in a low resoruce setting we employ classical machine learning algorithms to identify sample labels. 
An image patch is split into $n \times n$ blocks ($n=16$). Each block is clustered using deep features extracted from an autoencoder trained to reconstruct the original $n \times n$ block. \snm{Note that neither the original mask nor the image are downsampled to lower resolutions.}

Figure~\ref{fig:WSS} shows the architecture of the autoencoder feature extractor. The $n \times n$ patch is converted into a $n^{2}$ vector and input to the autoencoder. The encoder stage has two hidden layers, followed by the latent representation of the original patch (denoted as $z_{n^{2}}$). The decoder layer reconstructs the original patch from $z_{n^{2}}$. The feature vectors $z_{n^{2}}$ from all instances of a bag are used as input to a k-means clustering algorithm that assigns each patch to one of $4$ clusters. At this stage, we only separate the samples into different clusters without knowing their actual label. 
% The clustering objective function can be defined as 
% %
% \begin{equation}
%     \arg \min_s \sum_{i=1}^{k} \sum_{x\in S_i} \|\left\|x-\mu_i\right\|^{2}
% \end{equation}

As stated previously, according to the Gleason grading criteria, the two Gleason grades correspond to the cells with the top two largest areas.  
% Thus, for each bag (image), the clusters with the highest and second-highest \snm{number of samples} are assigned the corresponding labels of the image, while the label of the other two clusters is unknown.
\snm{The clusters with the highest and second highest number of samples correspond to those pixels/regions assigned the two Gleason scores for the particular image}. %
Note that we group the features onto $4$ clusters for all images despite the fact that many images will have one or two distinct labels. This is due to two factors: 1) it reduces computation time as we do not have to find the optimal number of clusters for each image; 2) those samples which have been assigned labels according to the Gleason grade indicate high label confidence.
For those images whose samples belong to, say, cluster $4$ and have been assigned cluster $1$, this discrepancy is corrected in the later stage of label refinement and semi-supervised learning.

Repeating the above steps for all images, we have a mix of `labeled' and `unlabeled' samples. `Labeled' samples refer to those that have been assigned labels according to the image's Gleason score (i.e., samples from the two largest clusters) and `unlabeled' samples are from the other two clusters. The labeled samples have relatively high confidence and can be relied upon to train a robust and accurate classifier. \snm{Our next step is to remove those samples which may have been erroneously assigned to the `labeled' clusters. For this purpose we train a supervised RF classifier to predict sample class using only the labeled samples, identify samples with low prediction confidence and remove them from the `labeled’ training set.} RF is used because of its low memory requirement and ability to output confidence of predictions. The low confidence samples are then moved to the unlabeled set. We observe approximately  $1.5\%$ samples to have low confidence, indicating the robust nature of our clustering approach.

\subsubsection{\textbf{Semi Supervised Classification}}
\label{met:SSL}

% After removing low-confidence samples from the training set we have a set of labeled and unlabeled samples. Here we use a semi-supervised RF classifier using the following objective function \cite{RF}
% \begin{equation}
%     semi sueprvised classifer
% \end{equation}

% After training is complete we have the labels of all instances (samples).

We adopt the maximum margin approach to semi supervised random forests \cite{RF-SSL} to determine labels of unlabeled samples. 
Let $\textbf{g(x)} = [g_1(\textbf{x}),\cdots,g_K(\textbf{x})]^{T}$ be a multivalued
function. $\textbf{g(x)}$ is called a margin vector, if
\begin{equation}
    \sum_{i=1}^{K} g_i(\textbf{x})=0
\end{equation}

We can define the margin for the $i^{th}$
class as $g_i(\textbf{x})$ and the true margin as $g_y(\textbf{x})$. A loss function $\ell(g_y(\textbf{x}))$ is defined to be a margin
maximizing loss if $\ell(g_y(\textbf{x}))\leq 0$ for all values of $g_y$.
Therefore, an optimization based on this kind of loss functions will maximize the true margin.
In the absence of a label there is no known true margin. Therefore the margin for unlabeled samples is defined as:
\begin{equation}
    m_u(\textbf{x}_u) = \max_{i\in \mathcal{Y}} g_i(x_u)=0
\end{equation}
where $\mathcal{Y}$ is the set of labels.
Note that the predicted label for an unlabeled sample, $x_u$,
is $C(x) = \arg \max_{i\in Y} g_i(x_u)$. 

Similar to the traditional regularization-based
semi-supervised learning methods, we also regularize
the loss for the labeled samples with a loss over
the unlabeled samples. %Based on the definition of the
% margin for unlabeled samples, we use the same loss
% function used to grow the trees in a forest also to be
% the loss for the unlabeled samples.
The overall loss function is 
% \begin{equation}
%     \begin{split}
%         \mathcal{L}(g)=\frac{1}{\left|\mathcal{X}_l\right|} \sum_{(x,y)\in \mathcal{X}_l} \ell (g_y(x)) + \\ 
%         \frac{\alpha}{\left|\mathcal{X}_u\right|} \sum_{x\in \mathcal{X}_u} \ell (m_u(x)), 
%     \end{split}
% \end{equation}
%
\begin{equation}
        \mathcal{L}(g)=\frac{1}{\left|\mathcal{X}_l\right|} \sum_{(x,y)\in \mathcal{X}_l} \ell (g_y(x)) + 
        \frac{\alpha}{\left|\mathcal{X}_u\right|} \sum_{x\in \mathcal{X}_u} \ell (m_u(x_u)), 
\end{equation}
where $\mathcal{X}_l$ denotes set of labeled samples, $\mathcal{X}_u$ denotes the unlabeled samples, and  $\alpha=0.9$ defines the contribution rate of unlabeled
samples. Training only on labeled data keeps the loss convex, while  additional unlabeled data makes it non-convex. Hence Deterministic Annealing is used to obtain the global optimum and we refer  to \cite{RF-SSL} for further details. The output of this step is a label for each instance ($n\times n$ block)

\subsubsection{\textbf{Segmentation Using Standard Network}}

The assigned labels are mapped back to the  $n\times n$ patches to obtain a segmentation map. Using these segmentation maps as the ground truth, we train a UNet++ network \cite{UNet++} to predict the segmentation labels of a new image. The UNet++ network has redesigned skip connections, and the loss function \snm{as defined in \cite{UNet++}} is a combination of binary cross-entropy and Dice loss,
\begin{equation}
L(Y,\hat{Y}) = -\frac{1}{N} \sum_{b=1}^{N} \left(\frac{1}{2}\cdot Y_b \cdot \log\hat{Y_b} + \frac{2Y_b \cdot\hat{Y_b}}{Y_b+\hat{Y_b}} \right)    
\end{equation}
where $\hat{Y_b},Y_b$ denote the flattened predicted probabilities and the ground truths of $b^{th}$ image respectively, and $N$ indicates the batch size.

\begin{figure}[!htbp]
\centering
\begin{tabular}{cccc}
\includegraphics[height=2.0cm, width=1.9cm]{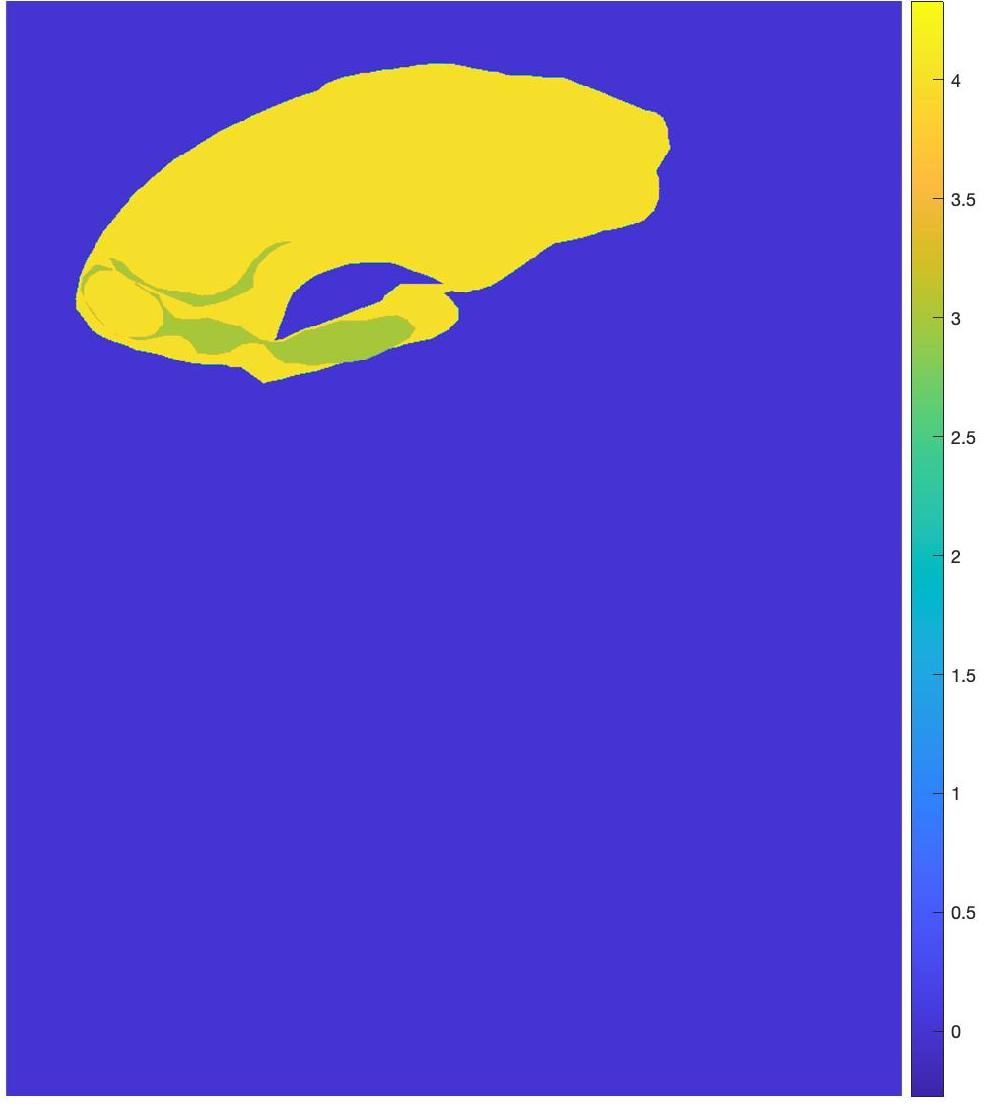} &
\includegraphics[height=2.0cm, width=1.9cm]{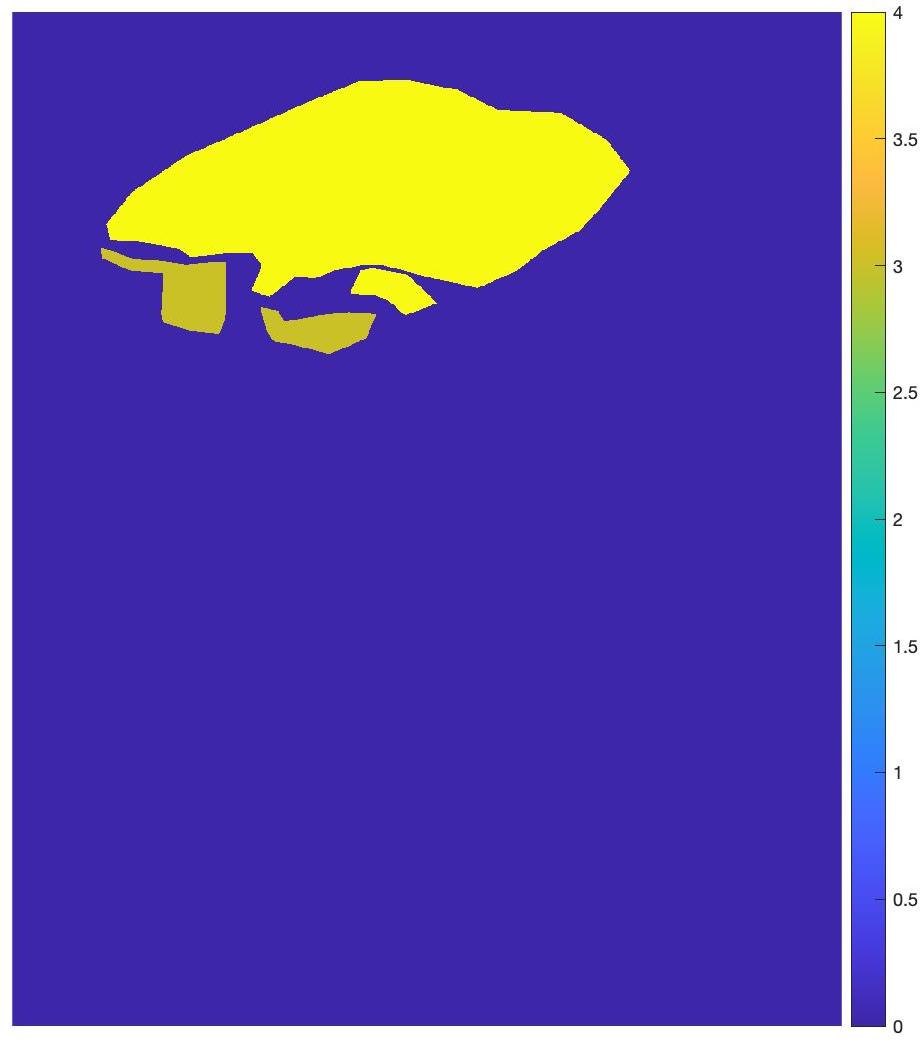} & 
\includegraphics[height=2.0cm, width=1.9cm]{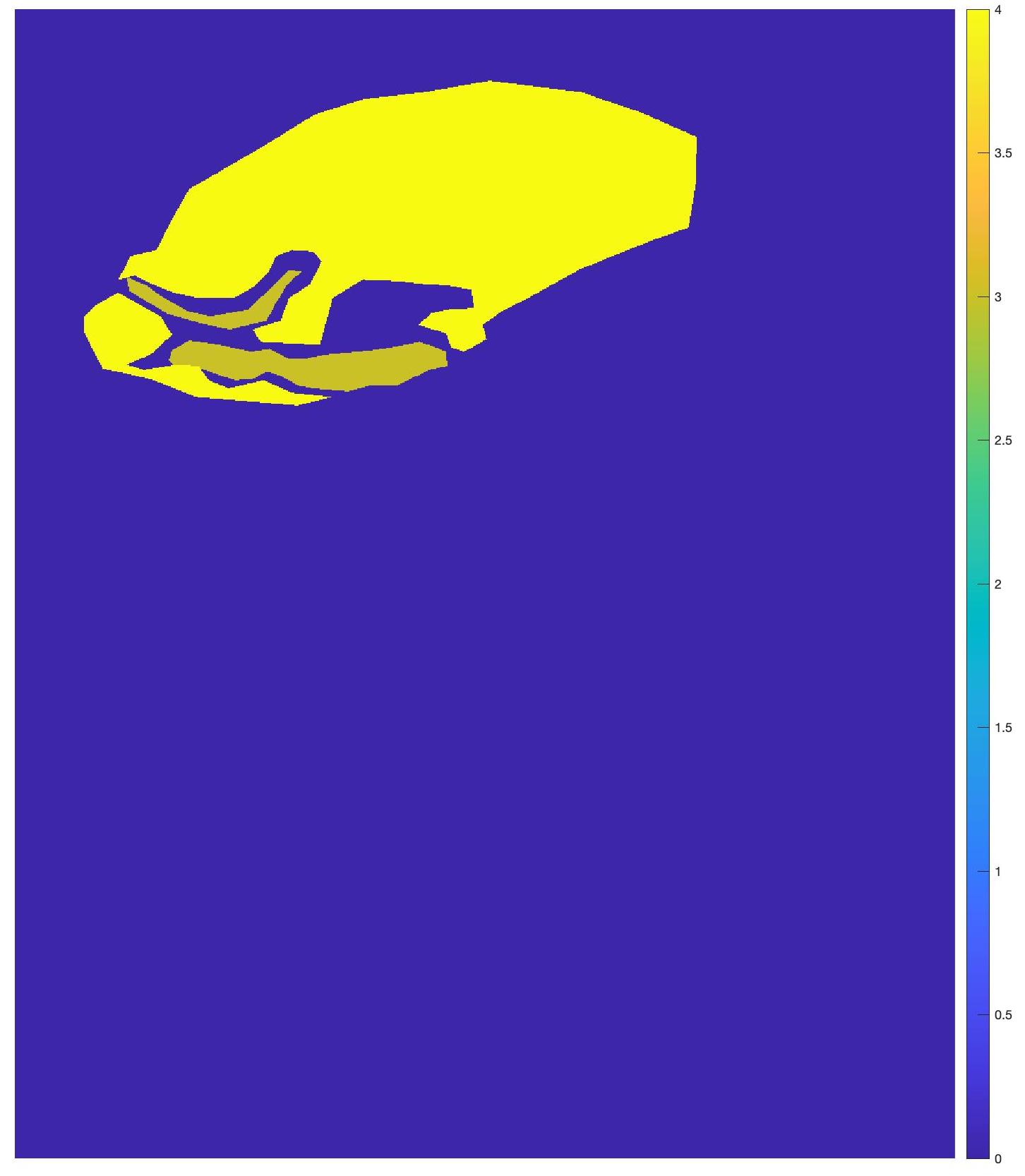} & 
\includegraphics[height=2.0cm, width=1.9cm]{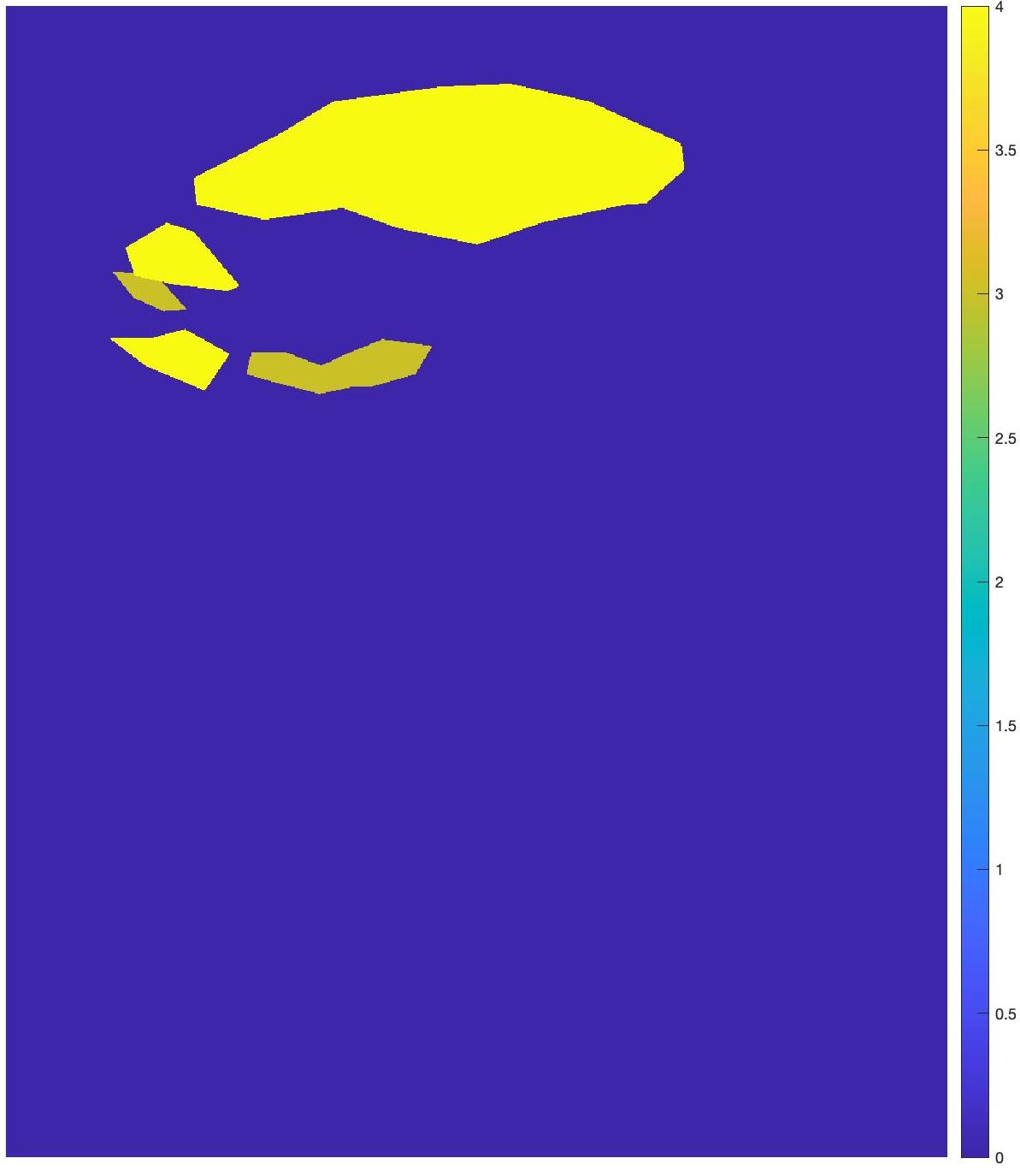}\\
%---------
\includegraphics[height=2.0cm, width=1.9cm]{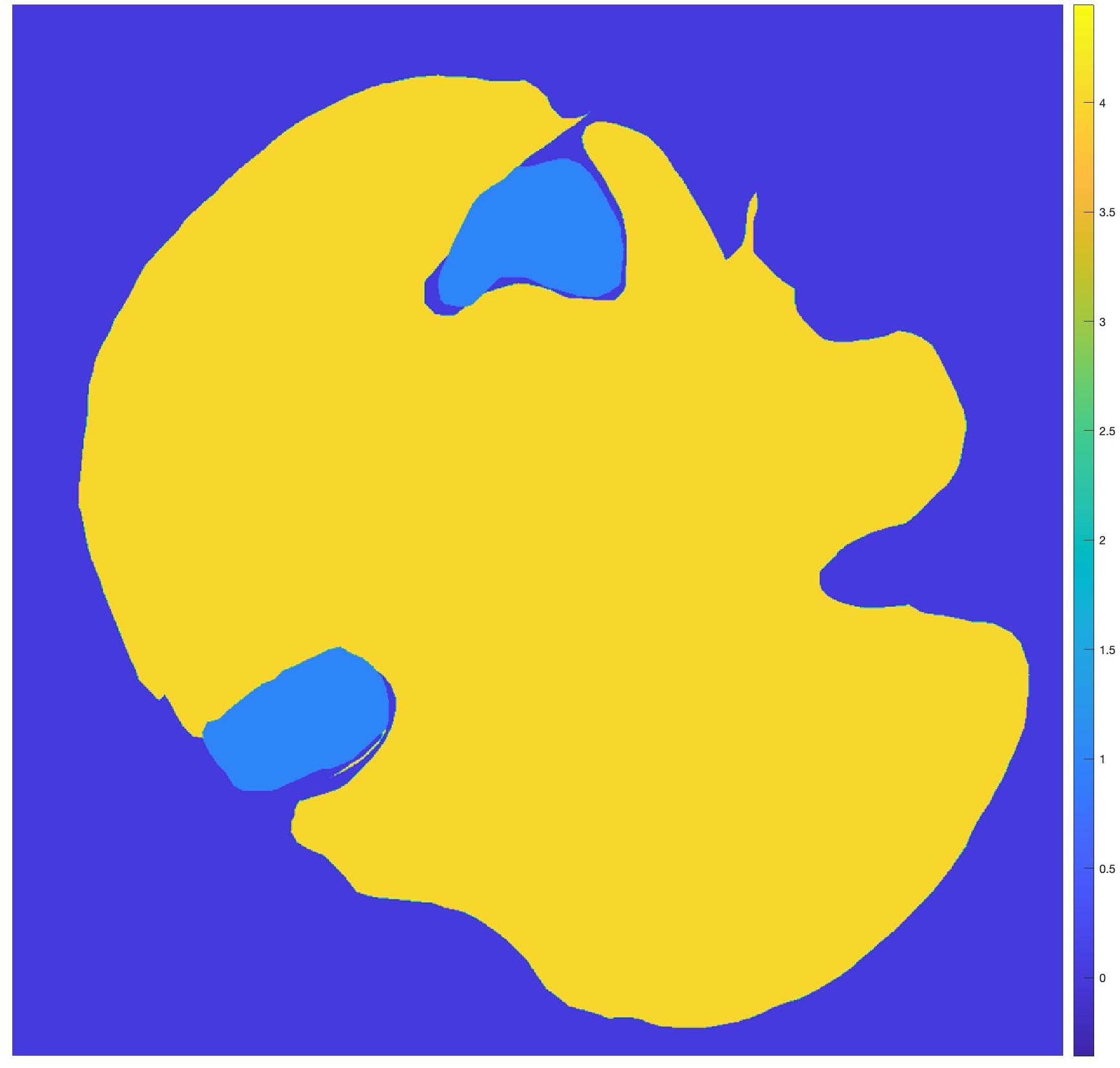} &
\includegraphics[height=2.0cm, width=1.9cm]{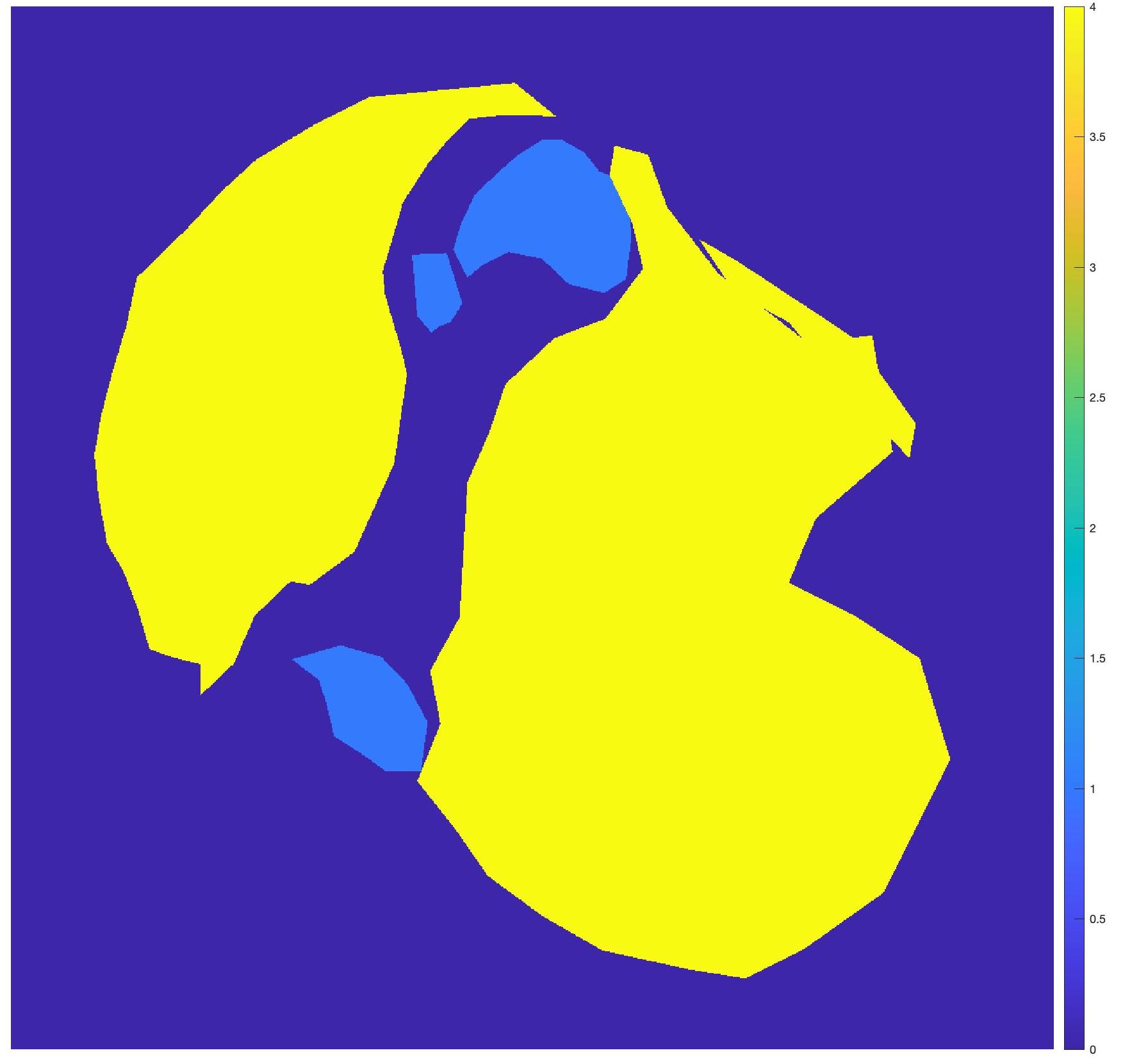} & 
\includegraphics[height=2.0cm, width=1.9cm]{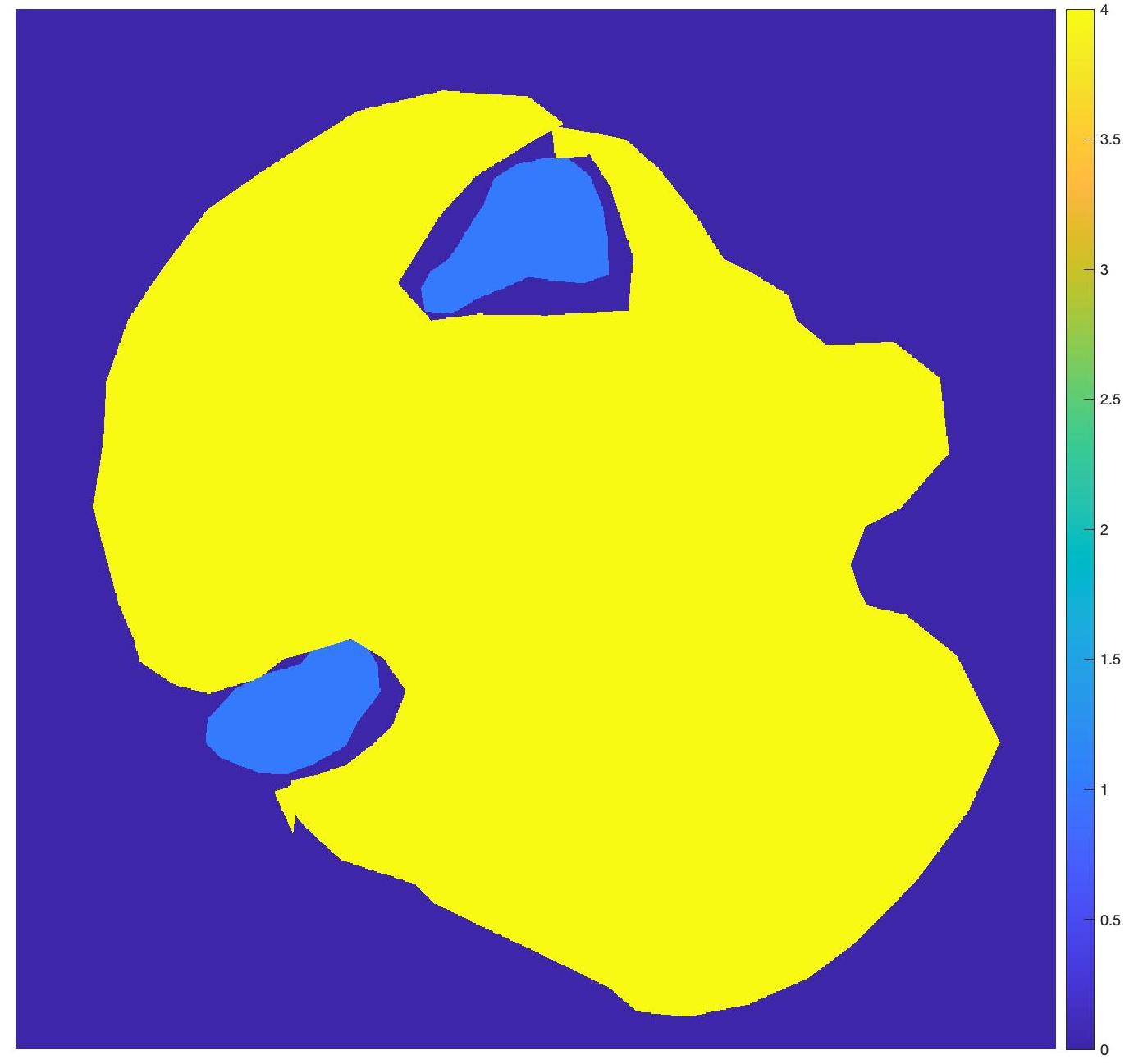} & 
\includegraphics[height=2.0cm, width=1.9cm]{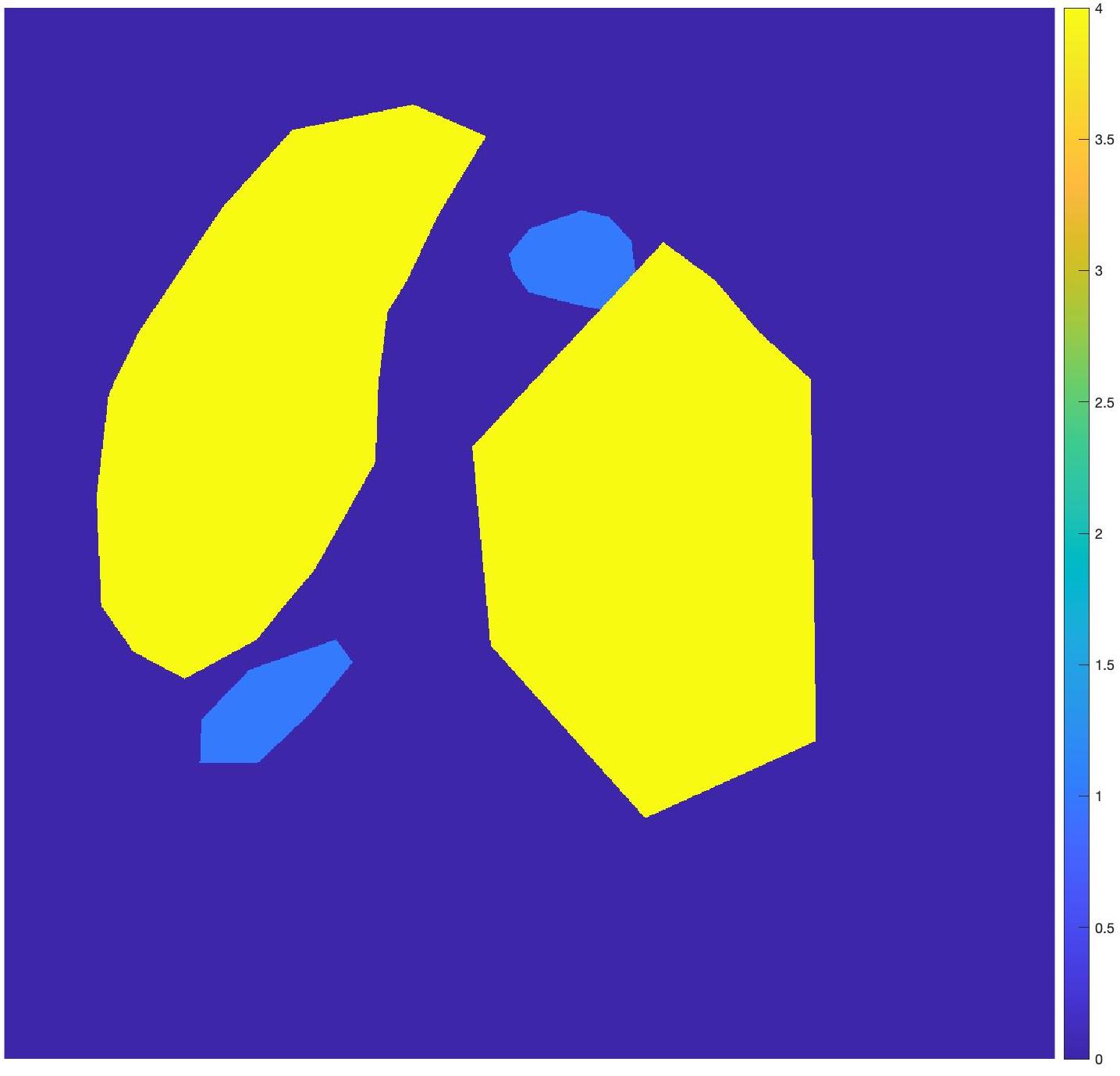}\\
(a) & (b) & (c) & (d)\\
\end{tabular}
\caption{Demonstration of UNet++'s robustness to noisy labels. (a) manual ground truth segmentation label map; (b) initial noisy label map before refinement; (c) final label map generated by our weakly-supervised approach using \textbf{UNet++}; (d) label map generated using \textbf{UNet}. The color bar indicates the corresponding labels.}
\label{fig:WSS_noise}
\end{figure}

\begin{figure}[h]
\centering
\begin{tabular}{ccc}
\includegraphics[height=2.4cm, width=2.6cm]{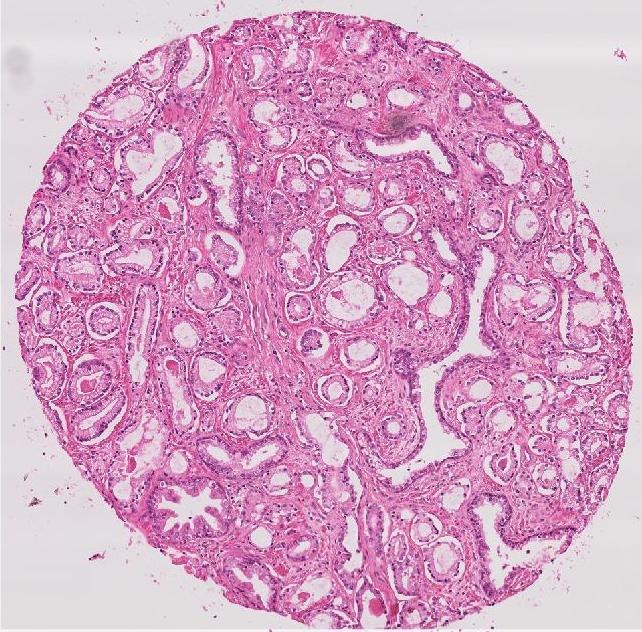} &
\includegraphics[height=2.4cm, width=2.6cm]{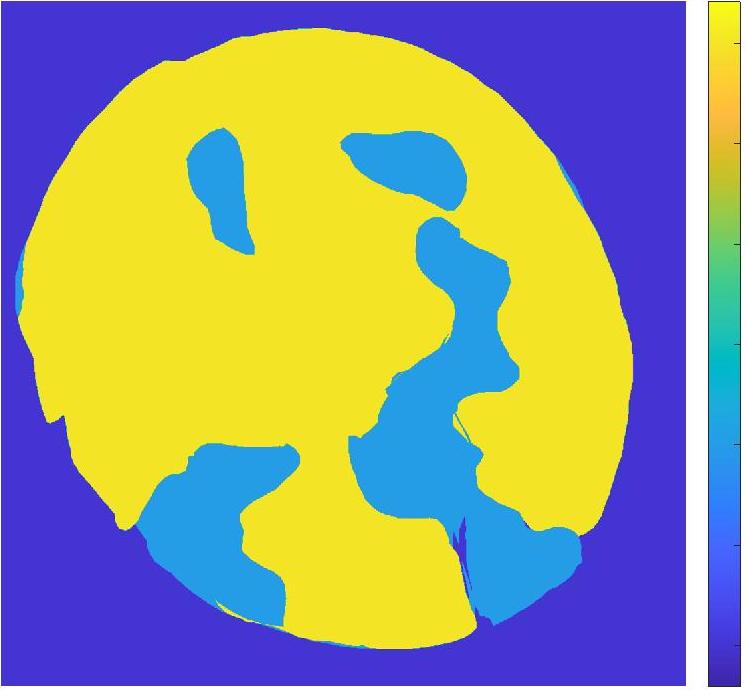} & 
\includegraphics[height=2.4cm, width=2.6cm]{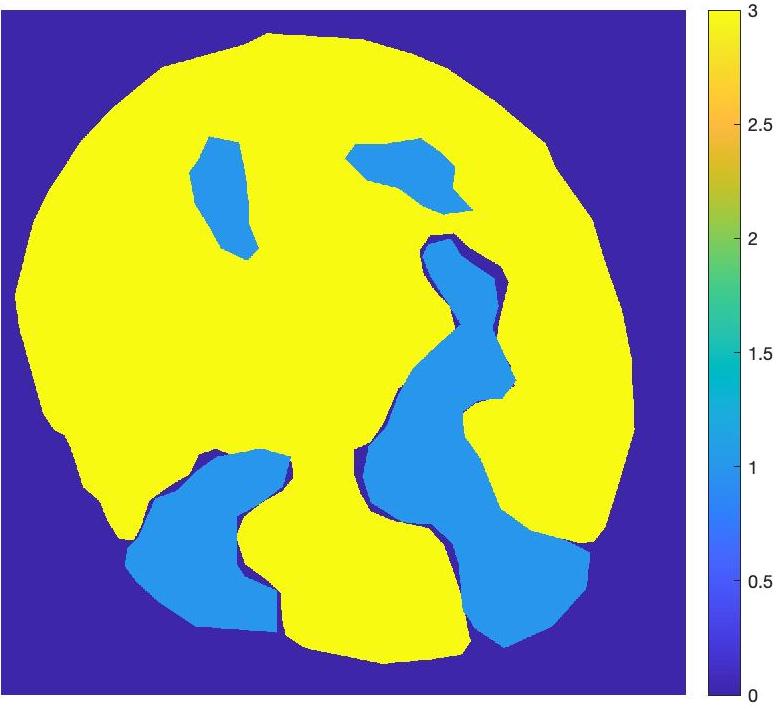} \\
%---------
\includegraphics[height=2.4cm, width=2.6cm]{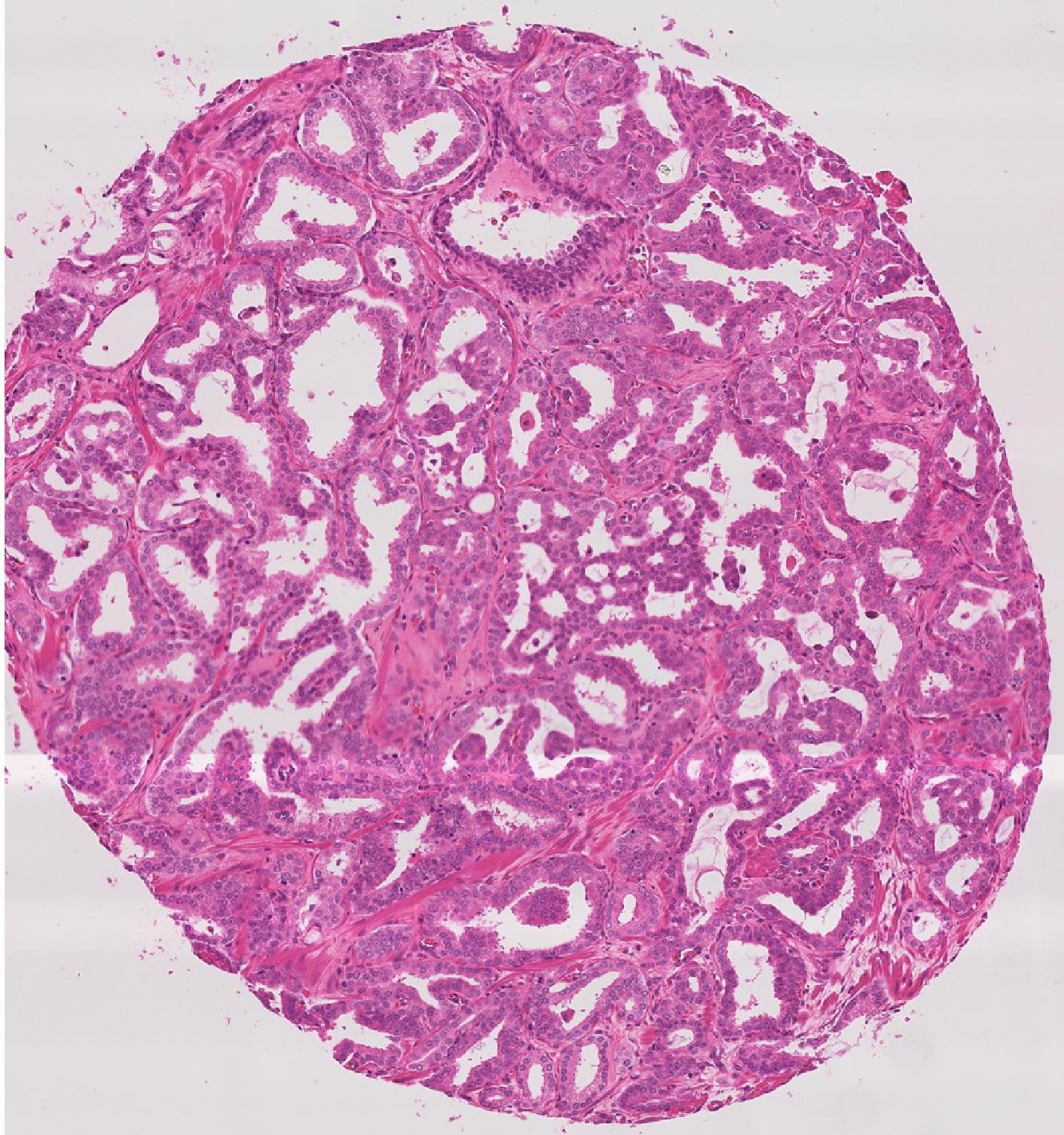} &
\includegraphics[height=2.4cm, width=2.6cm]{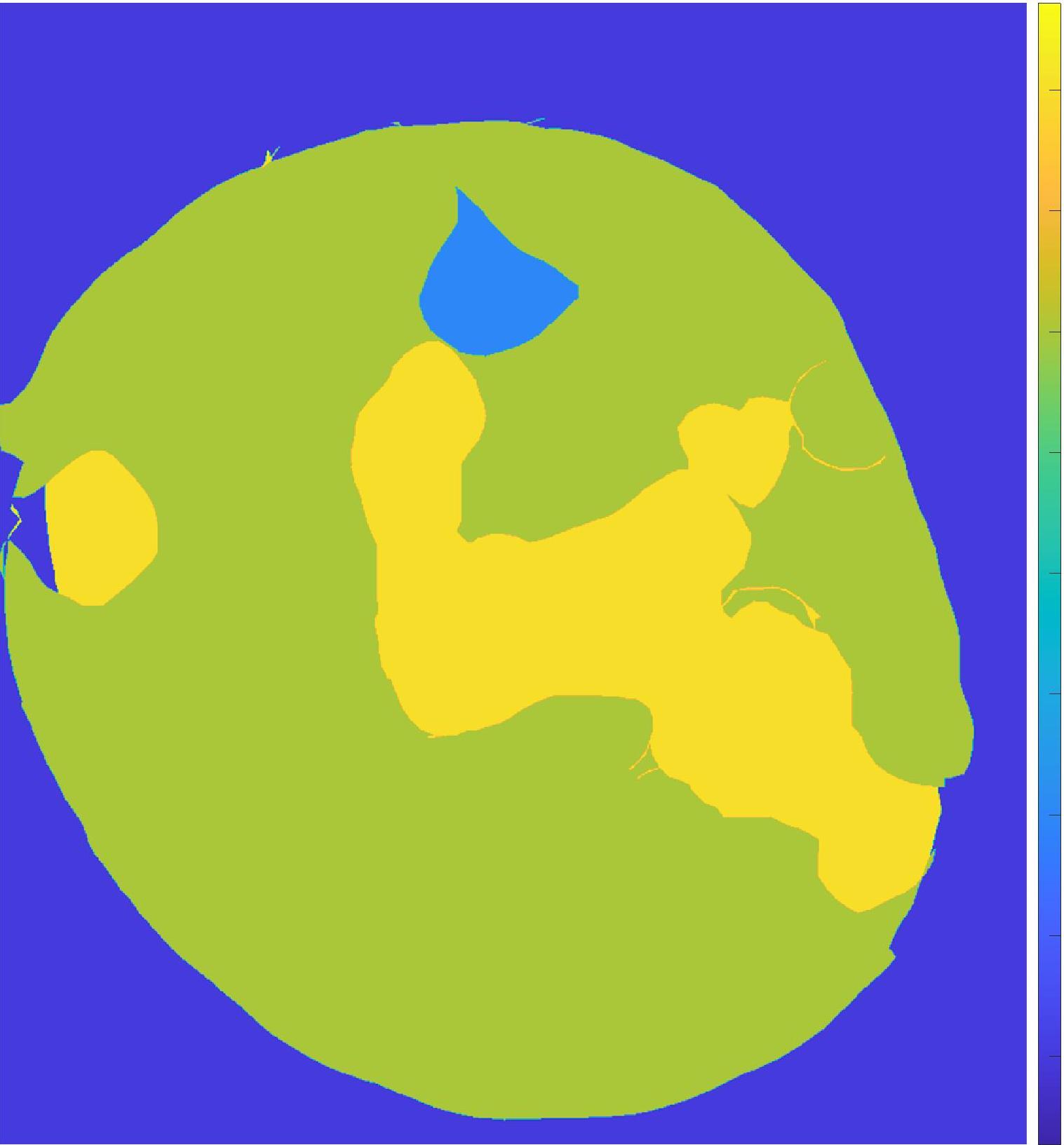} & 
\includegraphics[height=2.4cm, width=2.6cm]{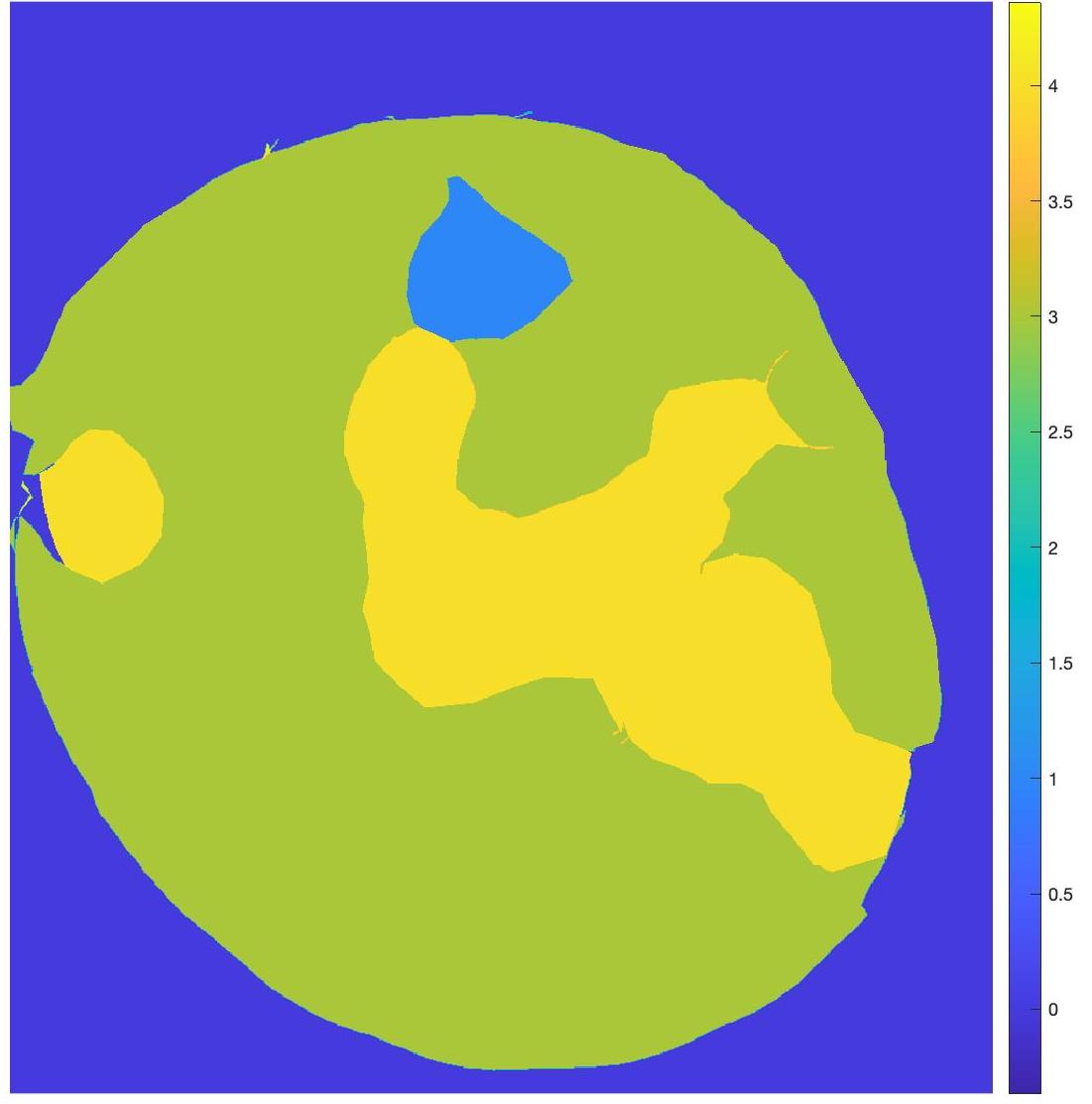} \\
(a) & (b) & (c) \\
\end{tabular}
\caption{Results for weakly supervised segmentation compared to ground-truth maps. (a) original histopathology image; (b) manual ground truth segmentation label map; (c) label map generated by our weakly-supervised approach. The color bar indicates the corresponding labels.}
\label{fig:WSS_out}
\end{figure}

Ground truth maps obtained from the semi-supervised learning step has labels for each $n\times n$ block. \snm{Consequently, there will be cases where a pixel outside the actual structure of interest shares its label since it is part of the $n\times n$ block. This erroneous labeling is referred as label noise. Our experimental results show that the UNet++ network is more robust to such label noise than the conventional UNet. 
% We ran a pair of experiments on small dataset where UNet++ was trained on manual labels and on WSSS obtained labels. The average Dice Metric for the test set was $0.932$ (manual labels) and $0.911$ (WSSS labels). The corresponding numbers for UNet was $0.914$ and $0.871$. Thus, not only does UNet++ achieves higher performance but the difference in Dice values between manual and WSSS labels ($0.021$) is lower than UNet ($0.043$). This demonstrates the greater robustness of UNet++ in overcoming label noise. 
Figure~\ref{fig:WSS_noise} shows comparative results of using UNet and UNet++ in obtaining the weak segmentation map. UNet++ proves to be more robust to initial label noise.  } %Hence there is noise in the labels which is removed by using the  trained UNet++ to predict the pixel wise labels of the training images. We observe that the UNet++ does a good job of refining the segmentation map by removing noisy \snm{classification labels. -= needs more explanation}, much better than conventional UNet \cite{Unet}.
Figure~\ref{fig:WSS_out} shows example images, their corresponding expert annotated manual segmentation maps, and the maps obtained using our method. A very high degree of agreement exists between weak supervision generated maps and the ground truth, with a Dice Metric of $0.962$ validating this observation. \snm{Without the label refinement step we obtain a Dice Metric of $0.941$, highlighting its contribution in improving segmentation performance. Using UNet we obtain a Dice metric of $0.930$ and $0.907$ (without label refinement), which indicates superior capacity of UNet++ to overcome label noise.} Thus we repose high confidence in the weakly supervised generated label maps as representing the correct labels and are reliable enough to model inter-label geometric relationship (as described in the next step).

% \begin{figure}[h]
% \centering
% \begin{tabular}{ccc}
% \includegraphics[height=2.4cm, width=2.6cm]{NoisyLabels_Manual1.jpg} &
% \includegraphics[height=2.4cm, width=2.6cm]{NoisyLabels_Noise1.jpg} & 
% \includegraphics[height=2.4cm, width=2.6cm]{NoisyLabels_WSS1.jpg} \\
% %---------
% \includegraphics[height=2.4cm, width=2.6cm]{NoisyLabels_Manual2.jpg} &
% \includegraphics[height=2.4cm, width=2.6cm]{NoisyLabels_Noise2.jpg} & 
% \includegraphics[height=2.4cm, width=2.6cm]{NoisyLabels_WSS2.jpg} \\
% (a) & (b) & (c) \\
% \end{tabular}
% \caption{Demonstration of UNet++'s robustness to noisy labels. (a) manual ground truth segmentation label map; (b) initial noisy label map before refinement; (c) final label map generated by our weakly-supervised approach. The color bar indicates the corresponding labels.}
% \label{fig:WSS_noise}
% \end{figure}

\subsection{Geometry Aware Shape Generation}

Let us denote an input image as $x$, the corresponding weakly supervised segmentation masks as $s_x$ and its label (Gleason score) as $l_x$. Our method learns to generate a new segmentation map from the base mask.  
  The first stage is a spatial transformer network (STN) \cite{STN} that transforms the base mask with different attributes of location, scale and orientation. 
 The transformations used to obtain new segmentation mask $s_x^{'}$ are applied to $x$ to get corresponding transformed image $x^{'}$.
Since the primary aim of our approach is to learn contours and other shape specific information of anatomical regions, a modified DenseUNet architecture as the generator network effectively captures hierarchical shape information and easily introduces diversity at multiple levels of image generation. %at different levels of image abstraction. 

The generator $\textbf{G}_g$ takes input $\textbf{s}_x$ and a desired label vector of output mask $c_g$ to output an affine transformation matrix $\textbf{A}$ via a STN, i.e., $\textbf{G}_g$($\textbf{s}_x, c_g) = \textbf{A}$.  
$\textbf{A}$ is used to generate  $s_x^{'}$  and  $x^{'}$. The discriminator $\textbf{D}_{class}$ determines whether output image preserves the desired label $c_g$ or not. 
The discriminator $\textbf{D}_g$ is tasked with ensuring that the generated masks are realistic. Let the minimax criteria between $\textbf{G}_g$ and $\textbf{D}_g$ be $\min_{\textbf{G}_g} \max_{\textbf{D}_g} \textbf{L}_g(\textbf{G}_g,\textbf{D}_g)$. The loss function $\textbf{L}_g$ is,% has three components
 \begin{equation}
     L_g=L_{adv} + \lambda_1{L}_{class} + \lambda_2{L}_{shape}
 \label{eqn:Tloss}
 \end{equation}
 where 1) $\textbf{L}_{adv}$ is an adversarial loss to ensure $\textbf{G}_{g}$ outputs realistic deformations; 2) $\textbf{L}_{class}$ ensures generated image has same label as $x$ and preserves it's Gleason grade; and 3) $\textbf{L}_{shape}$ ensures new masks have realistic shapes. $\lambda_1=0.95,\lambda_2=1$ balance each term's contribution. 

\paragraph{\textbf{Adversarial loss}- $\textbf{L}_{adv}(\textbf{G}_g,\textbf{D}_g)$}%\linebreak

The STN outputs $\widetilde{A}$, a prediction for $\textbf{A}$ conditioned on $\textbf{s}_x$, and a new semantic map $\textbf{s}_x \oplus \widetilde{A}(\textbf{s}_x)$ is generated. % using  composing a transformed box onto the input.
 $\textbf{L}_{adv}$ is  defined as:
% %
\begin{equation}
\begin{split}
L_{adv}(G_g,D_g) =\mathbb{E}_x\left[\log D_{g} (\textbf{s}_x \oplus \widetilde{A}(\textbf{s}_x)) \right] \\
+ \mathbb{E}_{\textbf{s}_x} \left[\log (1-D_{g} (\textbf{s}_x \oplus \widetilde{A}(\textbf{s}_x))) \right] ,
\end{split}
\label{eq:D2}
\end{equation}
% %

\paragraph{\textbf{Classification Loss}- $\textbf{L}_{class}$}

The affine transformation $\widetilde{A}$ is applied to the base image $\textbf{x}$ to obtain  the generated image $\textbf{x}^{'}$. We add an auxiliary classifier \snm{(to ensure the output image has the desired label)} when optimizing both $\textbf{G}_g$ and $\textbf{D}_g$ and define the classification loss as,
\begin{equation}
L_{class} = \mathbb{E}_{\textbf{x}^{'},c_g} [-\log D_{class}(c_g|x')],
\label{eq:sgan2}
\end{equation}
where the term $D_{class}(c_g|x')$ represents a probability distribution over classification labels computed by $D$.

\paragraph{\textbf{Shape Loss}-$\textbf{L}_{shape}$} \textbf{Self Supervised Modeling of Inter-Label Geometric Relationships} %\linebreak

We propose a novel approach to model the relationship between different anatomical labels. In \cite{Mahapatra_CVPR2020} we achieve this by training a conditional classifier to output the label of an image region given another labeled region. However we find that use of a self-supervised learning based approach is much more effective in learning contextual information than our previous approach. Here we describe our novel approach in modeling the inter-label relationship.

 Given a dataset of masks $S = \{s_1 , s_2 , \cdots, s_N \}$ consisting of N label masks, a new dataset $\hat{S} = f(S)$  is generated, where $f$ is a function altering the original label maps by arbitrarily masking labeled regions or swapping patches. A CNN is trained to predict the intensity values of the altered pixels. \snm{Here the intensity values correspond to the mask's Gleason labels. The task here is to reconstruct the masks when part of it has been altered. By learning to predict the missing labels the network implicitly learns the geometrical relationship between different anatomical labels (Gleason grades).}
%  By successfully reconstructing the original mask, the CNN thus learns the geometrical relationship between regions having different labels . 
 
%   A CNN is learned to approximate the function $g$ to model the mapping $\hat{x}_i \leftarrow x_i$ , i.e. $x_i = g(\hat{x}_i) = f^{-1} (\hat{x}_i)$ , where $i = 1 , 2 , \cdots, N$. Fig. 2 demonstrates this process on exemplar

% With the help of context restoration as a pre-text task we are able to implicitly learn the geometrical relationship between different labeled regions. 
We train an Encoder-Decoder  style network similar to UNet where the input is the altered mask and the output is the original mask, and use a $L_2$ loss term. We call this network as the Shape Restoration Network (ShaRe-Net). To compute $L_{shape}$ we obtain the feature maps of the generated mask ($SN(G_g(\textbf{s}_x))$ and original mask $SN(\textbf{s}_x)$ using all layers of ShaRe-Net, and calculate the mean square error values  between them.
\begin{equation}
L_{shape} = \frac{1}{N} \sum_{i}^{N} \left( SN(\textbf{s}_x) - SN(G_g(\textbf{s}_x)) \right)^{2}
\label{eq:sgan3}
\end{equation}
$SN(\cdot)$ denotes the input processed through ShaRe-Net.
%  
% The probability value is determined from a pre-trained modified VGG16 architecture to compute $L_{shape}$ where the input has two separate maps corresponding to the label pair. Each map's foreground has only the region of the corresponding label and other labels considered background.  The conditional probability between the pair of label maps enables the classifier to implicitly capture geometrical relationships and volume information between the label pair without the need to define explicit features. The geometric relation between different labels will vary for infected and non-infected cases, which is effectively captured by our approach.

%
% To get the pre-trained VGG16 network we used a separate dataset of $24$ images with its WSS generated segmentation maps. %manually annotated images. %In Section~\ref{expt:ablation2} we analyze the effect of using manually annotated images versus using the segmentation maps obtained from the WSS step.

% Since we are proposing a weakly supervised segmentation based approach and work with the assumption that we do not have manually labeled masks, we 
%  

\subsection{Sample Diversity From Uncertainty Sampling}

Our approach based on uncertainty sampling is inspired from \cite{PhiSeg}. However we also introduce the  novelty of using a Dense UNet architecture \cite{DenseUNet} instead of UNet. This allows us to reuse information from previous layers which introduces complementary information across layers. This enables the generation step to factor in the dependence across different levels of shape abstraction.
% \begin{enumerate}
%     \item We use dense connections with a  Dense UNet architecture \cite{DenseUNet} instead of UNet. This allows us to reuse information from previous layers which introduces a  level of complementary information across layers. This enables the generation step to factor in the dependence across different levels of shape abstraction.

% \end{enumerate}

The generated mask $s_x'$ is obtained by fusing $L$ levels of the generator $G_g$ (as shown in Figure~\ref{fig:workflow}), each of which is associated with a latent variable $z_l$. We use probabilistic uncertainty sampling to model conditional distribution of segmentation masks and use separate latent variables at multi-resolutions to factor inherent uncertainties.
The hierarchical approach introduces diversity at different stages and influences different features (e.g., low level features at the early layers and abstract features in the later layers). Denoting the generated mask as $s$ for simplicity, we obtain conditional distribution $p(s|x)$ for $L$ latent levels as:
\begin{equation}
\begin{split}
p(s|x) = \int p(s|z_1 , \cdots, z_L )p(z_1 |z_2 , x) \cdots \\ 
     p(z_{L-1} |z_L , x)p(z_L |x) dz_1 \cdots dz_L . 
\end{split}
\label{eq:prob1}
\end{equation}

Latent variable $z_l$  models diversity at resolution  $2^{-l+1}$ of the original image (e.g. $z_1$ and $z_3$ denote the original and $1/4$ image resolution). 
A variational approximation $q(z|s, x)$ approximates the posterior distribution $p(z|s,x)$  where $z=\{z_1 , . . . , z_L\}$. $ \log p(s|x) = L(s|x) + KL(q(z|s, x)||p(z|s, x))$, where $L$ is the evidence lower bound, and $KL(.,.)$ is the Kullback-Leibler divergence. %Since $KL(·, ·) \geq 0$, $L$ is a lower bound on the conditional log probability when the approximation $q$ exactly matches the posterior. 
 The prior and posterior distributions are parameterized as normal distributions $\mathcal{N}(z|\mu,\sigma)$. %Thus, we define

Figure~\ref{fig:workflow}  shows example implementation for $L=3$. We use $6$ resolution levels and $L = 4$ latent levels.
Figure~\ref{fig:workflow}  shows the latent variables $z_l$ forming skip connections such that information between the image and segmentation output goes through a sampling step. The latent variables \emph{are not mapped} to a 1-D vector to preserve the structural  relationship between them, and this substantially improves segmentation accuracy. 
$z_l$'s dimensionality is $r_x 2^{-l+1} \times r_y 2^{ -l+1}$, where $r_x$ , $r_y$ are image dimensions. %$z_l$  models image data at $2^{-l+1}$ of original resolution due to downsampling operations and transmits the learned representation to the latent space embedding above ($z_{l-1}$). 

%
%

% this file describes the experiments and results for the paper

\section{Experimental Results}
\label{sec:expt}

\subsection{Dataset Description}
 We use the public Gleason grading challenge dataset \footnote{https://gleason2019.grand-challenge.org/Home} \cite{GleasonData}.
  A total of 333  Tissue
Microarrays (TMAs) from 231 patients who had undergone radical prostatectomy
were used for this study. The TMAs were prepared
 following the standard procedures at the
Vancouver Prostate Centre in Vancouver, Canada and
was approved by the institutional Clinical Research Ethics
Board. The TMAs had been stained
in $H\&E$ and scanned at 40x magnification with a SCN400
Slide Scanner (Leica Microsystems, Wetzlar, Germany). The
digitized TMA images were loaded onto a Samsung Galaxy
Tab A6 Tablet (Samsung Electronics, Yeongtong, Suwon,
South Korea). An Android-based application was used for
annotation of the images with a stylus. Six pathologists were
asked to annotate the TMA images in detail. The pathologists
had $27, 15, 1, 24, 17$, and $5$ years of experience. Four
of the pathologists annotated all $333$ cores. The other two
pathologists annotated $191$ and $92$ of the cores.
Pixel-wise majority voting was used to construct the “ground truth
label”. %An example of a TMA core image and the detailed
%annotations provided by the pathologists is shown in Figure~\ref{fig:annotations}. 
The training set had $200$ TMAs while the validation set had $44$ TMAs. A separate test set consisting of $87$  TMAs from 60
other patients was used.
Patients were randomly assigned to training and test set.
 
%  \begin{figure}[h]
% \centering
% \begin{tabular}{c}
% \includegraphics[height=4.6cm, width=8.6cm]{Annotations.png} \\
% \end{tabular}
% \caption{Workflow of annotation generation taken from the challenge website at https:gleason2019.grand-challenge.org/Home } 
% \label{fig:annotations}
% \end{figure}

\subsection{Experimental Setup, Baselines and Metrics}

Our method  has the following steps: 1) Use the default training, validation, and test folds of the dataset. 2) Use training images to train the image generator. 3) Generate shapes from the training set and train UNet++ segmentation network \cite{UNet++} on the generated images. 4) Use trained UNet++ to segment test images. 5) Repeat the above steps for different data augmentation methods. 
%a NVIDIA Titan X GPU having $12$ GB RAM, 
Our model is implemented in PyTorch, on a NVIDIA TITAN X GPU.
 We trained all models using Adam optimiser \cite{Adam} with a learning rate of $10^{-3}$, batch-size of $16$ and using batch-normalisation. % .
 The values of parameters $\lambda_1$ and $\lambda_2$ in Eqn.~\ref{eqn:Tloss} were set by a detailed grid search on a separate dataset of $45$ TMAs  that was not used for training or testing. They were varied between $[0,1]$ in steps of $0.05$ by fixing $\lambda_1$ and varying $\lambda_2$ for the whole range. This was repeated for all values of $\lambda_1$. The best segmentation accuracy was obtained for $\lambda_1=0.97$ and $\lambda_2=0.93$, which were our final parameter values.

We denote our method as  GeoGAN$_{WSS}$ (Geometry Aware GANs using weakly supervised segmentation), and compare it's performance against other methods such as: 
\begin{enumerate}
    \item $DA$- conventional data augmentation consisting of  rotation, translation and scaling. 
    \item $DAGAN$ - data augmentation GANs of \cite{DAGAN}. 
    \item $cGAN$ - the conditional GAN based method of \cite{Mahapatra_MICCAI2018}.
    \item $HGAN$- pathology image augmentation method of \cite{HGAN}.
    \item the image enrichment method of \cite{Gupta}.
    \item GeoGAN$_{Manual}$ - Geometry Aware GANs using the manual segmentation maps for image synthesis.
    \item GeoGAN$_{Cond}$ - GeoGAN$_{WSS}$ using the conditional classifier of \cite{Mahapatra_CVPR2020} instead of self supervised learning for modeling geometrical relationship between labels.
    \end{enumerate}
%
%  %
Segmentation performance on the training set is evaluated in terms of Dice Metric (DM), Hausdorff Distance (HD) and Mean Absolute error (MAE). %DM of $1$ indicates perfect overlap and $0$ indicates no overlap, while lower values of MAE indicate better segmentation performance.
The ranking for the competition was based on a combination of Cohen's kappa and the F1-scores using the formula: 
\begin{equation}
% \begin{split}
    score= \text{Cohen's kappa} + \frac{F1_{macro} + F1_{micro}} {2}
% \end{split}
\label{eq:score}
\end{equation}

The overall score is the average of all test image scores. Rankings were based on test images. Since we do not have access to the manual segmentations of the test set, we cannot report their corresponding DM, HD and MAE values.
%

% \subsubsection{Ablation Studies}
\textbf{Ablation Experiments:}
The following variants of our method were used for ablation experiments: 
\begin{enumerate}
    \item  GeoGAN$_{noL_{class}}$- GeoGAN$_{WSS}$ without classification loss (Eqn.\ref{eq:sgan2}).
    \item GeoGAN$_{noL_{shape}}$- GeoGAN$_{WSS}$ without shape relationship  modeling term (Eqn.\ref{eq:sgan3}).
    \item GeoGAN$_{NoSamp}$ - GeoGAN$_{WSS}$ without  uncertainty sampling for injecting diversity. % to determine sampling's relevance to the final network performance. 
    %
    %The original and upsampled images are directly connected without any sampling step.
    %
    %
    % \item  GeoGAN$_{L_{class}}$ - GeoGAN using classification loss (Eqn.\ref{eq:sgan2}) and adversarial loss (Eqn.\ref{eq:D2}) to determine $L_{class}$'s  relevance to GeoGAN's performance.
    % \item GeoGAN$_{L_{shape}}$ - GeoGAN using shape loss (Eqn.\ref{eq:sgan3}) and adversarial loss (Eqn.\ref{eq:D2}) to determine $L_{shape}$'s contribution to GeoGAN's performance.
    % \item GeoGAN$_{Samp}$ - GeoGAN using only uncertainty sampling and adversarial loss (Eqn.\ref{eq:D2}). This baseline quantifies the contribution of sampling to the image generation process. 
\end{enumerate}

\subsection{Segmentation Results on Gleason Training Data}
\label{expt:seg}

A suitable image augmentation method should capture the different complex relationships between various labels, with the generated images leading to improvement in segmentation accuracy. 
Table~\ref{tab:GleasonTr} shows the average DM, HD, and MAE for different augmentation methods on the Gleason challenge training dataset. Table~\ref{tab:GleasonTr} also shows the $p$ values comparing the results of all methods with GeoGAN$_{WSS}$. Results of GeoGAN$_{Manual}$ denote the best performance obtained with a given network since they are trained on the clinician provided manual segmentation maps. GeoGAN$_{WSS}$'s results show that the WSSS component is very accurate in obtaining semantic segmentation and can be used effectively where manual segmentation maps are unavailable.

\begin{figure*}[!t]
\centering
\begin{tabular}{ccccccc}
\includegraphics[height=2.5cm, width=2.2cm]{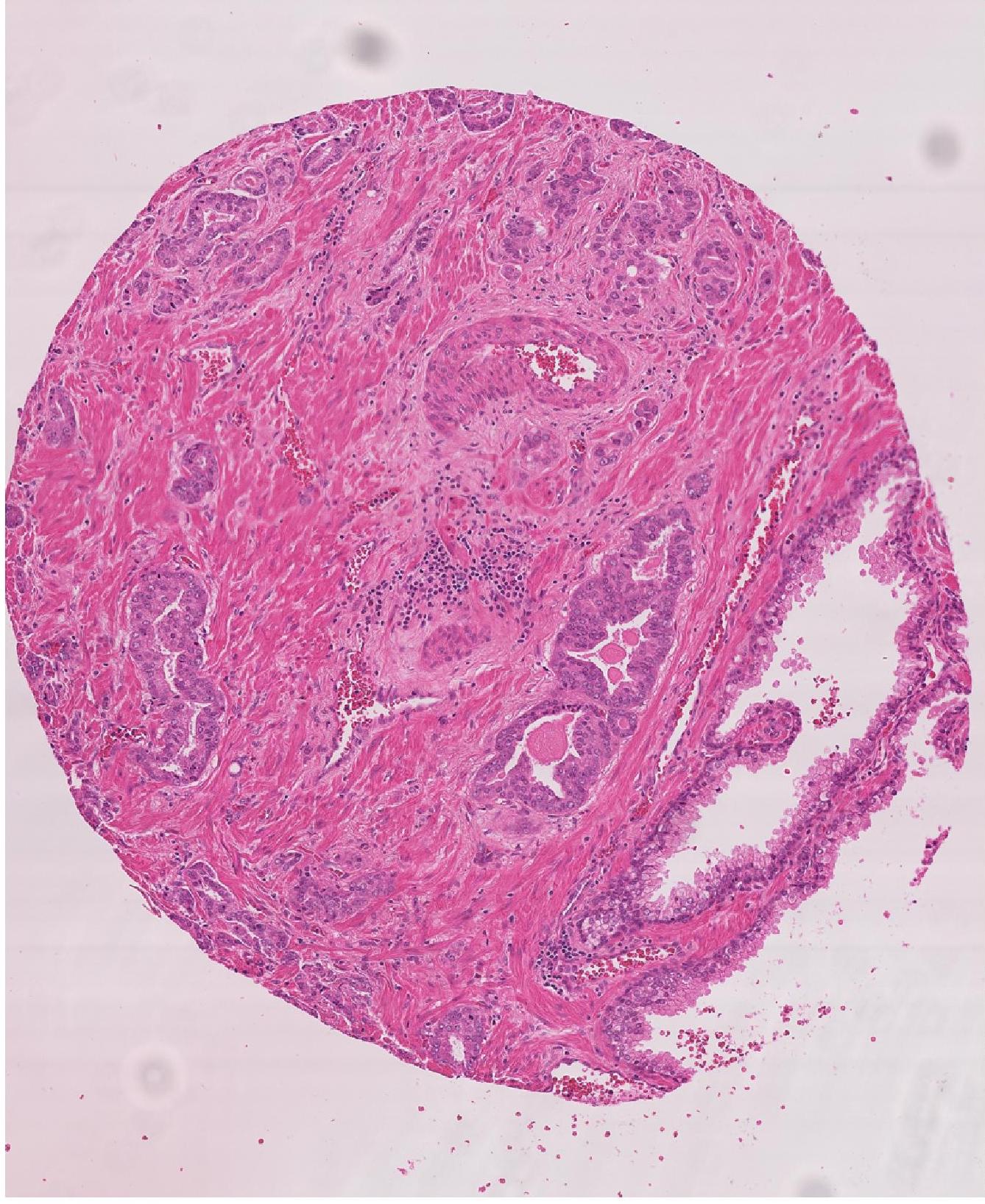} & 
\includegraphics[height=2.5cm, width=2.2cm]{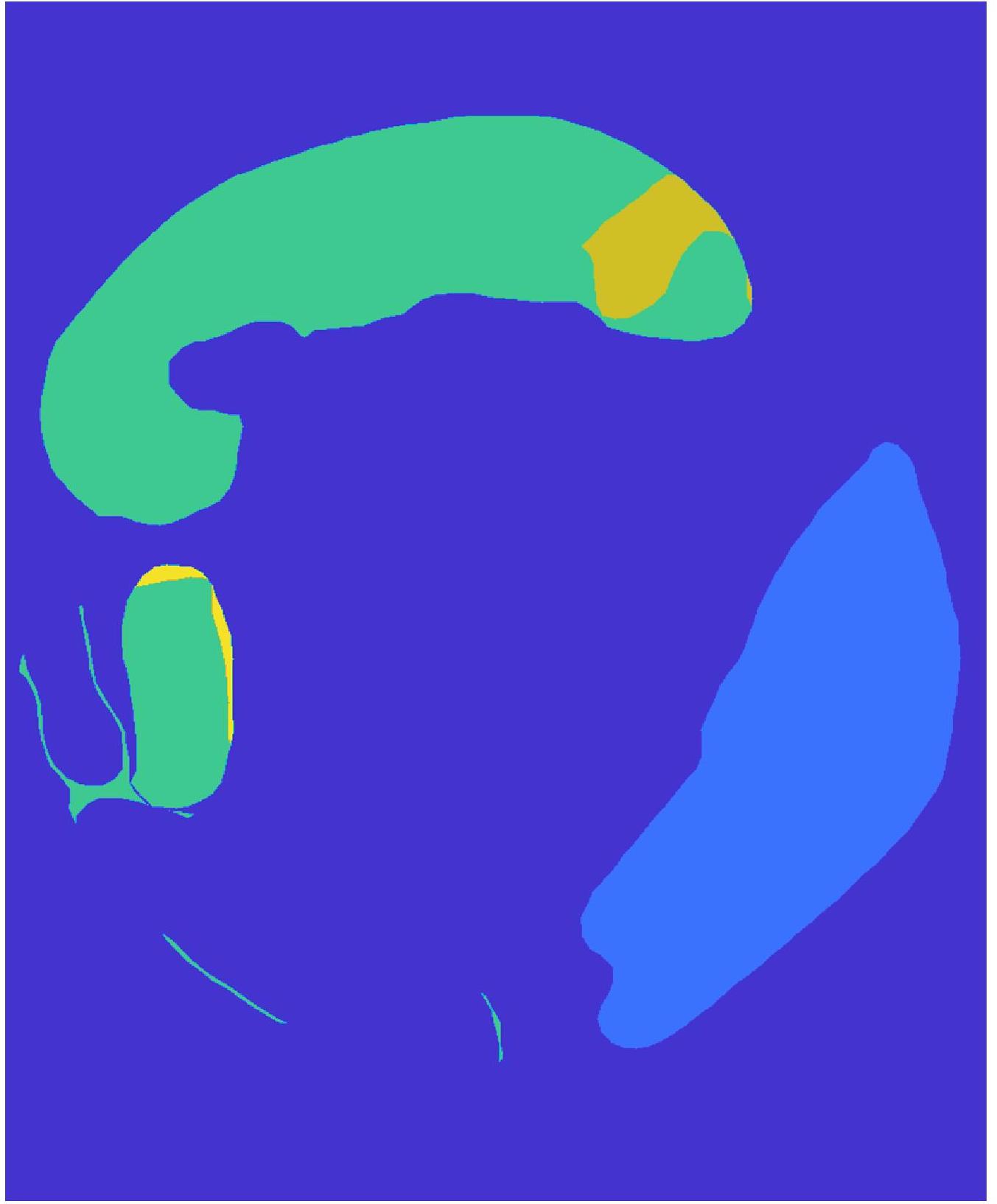} & 
\includegraphics[height=2.5cm, width=2.2cm]{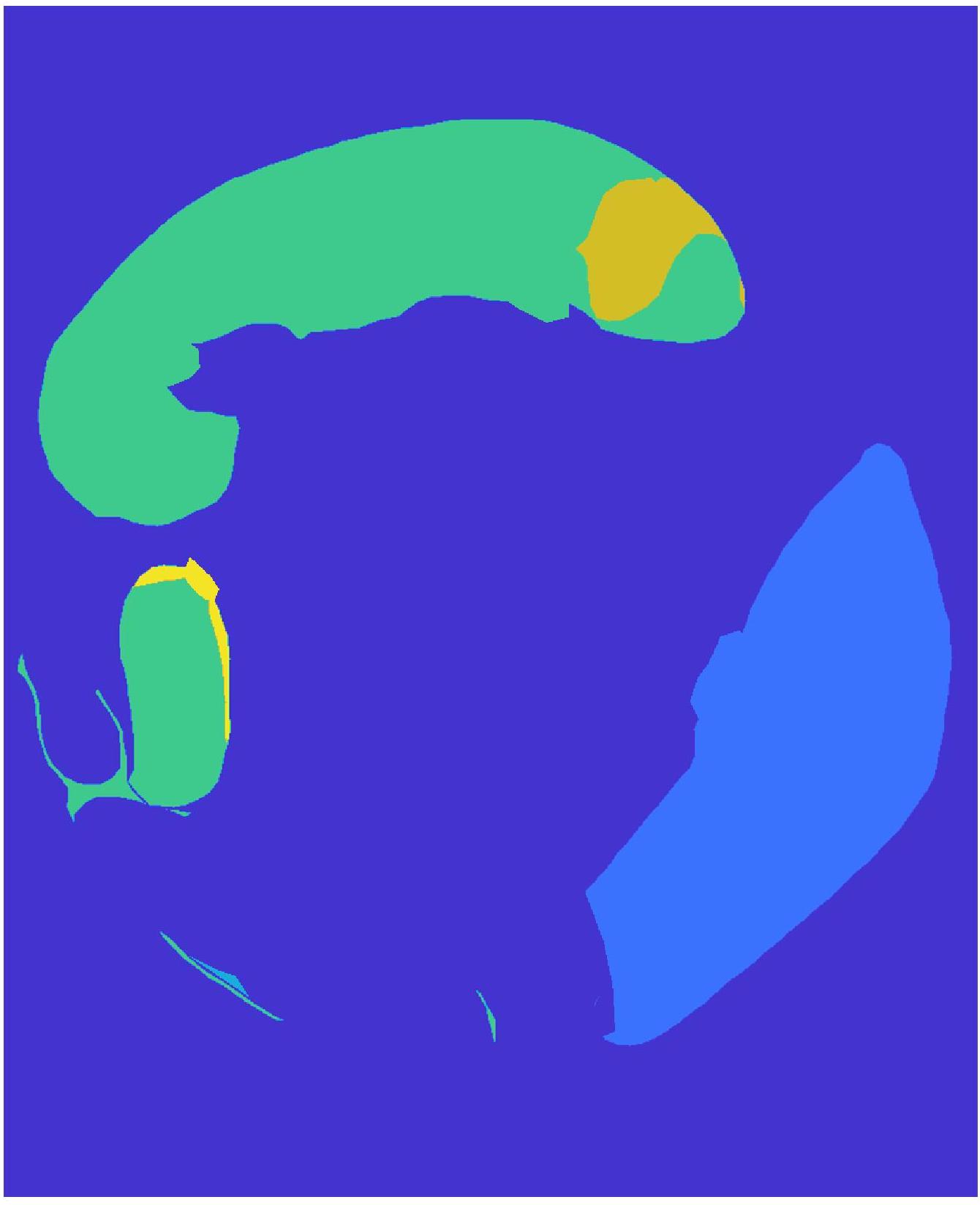} & 
\includegraphics[height=2.5cm, width=2.2cm]{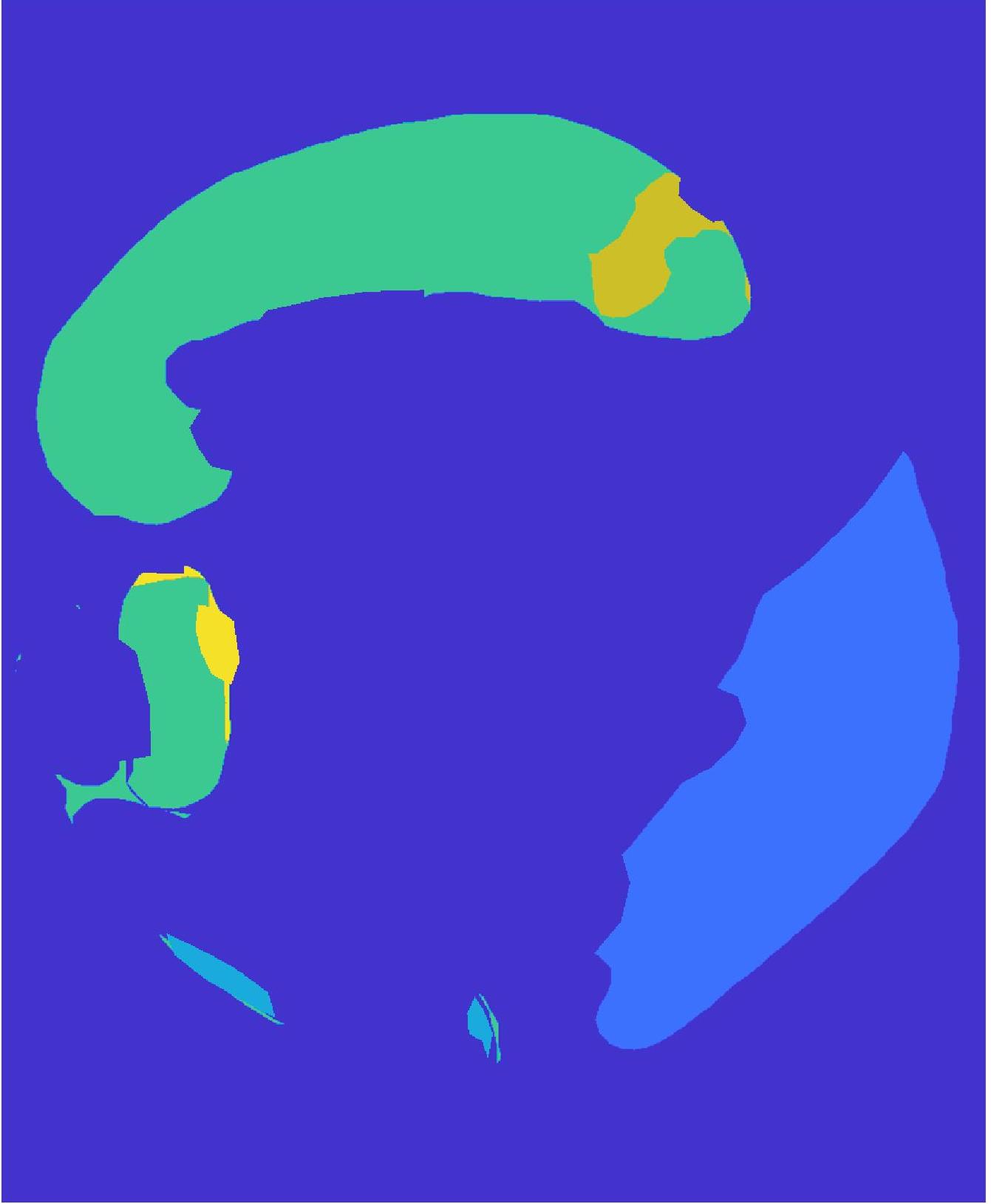} & 
\includegraphics[height=2.5cm, width=2.2cm]{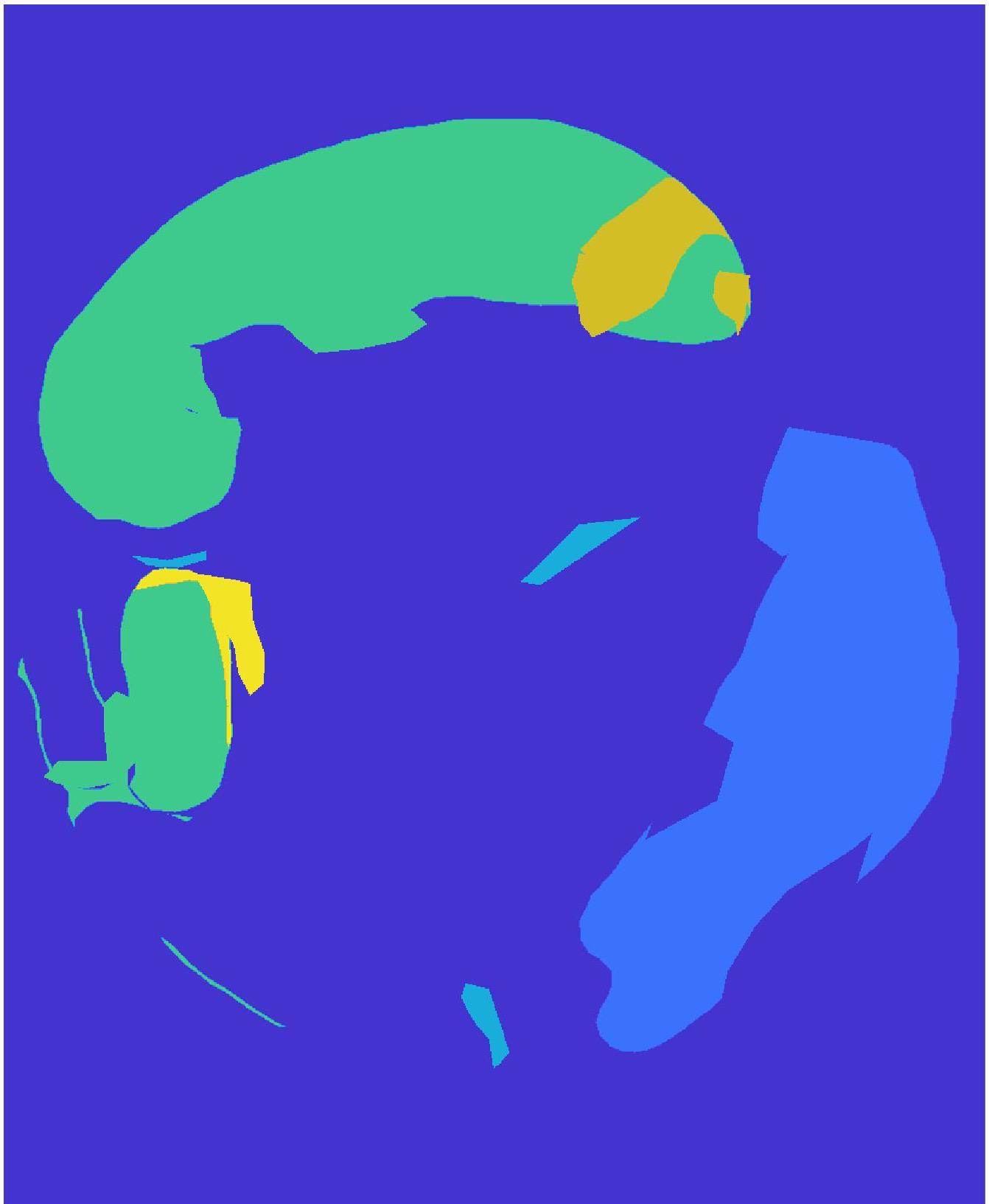} &
\includegraphics[height=2.5cm, width=2.2cm]{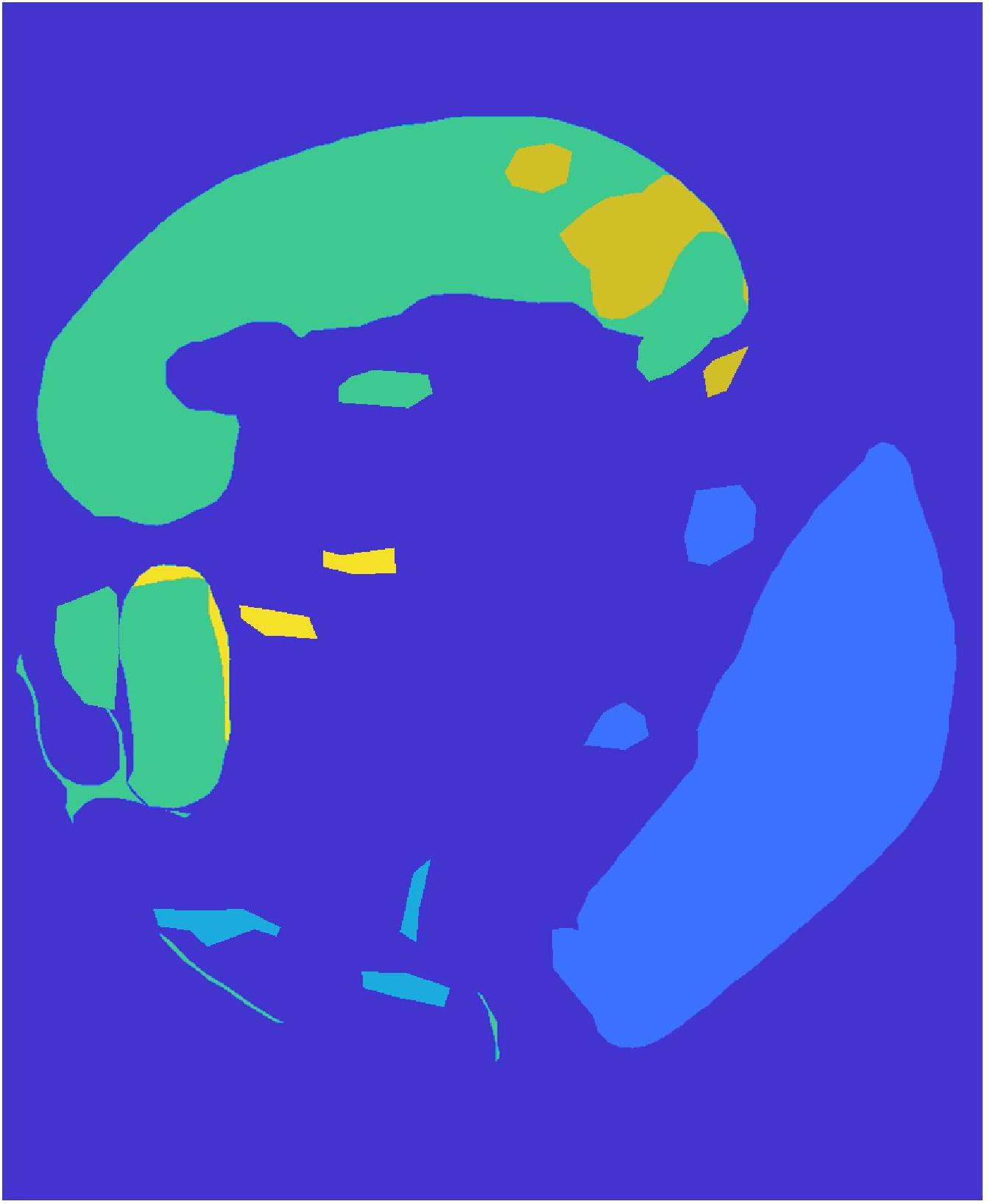} & 
\includegraphics[height=2.5cm, width=2.2cm]{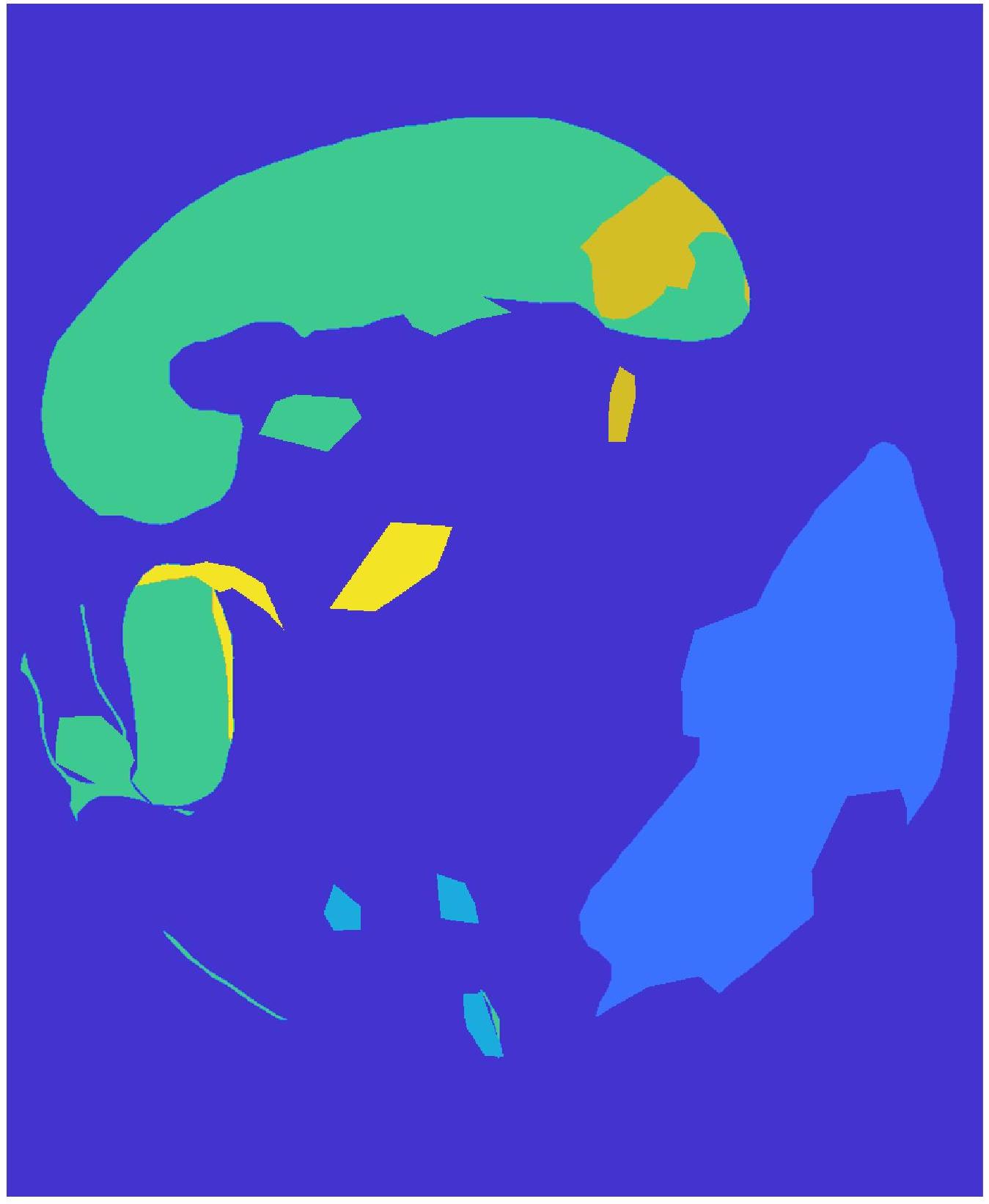} \\
%---------
\includegraphics[height=2.5cm, width=2.2cm]{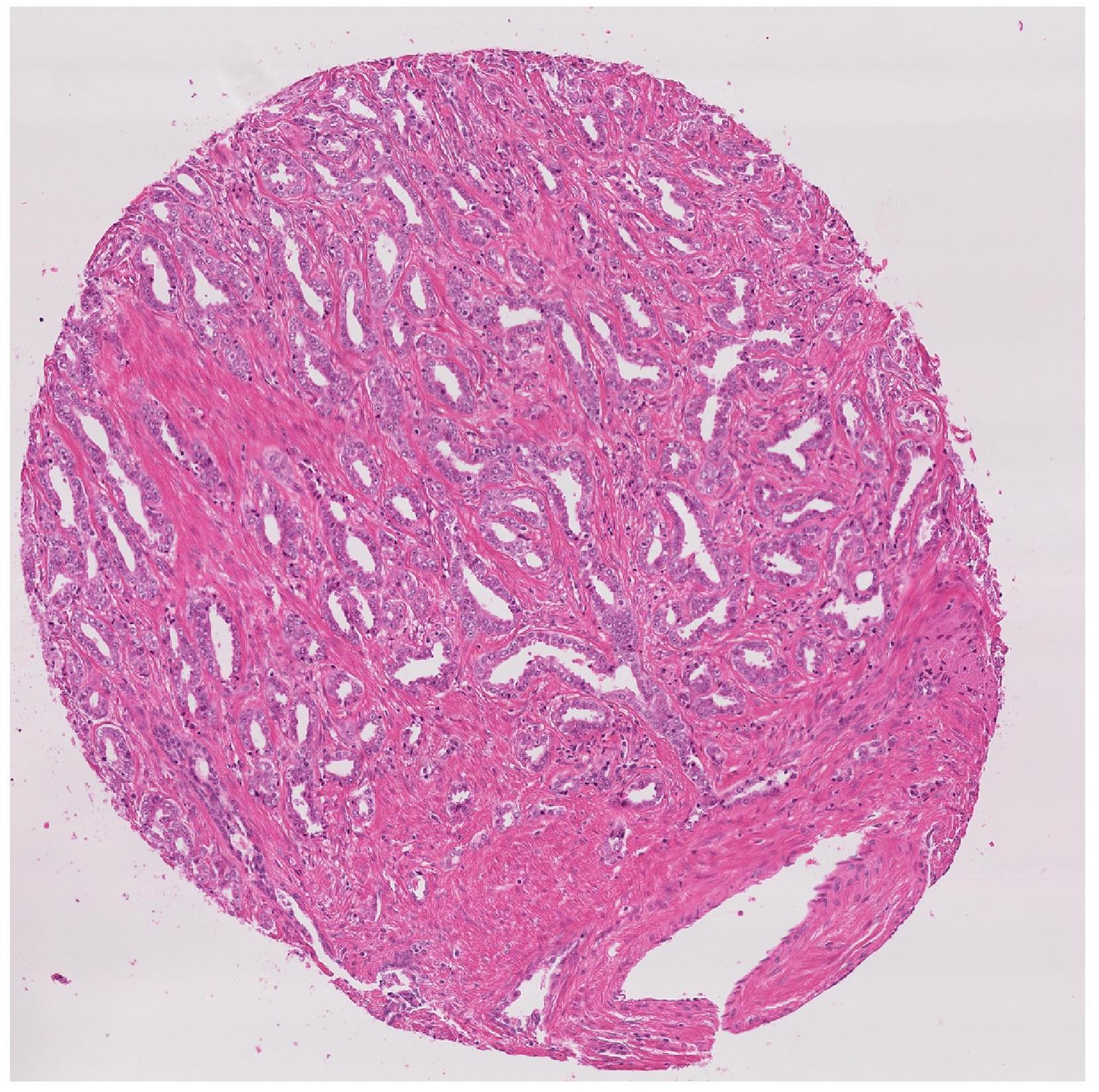} & 
\includegraphics[height=2.5cm, width=2.2cm]{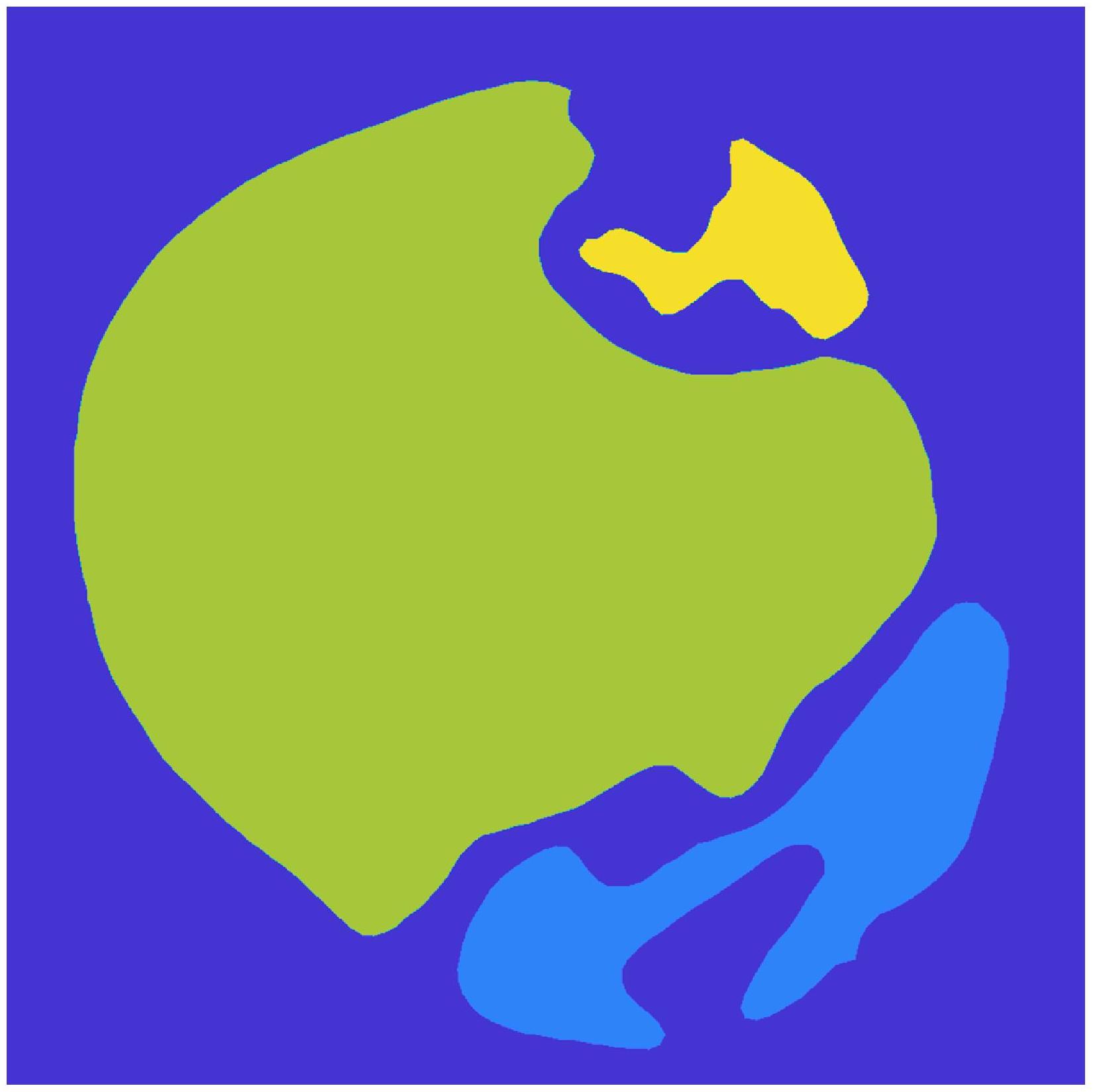} & 
\includegraphics[height=2.5cm, width=2.2cm]{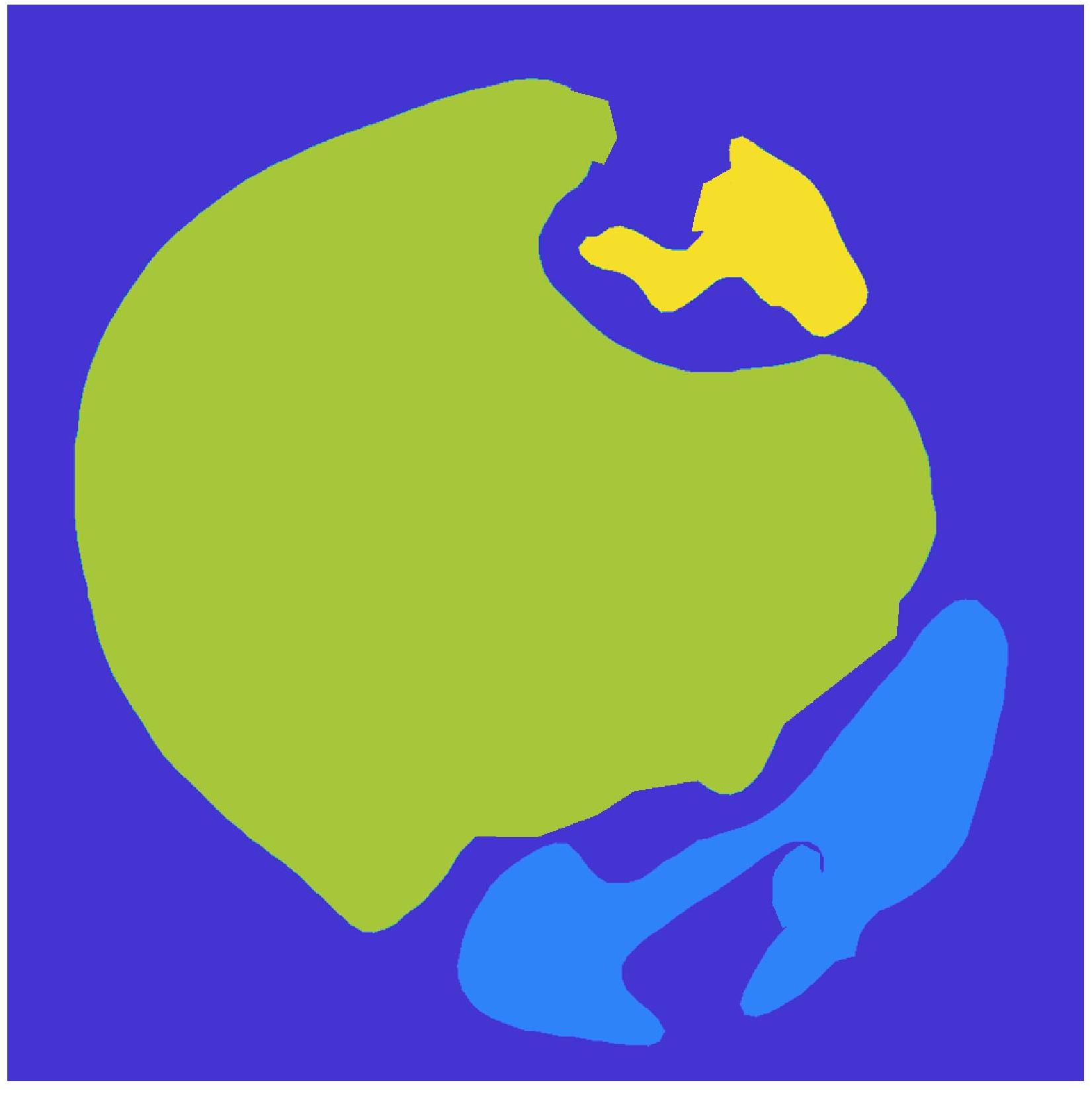} & 
\includegraphics[height=2.5cm, width=2.2cm]{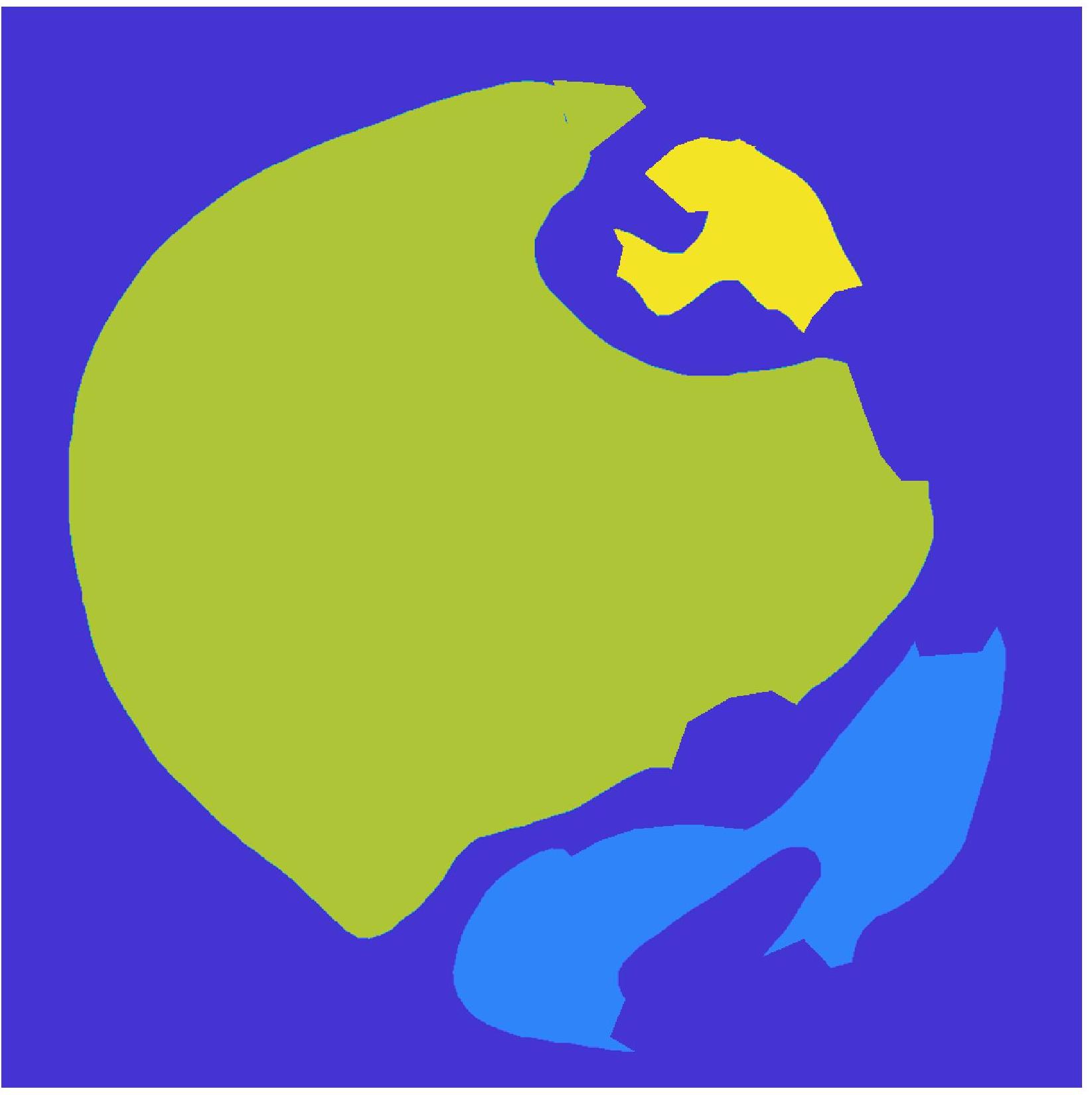} & 
\includegraphics[height=2.5cm, width=2.2cm]{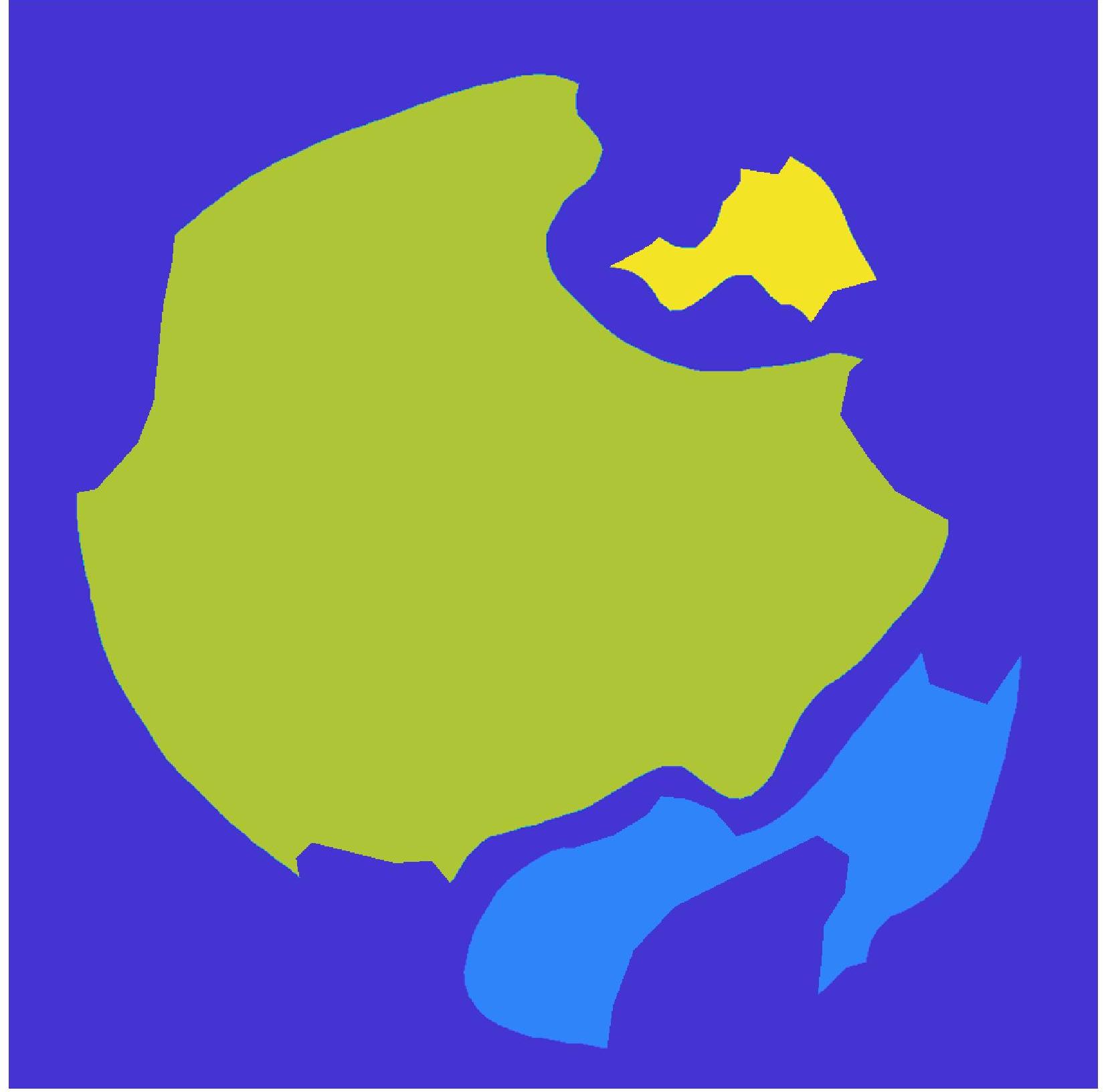} &
\includegraphics[height=2.5cm, width=2.2cm]{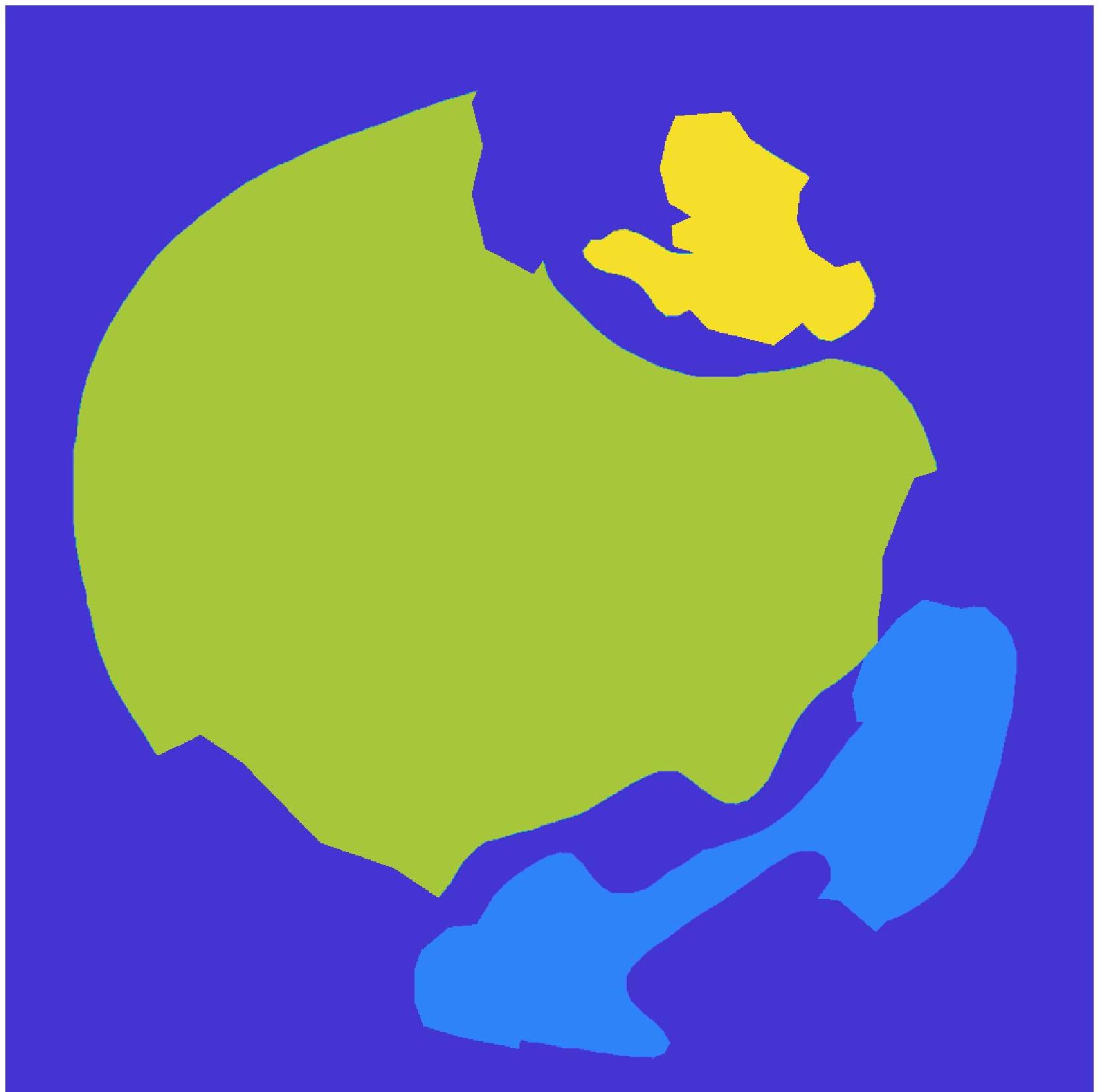} & 
\includegraphics[height=2.5cm, width=2.2cm]{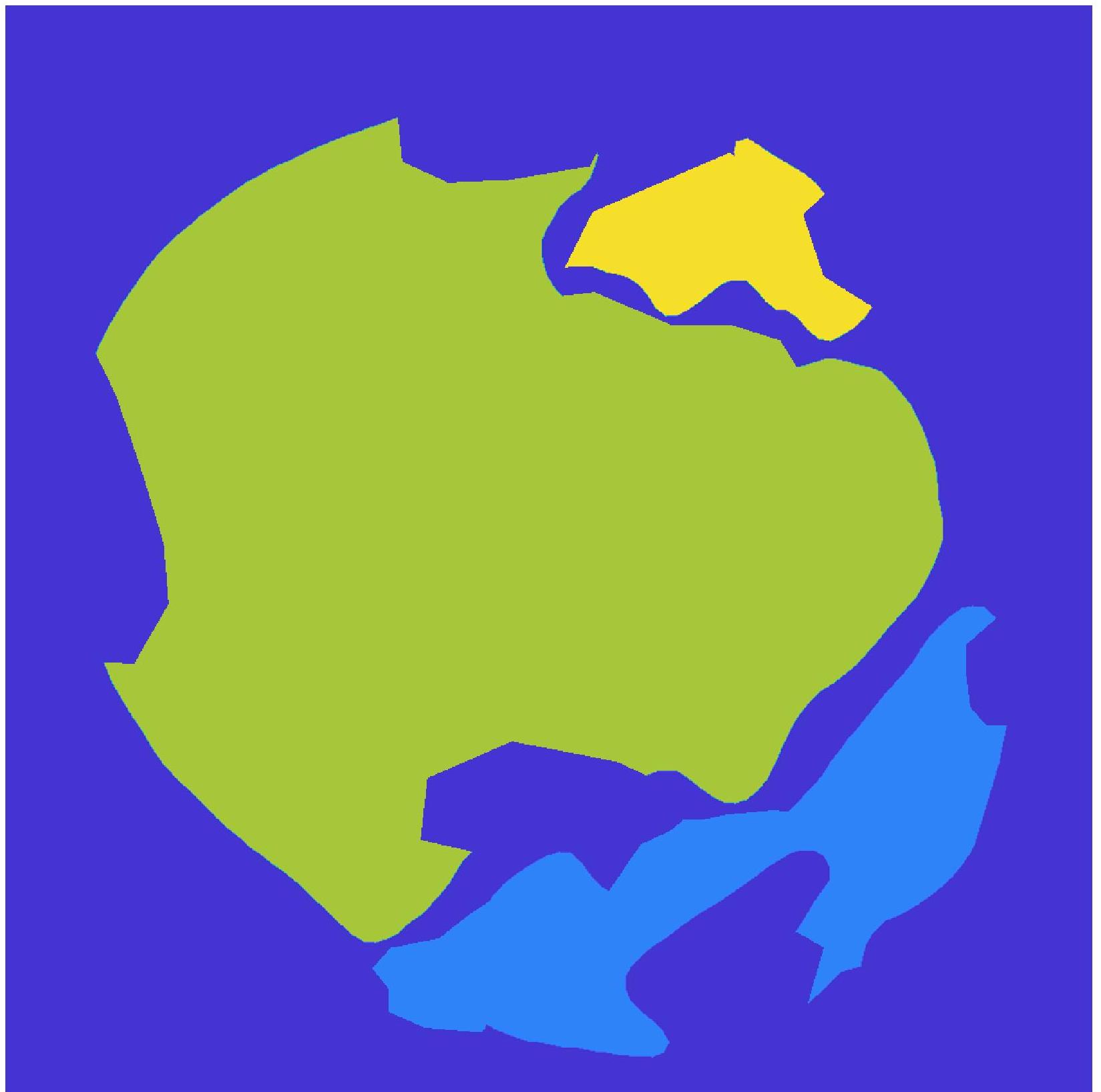} \\
(a) & (b) & (c) & (d) & (e) & (f) & (g) \\
\end{tabular}
\caption{Segmentation results on the Gleason training dataset: (a) original  images; (b) manual segmentation masks. Segmented masks using data generated by: (c) GeoGAN$_{WSS}$; (d)  \cite{HGAN}; (e) \cite{Gupta}; (f) $DAGAN$; (g) $cGAN$. Rows correspond to different images.}
\label{fig:segout1}
\end{figure*}

Figure~\ref{fig:segout1} shows the segmentation results using a UNet++ trained on images from different image synthesis methods. Figure~\ref{fig:segout1} (a) shows the test image and Figure~\ref{fig:segout1} (b) shows the manual mask. Figures~\ref{fig:segout1} (c)-(f) show, respectively, the segmentation masks obtained by GeoGAN$_{WSS}$, \cite{HGAN},\cite{Gupta}, $DAGAN$ and $cGAN$.
%
% Our method outperforms baseline conventional data augmentation and other competing methods by a significant margin. 
GeoGAN$_{WSS}$'s DM is higher than the DM value of the best performing method. Our results clearly show that modeling geometrical features leads to better performance than state of the art segmentation network architectures. 

\begin{table*}[!htbp]
 \begin{center}
 \caption{Segmentation results on the training dataset for the Gleason challenge using UNet++. Mean and standard deviation (in brackets) are shown. The best results per metric are shown in bold. $p$ values are with respect to $GeoGAN_{WSS}$}
\begin{tabular}{|c|c|c|c|c|c|c|c|c|}
\hline 
\multicolumn{9}{|c|}{Results for UNet++ Architecture} \\ \hline
{} & \multicolumn{5}{|c|}{Comparison approaches} & \multicolumn{3}{|c|}{Proposed}\\ \hline
{} & {DA}  & {DAGAN}  & {cGAN} & {\cite{Gupta}} & {\cite{HGAN}} & {GeoGAN$_{Cond}$} & {GeoGAN$_{WSS}$} & {GeoGAN$_{Manual}$}\\ \hline
{DM} & {0.843(0.08)} & {0.891(0.11)} & {0.881(0.13)} & {0.895(0.06)} & {0.901(0.07)} & {0.918(0.05)} & {0.937(0.06)} & {\textbf{0.942(0.04)}} \\ \hline
{p} & {0.0002} & {0.006} & {0.004}  & {0.0005} & {0.0003} & {0.01} & {-} & {0.11} \\ \hline
{MAE} & {0.088(0.015)} & {0.072(0.014)} & {0.079(0.015)}  & {0.068(0.016)} & {0.061(0.018)} & {0.052(0.018)} & {0.030(0.012)} & {\textbf{0.023(0.01)}} \\ \hline
{HD} & {12.6(4.2)} & {10.9(3.6)} & {11.1(3.8)}  & {10.2(3.7)} & {9.3(3.2)} & {8.8(2.7)} & {8.4(2.4)} & {\textbf{7.9(2.2)}}\\ \hline
\end{tabular}
\label{tab:GleasonTr}
\end{center}
\end{table*}

\begin{table*}[!htbp]
 \begin{center}
 \caption{Segmentation results on the training dataset for the Gleason challenge using PSPNet. Mean and standard deviation (in brackets) are shown. The best results per metric are shown in bold. $p$ values are with respect to $GeoGAN_{WSS}$ 
}
\begin{tabular}{|c|c|c|c|c|c|c|c|c|}
\hline 
\multicolumn{9}{|c|}{Results for PSPNet Architecture} \\ \hline
{} & \multicolumn{5}{|c|}{Comparison approaches} & \multicolumn{3}{|c|}{Proposed}\\ \hline
{} & {DA}  & {DAGAN}  & {cGAN} & {\cite{Gupta}} & {\cite{HGAN}} & {GeoGAN$_{Cond}$} & {GeoGAN$_{WSS}$} & {GeoGAN$_{Manual}$}\\ \hline
{DM} & {0.862(0.10)} & {0.907(0.13)} & {0.909(0.11)} & {0.917(0.05)} & {0.924(0.08)} & {0.937(0.08)} & {0.958(0.08)} & {\textbf{0.963(0.05)}} \\ \hline
{p} & {0.0003} & {0.005} & {0.003}  & {0.0007} & {0.0002} & {0.009} & {-} & {0.098} \\ \hline
{MAE} & {0.081(0.016)} & {0.068(0.016)} & {0.073(0.012)}  & {0.063(0.017)} & {0.048(0.012)} & {.034(0.012)} & {0.026(0.011)} & {\textbf{0.021(0.009)}} \\ \hline
{HD} & {12.1(4.3)} & {10.7(3.7)} & {10.9(3.2)}  & {10.1(3.4)} & {9.0(3.0)} & {8.7(2.9)} & {8.2(2.6)} & {\textbf{7.8(2.3)}}\\ \hline
\end{tabular}
\label{tab:GleasonPSPNet}
\end{center}
\end{table*}

Table~\ref{tab:GleasonPSPNet} shows segmentation results when using a PSPNet architecture \cite{PSPNet} since the top ranked method\footnote{https://github.com/hubutui/Gleason} used a PSPNet architecture for segmentation.
GeoGAN's superior segmentation accuracy is attributed to its capacity to learn the geometrical relationship between different labels (through $L_{shape}$). Thus our attempt to model the intrinsic geometrical relationships between different labels could generate superior quality masks. %
Moreover, the PSPNet and UNet++ results demonstrate that our image augmentation method does equally well for different segmentation architectures. %It can be used for various applications.

\subsection{Ablation Studies.}
\begin{figure}[!htbp]
\centering
\begin{tabular}{ccc}
\includegraphics[height=2.5cm, width=2.5cm]{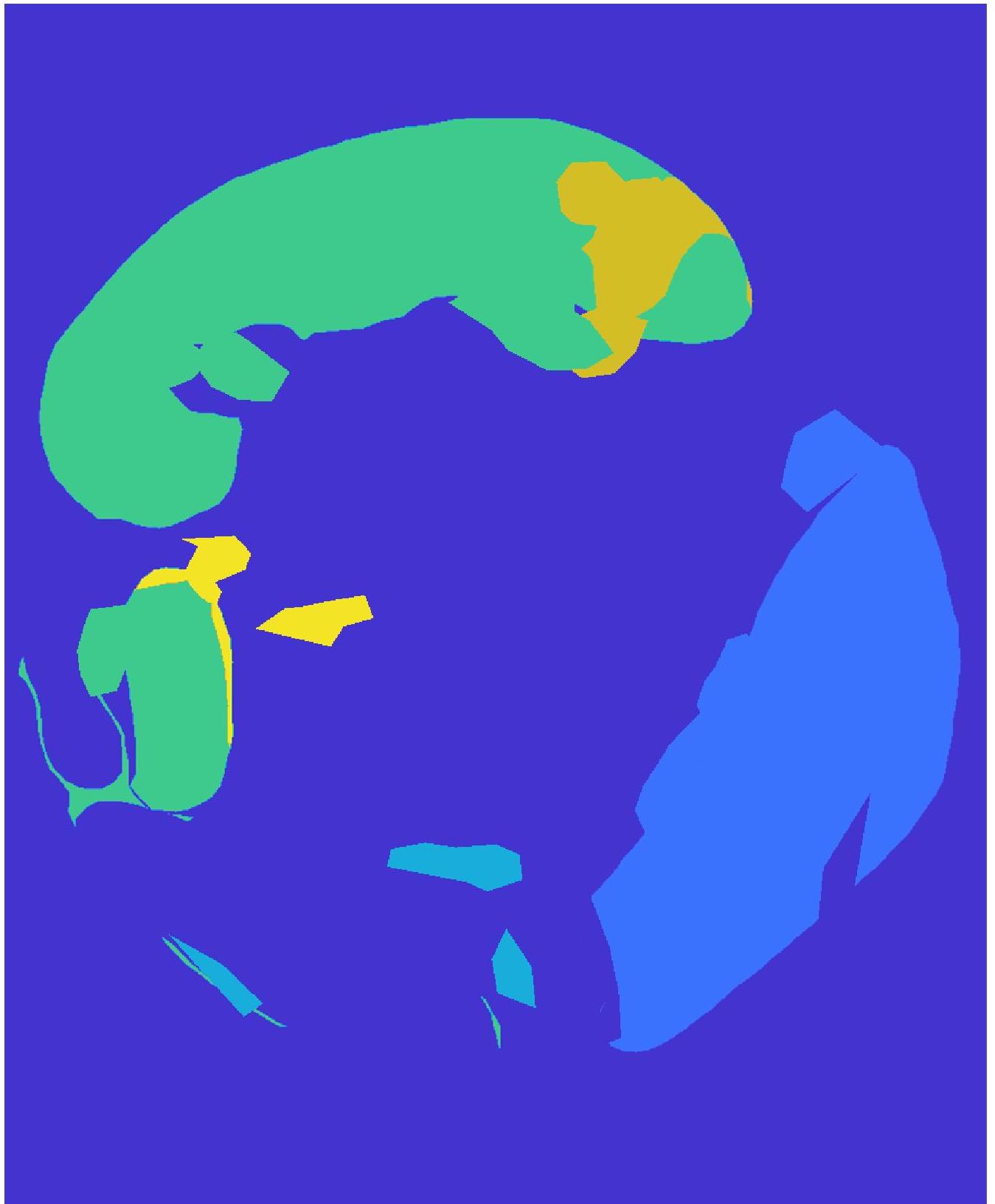} & 
\includegraphics[height=2.5cm, width=2.5cm]{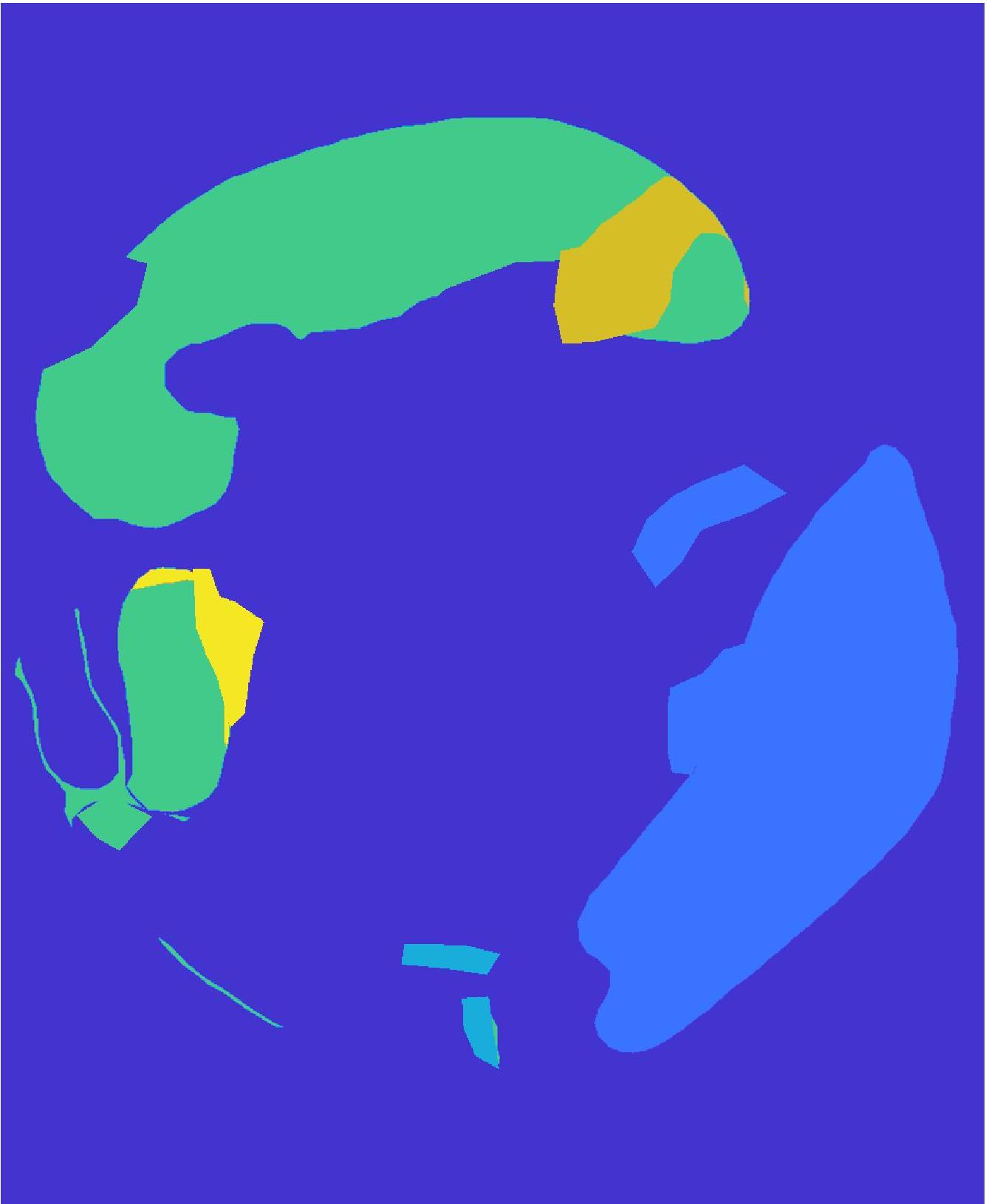} & 
\includegraphics[height=2.5cm, width=2.5cm]{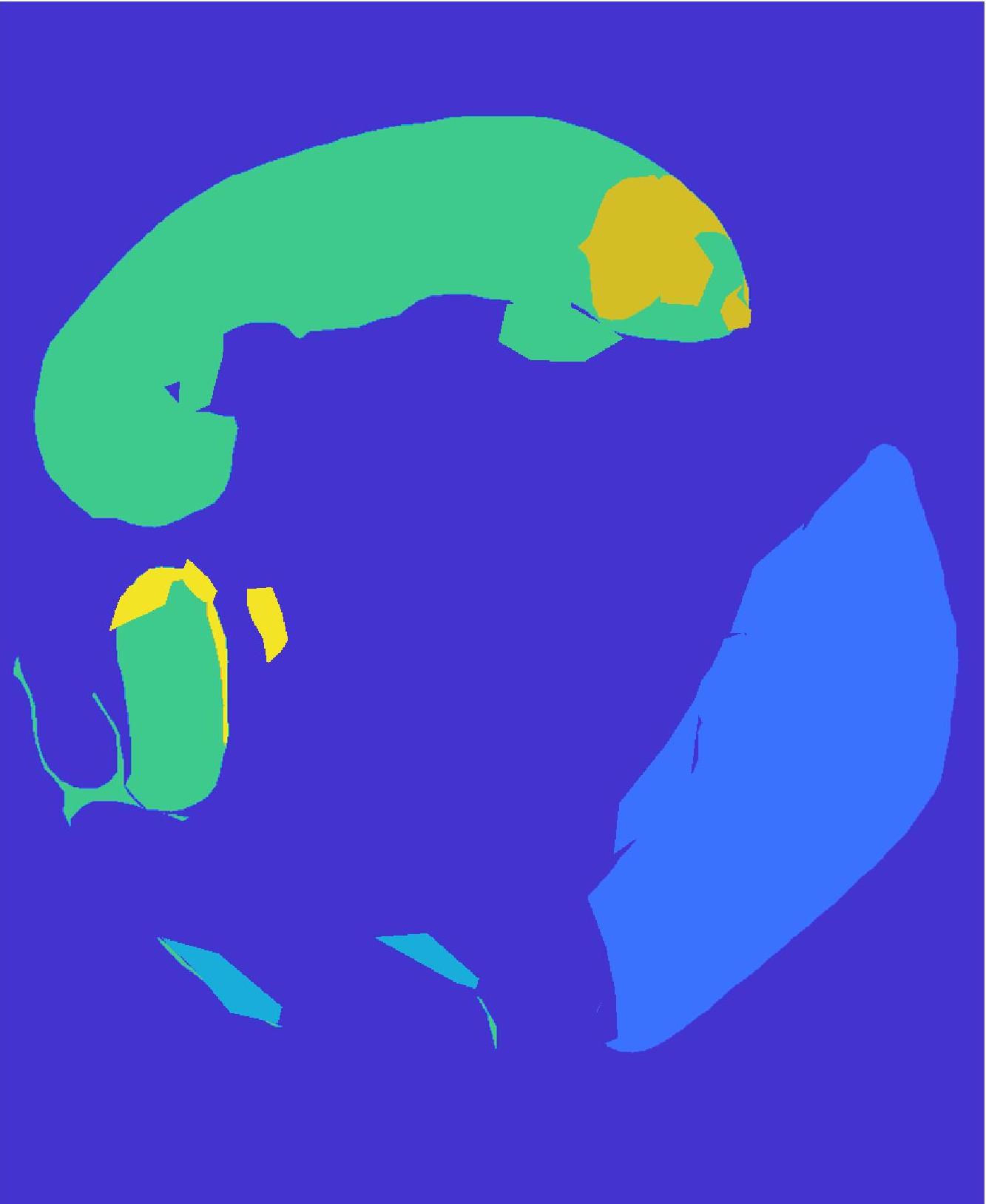} \\
%--------
\includegraphics[height=2.5cm, width=2.5cm]{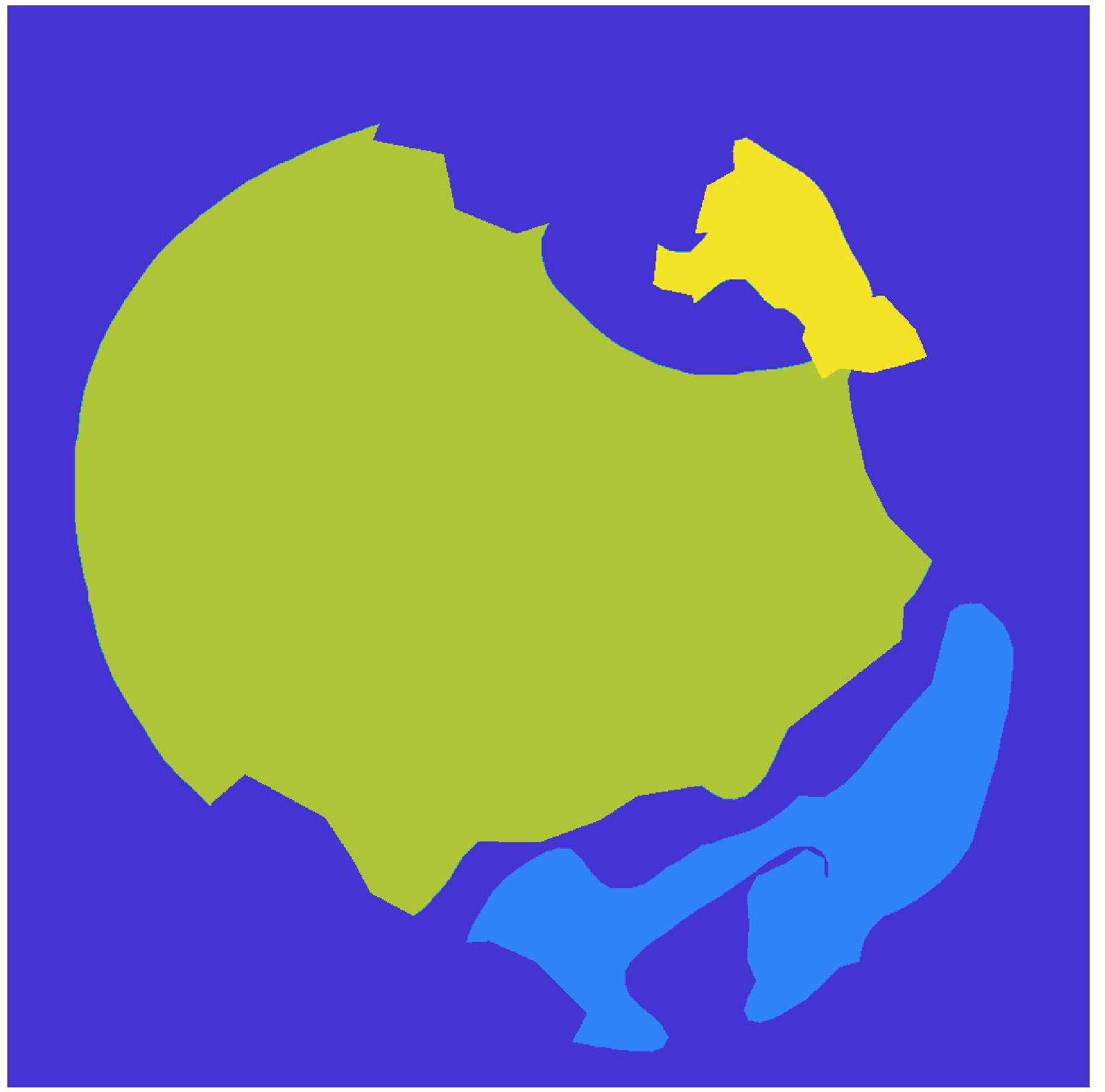} & 
\includegraphics[height=2.5cm, width=2.5cm]{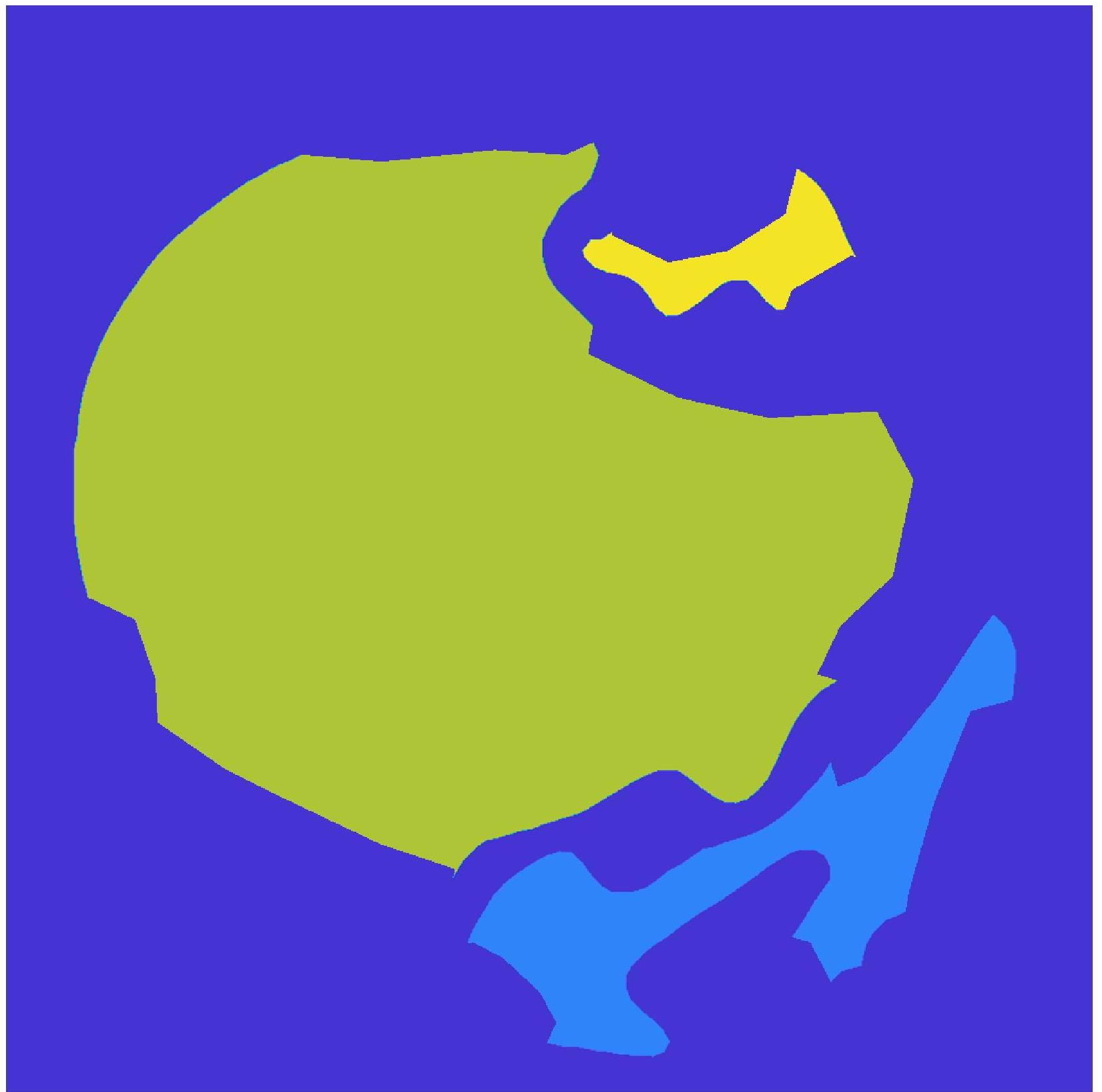} & 
\includegraphics[height=2.5cm, width=2.5cm]{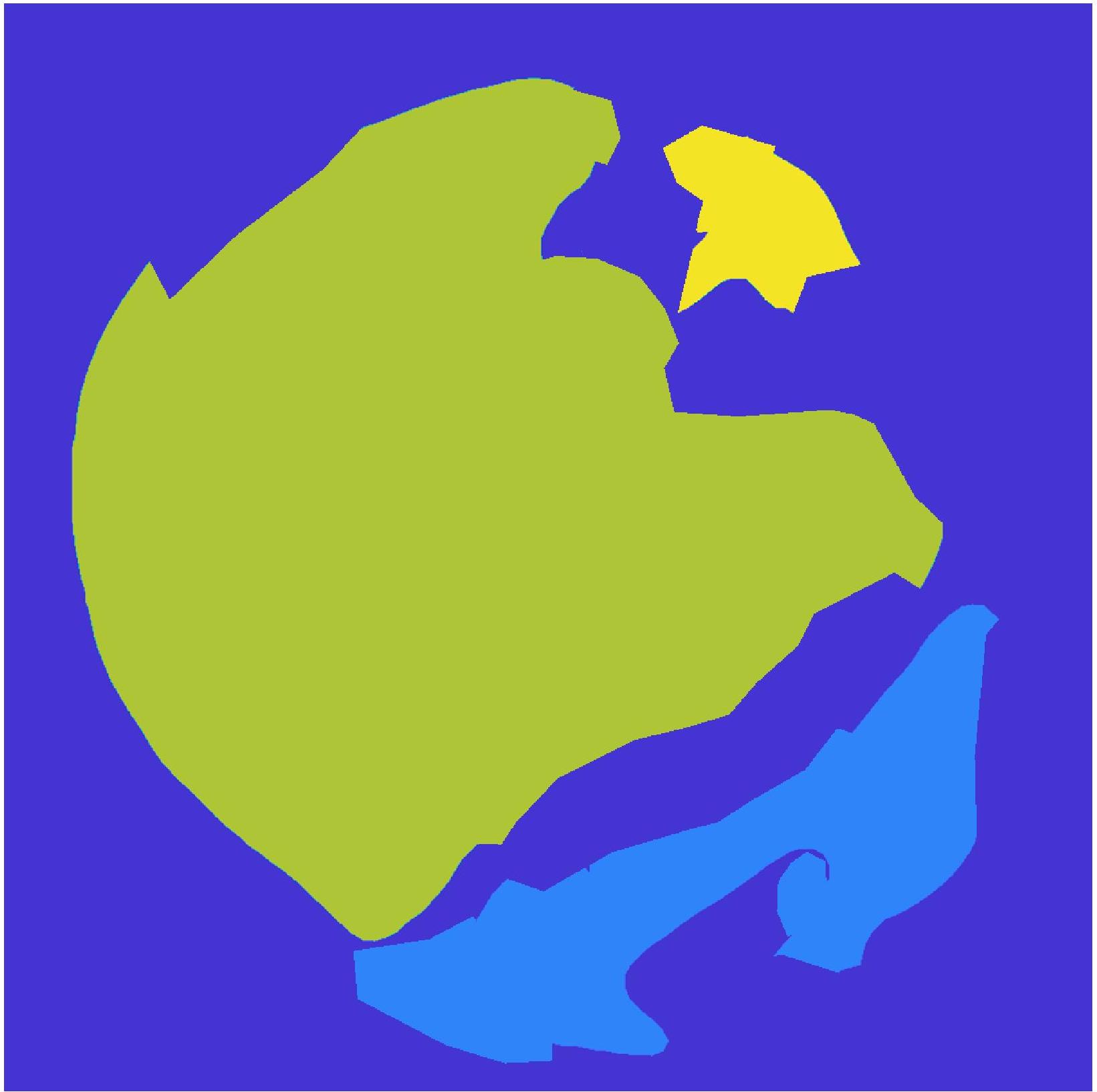} \\
 (a) & (b) & (c) \\
\end{tabular}
\caption{Segmentation results for ablation experiments on the training set: (a) $GeoGAN_{noL_{shape}}$; (b) $GeoGAN_{noL_{cls}}$; (c) $GeoGAN_{noSamp}$. Results correspond to the images shown in the two rows of Figure~\ref{fig:segout1}.}
\label{fig:segout1_Abl}
\end{figure}

Table~\ref{tab:Abl} summarizes the segmentation results for different ablation methods using UNet++ while  Figure~\ref{fig:segout1_Abl} shows the corresponding segmentation mask. The segmentation outputs are quite different from the ground truth and the one obtained by GeoGAN$_{WSS}$ (Figure~\ref{fig:segout1}). In some cases, the normal regions are included as pathological areas, while parts of the diseased regions are not segmented. Either case is undesirable for disease diagnosis and quantification. Thus, different components of our cost functions are integral to the method's performance. Excluding one or more of classification loss, geometric loss, and sampling loss adversely affects segmentation performance.

\begin{table}[!htbp]
 \begin{center}
 \caption{Mean and standard deviation (in brackets) of segmentation results from ablation studies on training dataset using the UNet++ architecture. HD is in mm.}
\begin{tabular}{|c|c|c|c|}
\hline 
% & \multicolumn{4}{|c|}{Comparison approaches}  \\ \hline
{} & {GeoGAN}  & {GeoGAN}  & {GeoGAN} \\ 
{} & {$_{noL_{cls}}$}  & {$_{noL_{shape}}$}  & {$_{noSamp}$}\\ \hline
{DM} & {0.891(0.07)} & {0.899(0.08)} & {0.895(0.06)} \\ \hline
{HD} & {9.6(3.1)} & {9.1(3.2)}  & {9.4(3.4)}   \\ \hline
{MAE} & {0.071(0.015)} & {0.065(0.016)} & {0.062}(0.016) \\  \hline
% \hline
% {} & {GeoGAN}  & {GeoGAN}  & {GeoGAN} \\ 
% {} & {$_{onlyL_{cls}}$}  & {$_{onlyL_{shape}}$}  & {$_{onlySamp}$}\\ \hline
% %
% {DM} & {0.824(0.08)} & {0.825(0.07)} & {0.818(0.06)} \\ \hline
% {HD} & {11.2(2.9)} & {11.1(3.0)}  & {12.5(2.8)}   \\ \hline
\end{tabular}
\label{tab:Abl}
\end{center}
\end{table}

\subsection{Gleason Challenge Results}

The ranking of different methods is based on the overall score (Eqn~\ref{eq:score}) \footnote{https://gleason2019.grand-challenge.org/Results}. The scores of GeoGAN$_{WSS}$ and other methods are summarized in Table~\ref{tab:Gleasonscore}. Our proposed method outperforms the top-ranked method, thus clearly demonstrating the effectiveness of geometrical modeling in generating informative images. This is particularly helpful in the absence of a large dataset of annotated images.

\begin{table}[!htbp]
\begin{center}
\caption{Results on the Gleason Challenge dataset for PCa detection based on overall score values.}
\begin{tabular}{|c|c|c|c|c|c|}
\hline 
{} & {GeoGAN$_{WSS}$}  & {Rank~1}  & {Rank~2} & {Rank~3} & {Rank~4} \\ \hline
{Score} & {0.8835} & {0.8451} & {0.7925} & {0.7896}  & {0.7780}\\ \hline
{} & {GeoGAN$_{Manual}$} & {Rank~5} & {Rank~6} & {Rank~7} & {Rank~8} \\ \hline
{Score}  & {0.8992} & {0.7597} & {0.7578} & {0.7160} & {0.7125}\\ \hline
\end{tabular}
\label{tab:Gleasonscore}
\end{center}
\end{table}

\subsection{Results On Additional Datasets}

\subsubsection{Segmentation Results on Glas Challenge Dataset}

We apply our method on the public GLAS segmentation challenge \cite{GlasReview}, which has manual segmentation maps of glands in $165$ $H\&E$ stained images derived from $16$ histological sections from different patients with stage $T3$ or $T4$ colorectal adenocarcinoma. The slides were digitized with a Zeiss MIRAX MIDI Slide Scanner having pixel resolution of $0.465\mu$m. The WSIs were rescaled to a pixel resolution of $0.620\mu$m (equivalent to $20\times$ magnification).
  $52$ visual fields from malignant and benign areas from the WSIs were selected to cover a wide variety of tissues. An expert pathologist graded each visual field as either ‘benign’ or ‘malignant’.
Further details of the dataset can be found at \cite{GlasReview}.

We generate augmented images using GeoGAN$_{WSS}$ and train a UNet++ segmentation network to obtain the final output.  The performance metrics - Dice Metric (DM), Hausdorff distance (HD), F1 score (F1)- for our results (including ablation studies) and top-ranked methods \cite{GlasResults,GlasReview} are summarized in Table~\ref{tab:GlasSeg}. \snm{ The numbers are taken from the challenge paper in \cite{GlasReview} (Table~2). Except for GeoGAN$_{Manual}$ (equivalent to fully supervised training), our method, GeoGAN$_{WSS}$, outperforms all other methods using a standard segmentation architecture.  The ablation study experiments' performance also demonstrates the benefits of including all components of our proposed method. Competing methods in the challenge have used conventional augmentation, whereas our image synthesis approach generates more informative images. 
}

\begin{figure*}[!t]
\centering
\begin{tabular}{ccccc}
\includegraphics[height=2.5cm, width=3.2cm]{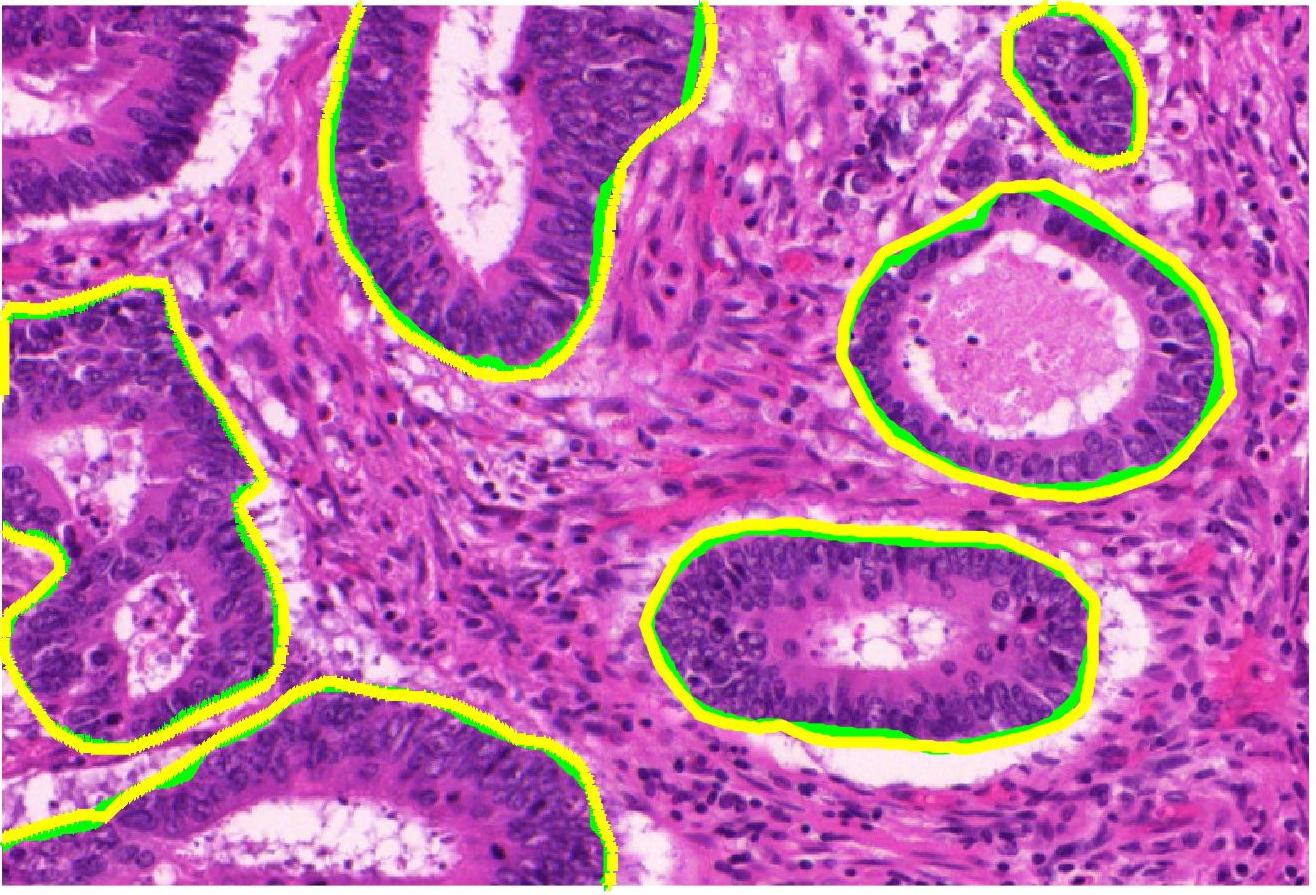} & 
\includegraphics[height=2.5cm, width=3.2cm]{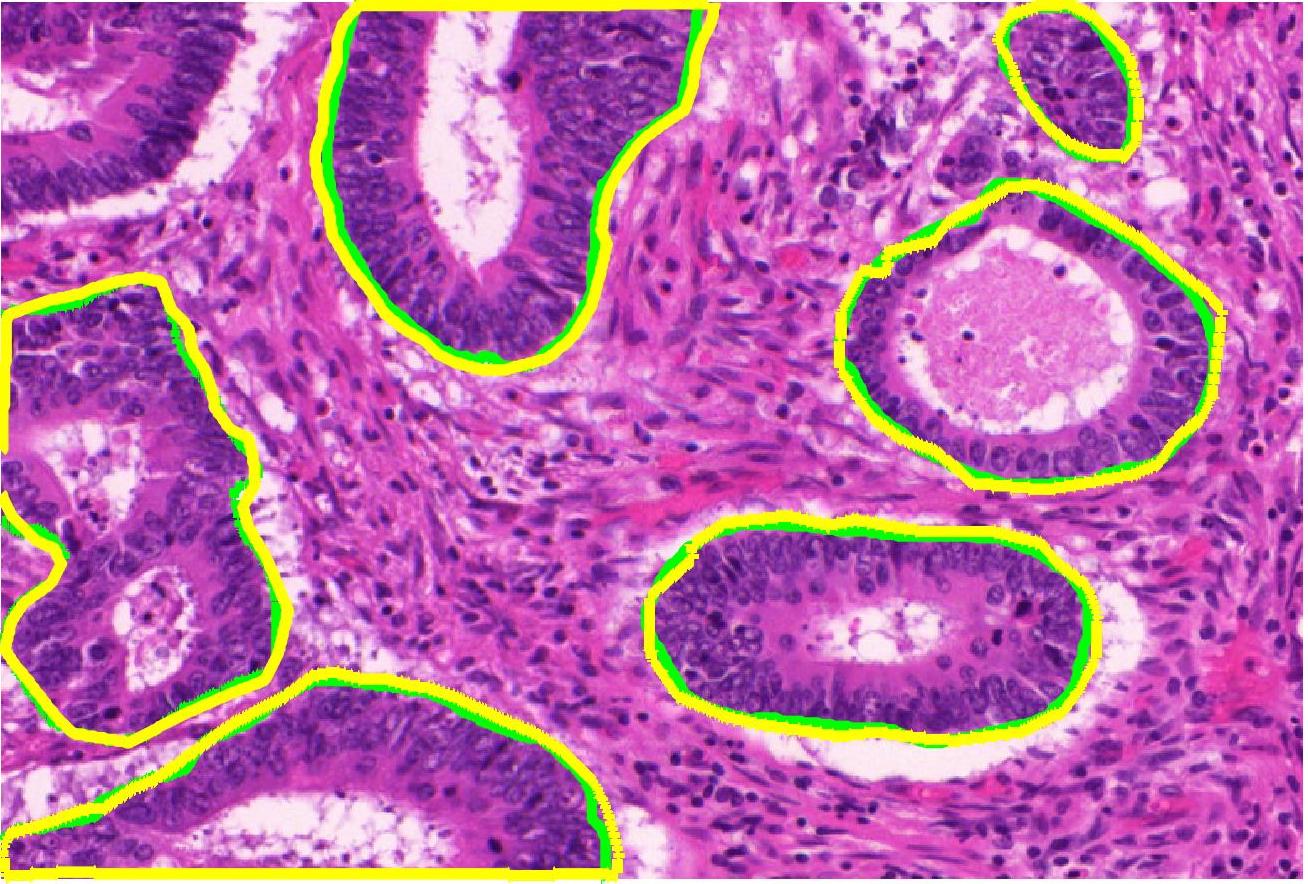} & 
\includegraphics[height=2.5cm, width=3.2cm]{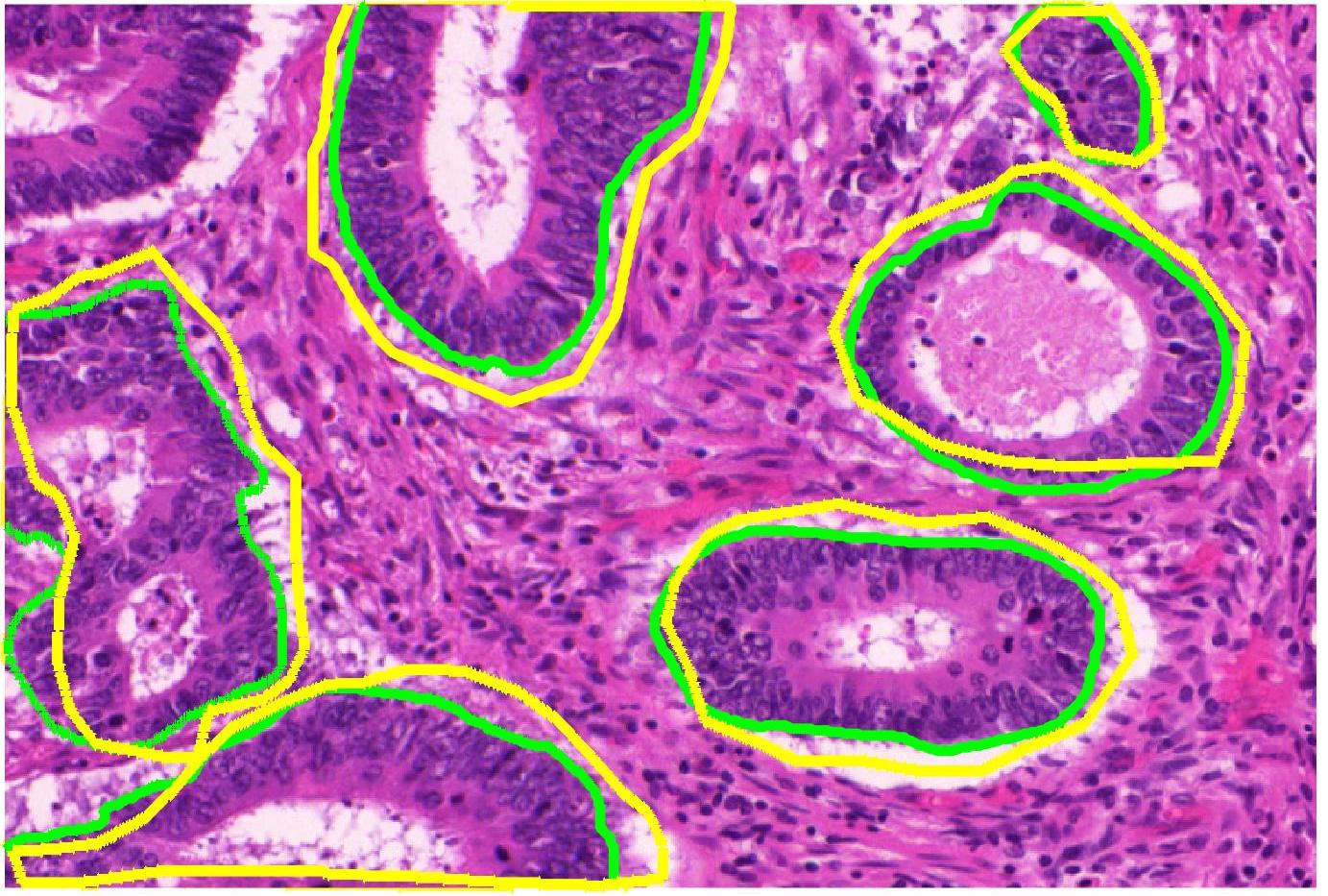} & 
\includegraphics[height=2.5cm, width=3.2cm]{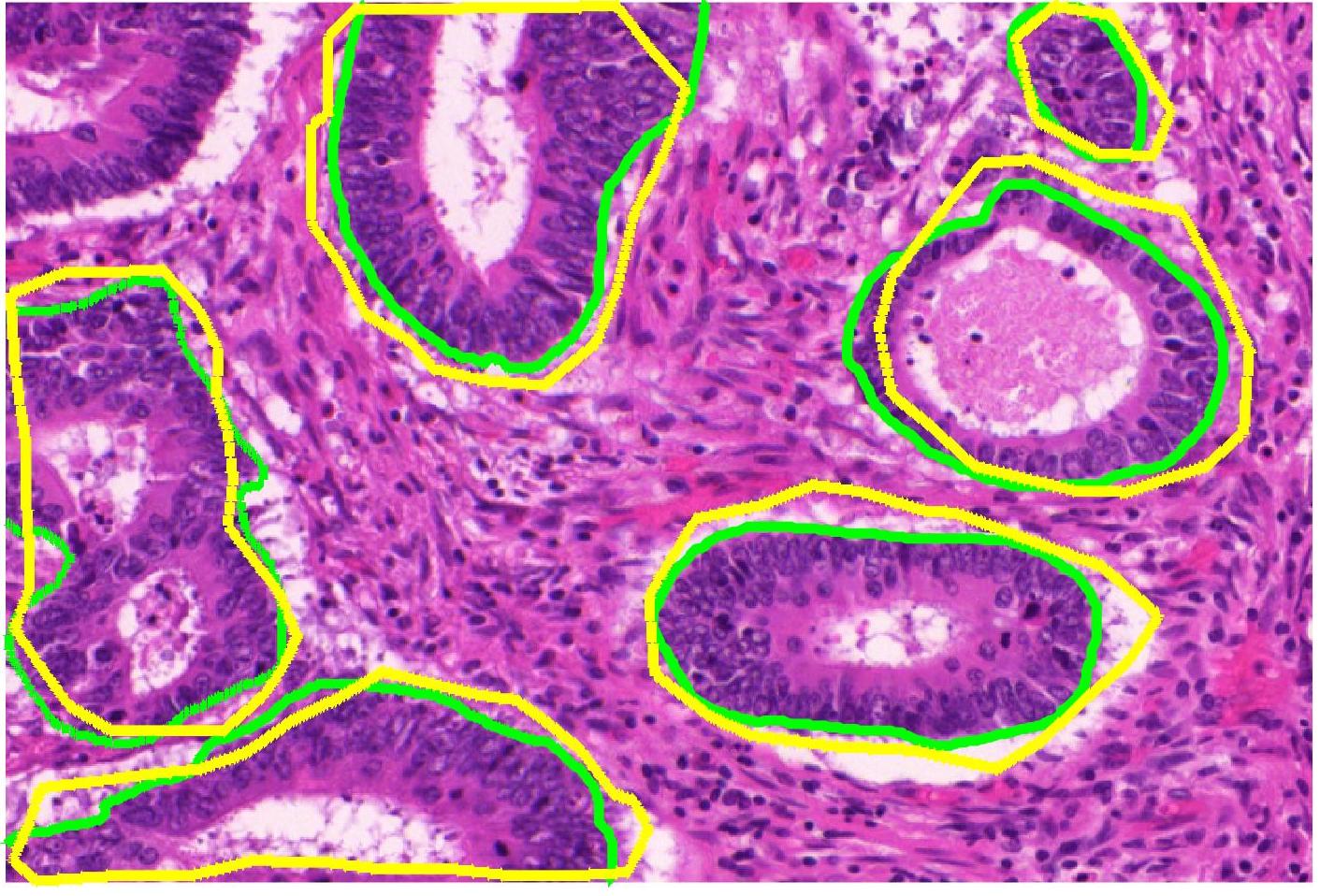} &
\includegraphics[height=2.5cm, width=3.2cm]{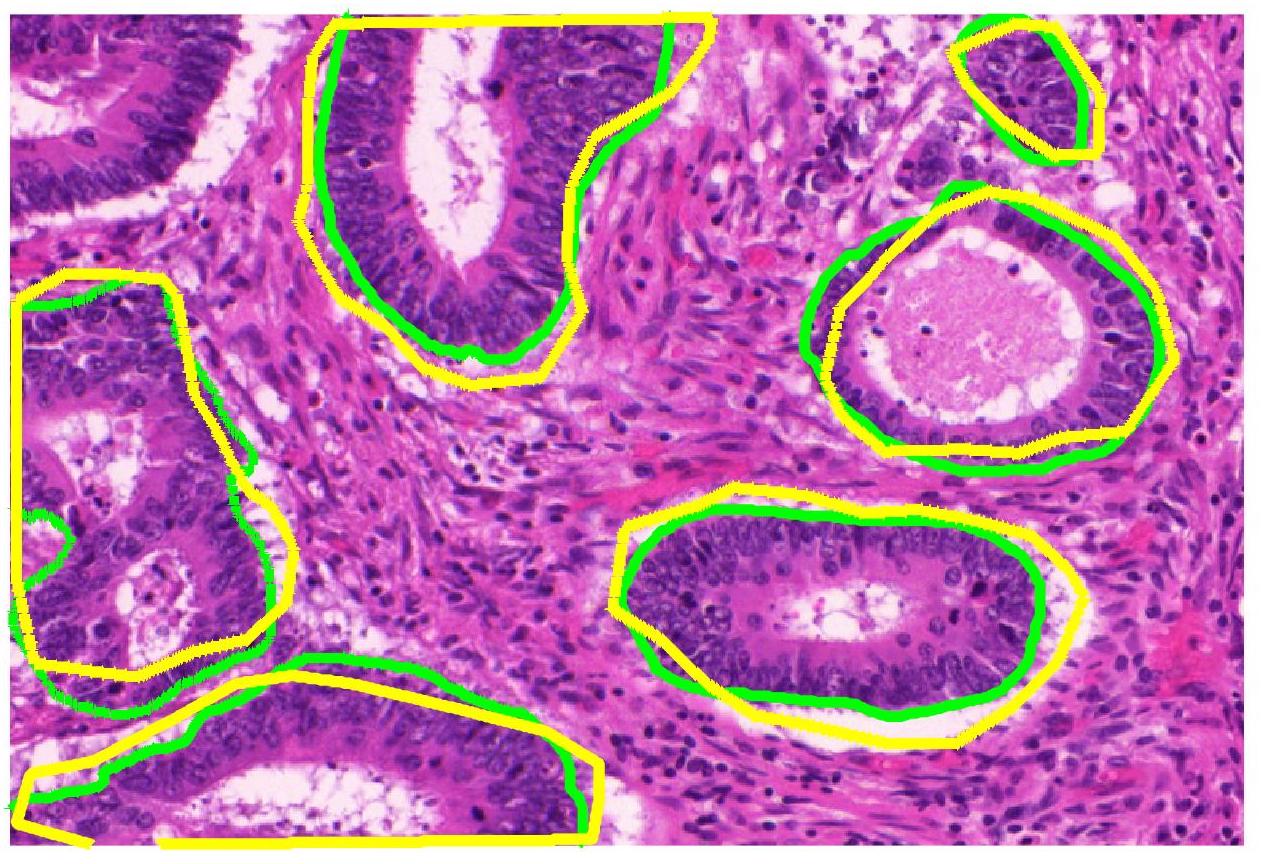} \\
%---------
% \includegraphics[height=2.5cm, width=2.2cm]{GlasRes2_Image.jpg} & 
\includegraphics[height=2.5cm, width=3.2cm]{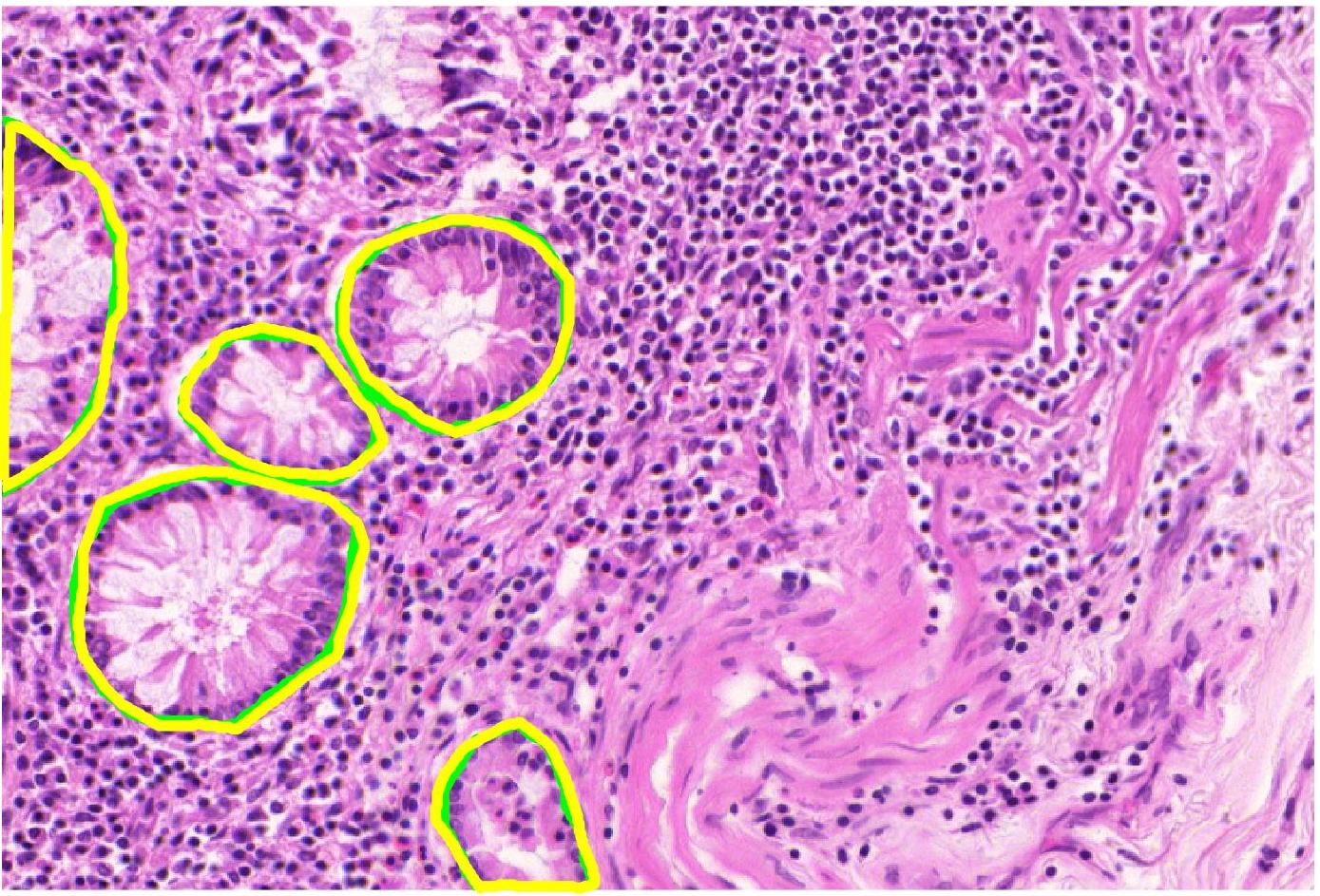} & 
\includegraphics[height=2.5cm, width=3.2cm]{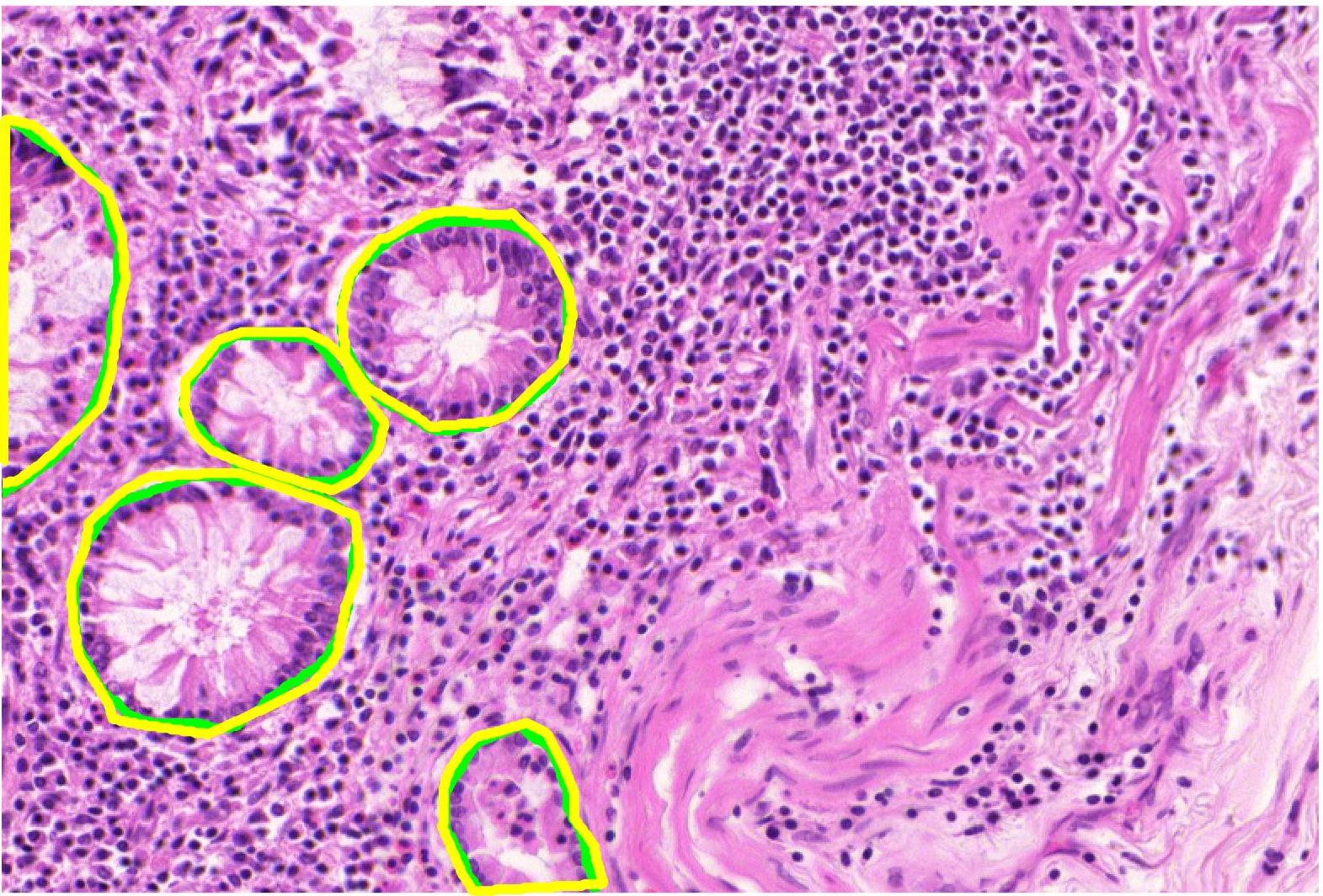} & 
\includegraphics[height=2.5cm, width=3.2cm]{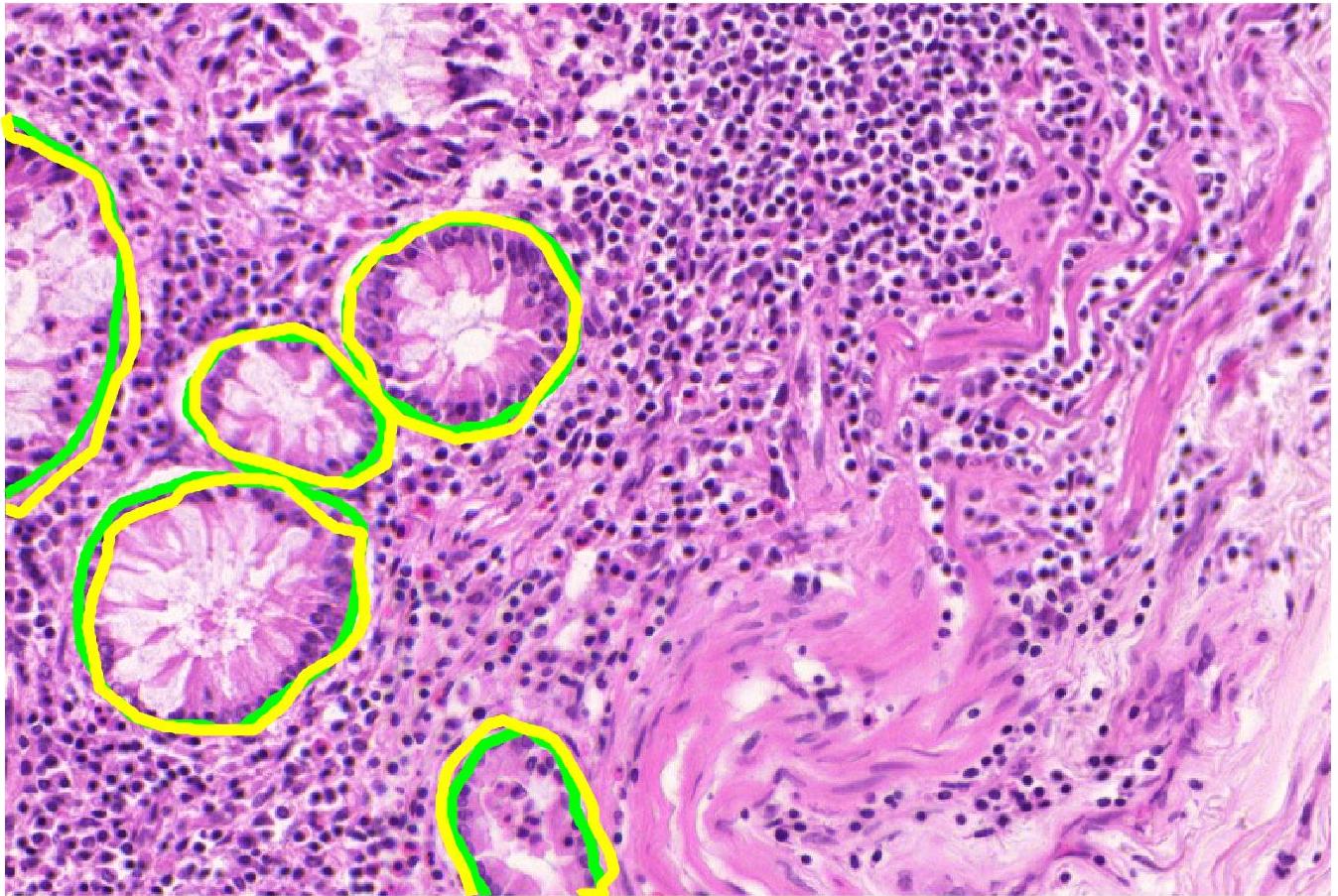} & 
\includegraphics[height=2.5cm, width=3.2cm]{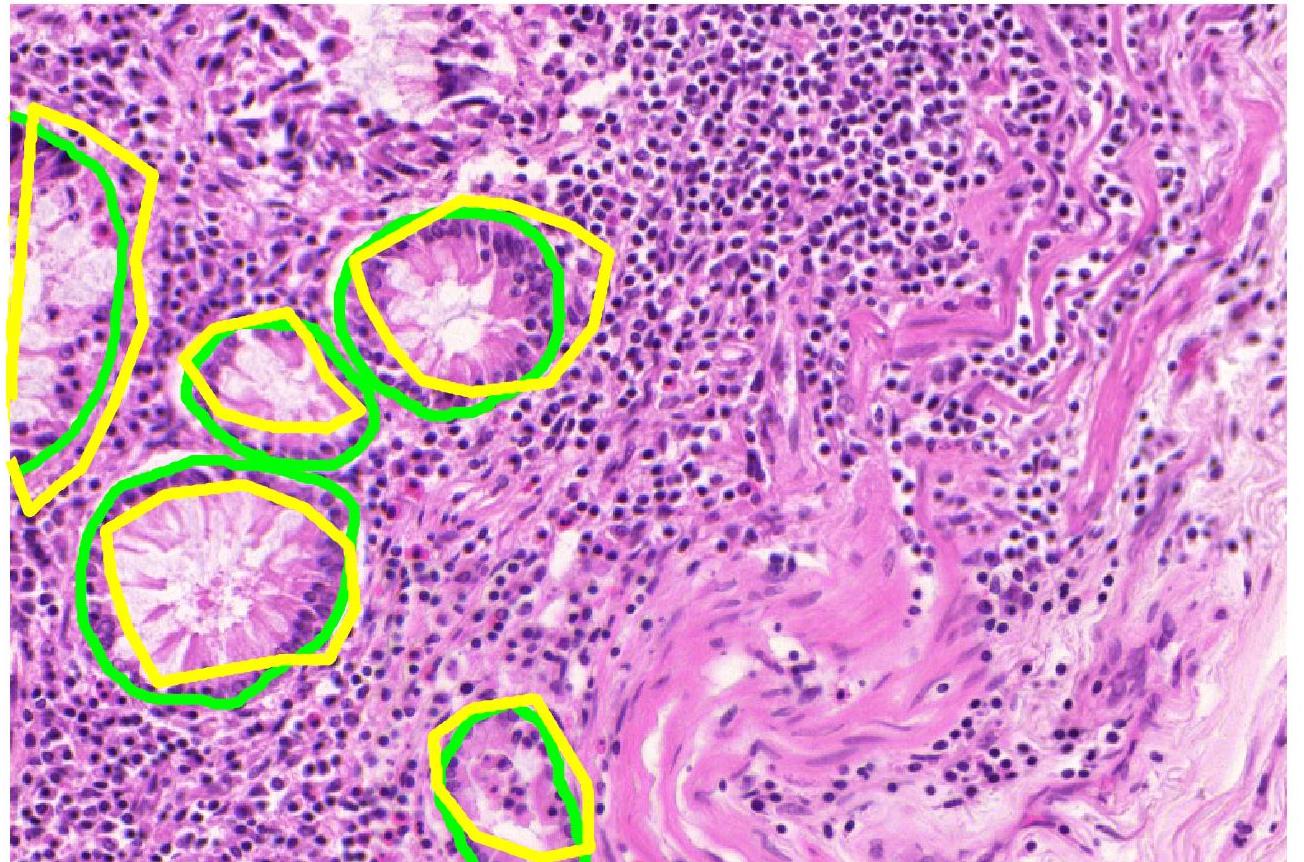} &
\includegraphics[height=2.5cm, width=3.2cm]{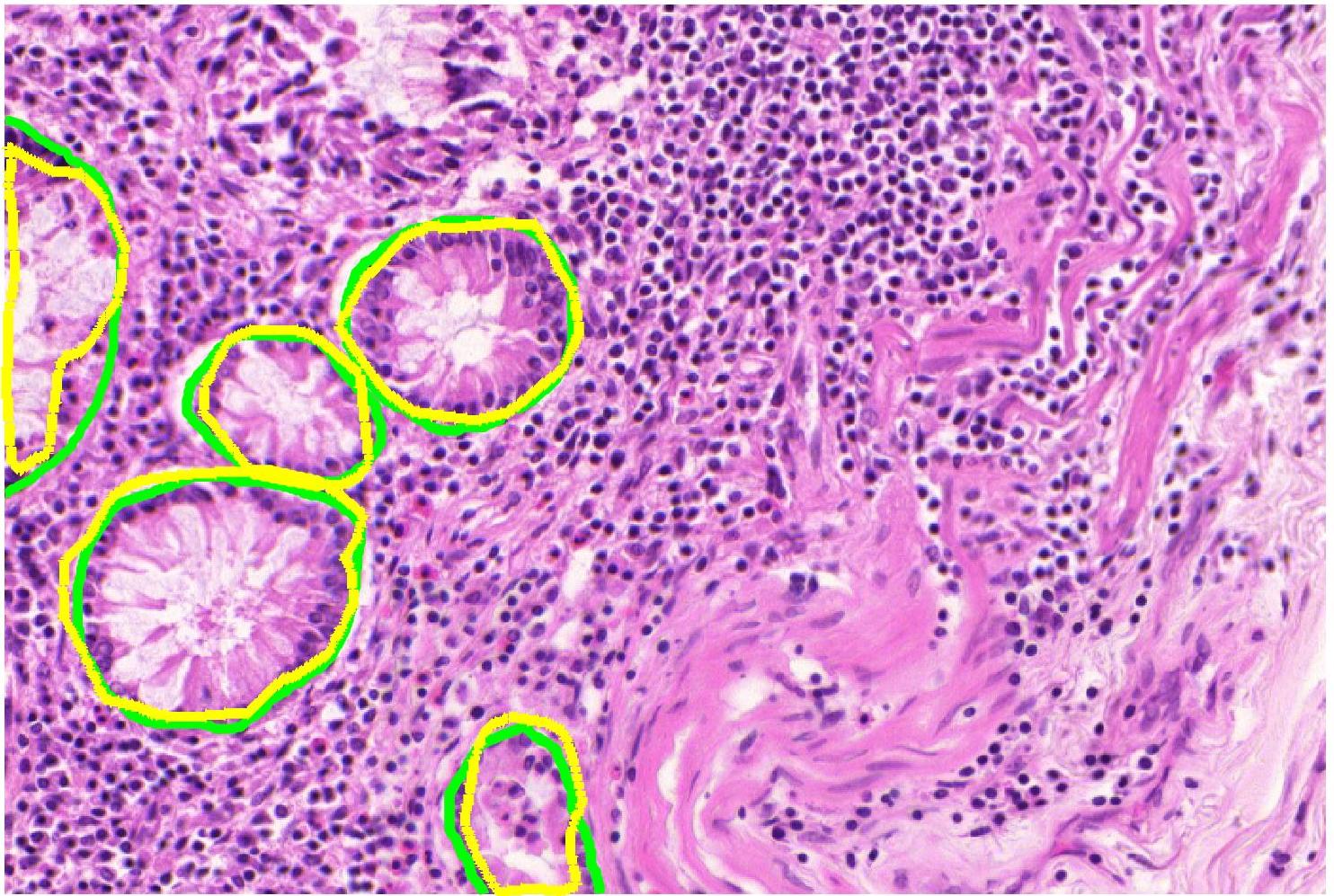} \\
(a) & (b) & (c) & (d) & (e) \\
\end{tabular}
\caption{Segmentation results from the GlaS Segmentation Challenge. Original image and contours of different segmentation methods are shown. The manual segmentation is shown as green contour and the algorithm output is shown in yellow for: (a) GeoGAN$_{Manual}$; (b) GeoGAN$_{WSS}$ (c) GeoGAN$_{noL_{cls}}$; (d) GeoGAN$_{noL_{Shape}}$; (e) GeoGAN$_{noSamp}$. Rows correspond to different images.}
\label{fig:Glas}
\end{figure*}

\snm{
Figure~\ref{fig:Glas} shows example segmentation outputs of GeoGAN$_{WSS}$ and other variants of our method used for ablation studies. The results clearly show that with the exclusion of our proposed method's different components, the segmentation performance degrades.
}

\begin{table*}[!htbp]
 \begin{center}
 \caption{Segmentation results on the GLas Segmentation challenge for GeoGAN$_{WSS}$ and the top $3$ ranked methods. $HD$ is in mm. Best results per metric in bold.}
\begin{tabular}{|c|c|c|c|c|c|c|c|c|c|c|c|c|c|c|}
\hline 
{} & \multicolumn{2}{|c|}{GeoGAN$_{WSS}$}  & \multicolumn{2}{|c|}{Glas Rank 1}  & \multicolumn{2}{|c|}{Glas Rank 2} & \multicolumn{2}{|c|}{GeoGAN$_{Manual}$} & \multicolumn{2}{|c|}{GeoGAN$_{noL_{cls}}$} & \multicolumn{2}{|c|}{GeoGAN$_{noL_{shape}}$} & \multicolumn{2}{|c|}{GeoGAN$_{noSamp}$} \\ \hline
{} & {Part A} & {Part B} & {Part A} & {Part B} & {Part A} & {Part B} & {Part A} & {Part B} & {Part A} & {Part B} & {Part A} & {Part B} & {Part A} & {Part B} \\ \hline
{F1} & {{0.9462}} & {{0.7586}} & {0.912} & {0.716} & {0.891} & {0.703} & {\textbf{0.9514}} & \textbf{0.7694} & {0.887} & {0.709} & {0.903} & {0.721} & {0.894} & {0.715}  \\ \hline
{DM} & {{0.9361}} & {{0.8237}} & {0.897} & {0.781} & {0.882} & {0.786} & \textbf{0.9483} & \textbf{0.8361} & {0.878} & {0.757} & {0.892} & {0.784} & {0.883} & {0.761} \\ \hline
{HD} & {{40.342}} & {{140.432}} & {45.418} & {160.347} & {57.413} & {145.575} & \textbf{38.560} & \textbf{139.1842} & {60.234} & {163.459} & {52.32} & {143.321} & {56.34} & {158.753} \\ \hline
\end{tabular}
\label{tab:GlasSeg}
\end{center}
\end{table*}

\subsubsection{Segmentation Results For CAMELYON16}

\snm{
We evaluate our method on the segmentation challenge part of the CAMELYON16 dataset \cite{Camelyon16} having $270$ WSIs for training and $130$ WSIs for
testing. The patches are of dimension $224 \times 224$ and were sampled from $10\times$ magnification WSIs (resolution of $0.972 \mu m/px$) and the receptive field of a patch covers $217.8 \mu$m $\times 217.8 \mu$m. Segmentation output is predicted for each patch and then fused to obtain the final WSI mask. At areas of overlap we   
 The results are summarized in Table~\ref{tab:cam16}. The baseline values are taken from \cite{ChengECCV20} which is a recent approach using teacher-student model and reports higher values than those reported on the challenge website. Our method outperforms the baseline DenseNet121 architecture and comes close to the state of the art results of \cite{ChengECCV20}. Note that the values are much higher than shown in the leaderboard\footnote{https://camelyon16.grand-challenge.org/Results/} since many improved methods have been proposed.
This indicates that our proposed augmentation approach does very well on different datasets 
}

\begin{table}[!htbp]
 \begin{center}
 \caption{Results on the CAMELYON16 dataset's  segmentation challenge for GeoGAN$_{WSS}$ and the state of the art results from \cite{ChengECCV20}.}
\begin{tabular}{|c|c|c|c|c|}
\hline 
{} & {DenseNet121} & {\cite{ChengECCV20}} & {$GeoGAN_{WSS}$} & {$GeoGAN_{Manual}$}\\ \hline
{DM} & {0.9268} & {0.9376} & {0.9311} & {0.9467}\\ \hline
{FROC} & {0.4112} & {0.3894} & {0.3982} & {0.3786}\\ \hline
%---------
% {}  & \multicolumn{2}{|c|}{Glas Rank 2} & \multicolumn{2}{|c|}{Glas Rank 3} \\ \hline
% {} & {Part A} & {Part B} & {Part A} & {Part B} \\ \hline
% %
% {F1} & {0.891} & {0.703} & {0.896} & {0.719} \\ \hline
% {DM} & {0.882} & {0.786} & {0.886} & {0.765} \\ \hline
% {HD} & {57.413} & {145.575} & {57.350} & {159.873} \\ \hline
\end{tabular}
\label{tab:cam16}
\end{center}
\end{table}

% \subsubsection{Results for TCGA Dataset}
% \label{met:tcga}

% \snm{
%  We also run classification experiments on the TCGA dataset used by \cite{NagpalGGL} which consists of deidentified, digitized whole-slide images of hematoxylin-and-eosin
% ($H\&E$)-stained formalin-fixed paraffin-embedded prostatectomy specimens. We test only on the `Gleason Grade sub dataset' consisting of 183 patient WSIs. We use our image synthesis method to augment the dataset and closely follow the different steps outlined in \cite{NagpalGGL}. We obtain a mean classification accuracy of $0.78$, which is higher than $0.7$ reported in \cite{NagpalGGL}. The mean classification accuracy of $29$ pathologists is $0.61$ 
% 
% Pathologists provided annotations of specific regions within a slide,
% outlining individual glands or regions and providing an associated label
% (non-tumor, or GP3, 4, or 5). We obtain a mean per pixel classification accuracy of $84.5$ which is considerably higher than that of \cite{NagpalGGL} ($0.77$). In this work the authors have presented results for per pixel accuracy and not dice metric values.
% }

\subsection{Comparative Analysis With Contrastive Learning Approaches}

\snm{
Contrastive learning approaches such as MoCo \cite{HeMoco} and SimCLR \cite{ChenSimCLR} are considered state-of-the-art for downstream task based self-supervised learning u. These methods have been mostly used for classification tasks. For a fair comparison with these methods, we use our data augmentation approach for image classification on the Gleason (using the provided Gleason grades as class labels) and CAMELYON16 dataset. %We do not use the Gleason grade images since there are no clinician provided ground truth \textbf{classification} labels for the images. 
}
\snm{
We follow the implementations provided by the authors, and the results are summarized in Table~\ref{tab:sslcomp}, where $ResNet$ indicates method using features derived from a ResNet \cite{ResNet} pre-trained with ImageNet. Both MoCo and SimCLR give similar performance, and the difference in their results is not statistically significant ($p=0.06$).
Our results are close to both methods and are statistically not very different from their results. The results indicate that our pre-text task-based approach is equally effective as MoCo and SimCLR for self-supervised learning. % compared to the supervised based approach for the purpose of classification of histopathology images. 
% While our approach shows similar results it must be k
% Since both of them are different approaches it is not straightforward to make a comparison.  similar to the 
}

% \begin{table}[!htbp]
%  \begin{center}
%  \caption{Results on Gleason 2019 and CAMELYON16 dataset's  classification challenge. Comparison with other self supervised learning approaches using contrastive loss. $p-$values are with respect to GeoGAN$_{WSS}$.}
% \begin{tabular}{|c|c|c|c|c|}
% \hline 
% \multicolumn{5}{|c|}{CAMELYON16 Results} \\ \hline
% {} & {MoCo \cite{HeMoco}} & {SimCLR \cite{ChenSimCLR}} & {GeoGAN$_{WSS}$} & {GeoGAN$_{Manual}$} \\ \hline
% %
% {Acc} & {0.9012} & {0.8965} & {0.8921} & {0.9032}  \\ \hline
% {AUC} & {0.9253} & {0.9181} & {0.9171} & {0.9231}\\ \hline
% {$p-$val} & {0.054} & {0.067} & {-} & {0.062}\\ \hline
% %---------
% \hline 
% \multicolumn{5}{|c|}{CGleason 2019 Results} \\ \hline
% {} & {MoCo \cite{HeMoco}} & {SimCLR \cite{ChenSimCLR}} & {GeoGAN$_{WSS}$} & {GeoGAN$_{Manual}$}\\ \hline
% %
% {Acc} & {0.9191} & {0.9071} & {0.9083} & {0.9165} \\ \hline
% {AUC} & {0.9347} & {0.9287} & {0.9292} & {0.9332}\\ \hline
% {$p-$val} & {0.062} & {0.057} & {-} & {0.066}\\ \hline
% \end{tabular}
% \label{tab:sslcomp}
% \end{center}
% \end{table}

\begin{table}[!htbp]
 \begin{center}
 \caption{Results on Gleason 2019 and CAMELYON16 dataset's  classification challenge. Comparison with other self supervised learning approaches using contrastive loss. $p-$values are with respect to GeoGAN$_{WSS}$.}
\begin{tabular}{|c|c|c|c|c|c|}
\hline 
\multicolumn{6}{|c|}{CAMELYON16 Results} \\ \hline
{} & {MoCo} & {SimCLR} & {GeoGAN} & {GeoGAN} & {ResNet}\\ %\hline
{} & {\cite{HeMoco}} & {\cite{ChenSimCLR}} & {$_{WSS}$} & {$_{Manual}$} & {}\\ \hline
{Acc} & {0.9012} & {0.8965} & {0.8921} & {0.9032} & {0.8323} \\ \hline
{AUC} & {0.9253} & {0.9181} & {0.9171} & {0.9231} & {0.8542}\\ \hline
{$p-$val} & {0.054} & {0.067} & {-} & {0.062} & {0.001}\\ \hline
%---------
\hline 
\multicolumn{6}{|c|}{Gleason 2019 Results} \\ \hline
% {} & {MoCo \cite{HeMoco}} & {SimCLR \cite{ChenSimCLR}} & {GeoGAN$_{WSS}$} & {GeoGAN$_{Manual}$} & {ResNet}\\ \hline
%
{} & {MoCo} & {SimCLR} & {GeoGAN} & {GeoGAN} & {ResNet}\\ %\hline
{} & {\cite{HeMoco}} & {\cite{ChenSimCLR}} & {$_{WSS}$} & {$_{Manual}$} & {}\\ \hline
{Acc} & {0.9191} & {0.9071} & {0.9083} & {0.9165} & {0.8124}\\ \hline
{AUC} & {0.9347} & {0.9287} & {0.9292} & {0.9332} & {0.8432}\\ \hline
{$p-$val} & {0.062} & {0.057} & {-} & {0.066} & {0.005}\\ \hline
\end{tabular}
\label{tab:sslcomp}
\end{center}
\end{table}

\subsection{\snm{Failure Cases}}
\begin{figure}[!htbp]
\centering
\begin{tabular}{ccc}
\includegraphics[height=2.5cm, width=2.5cm]{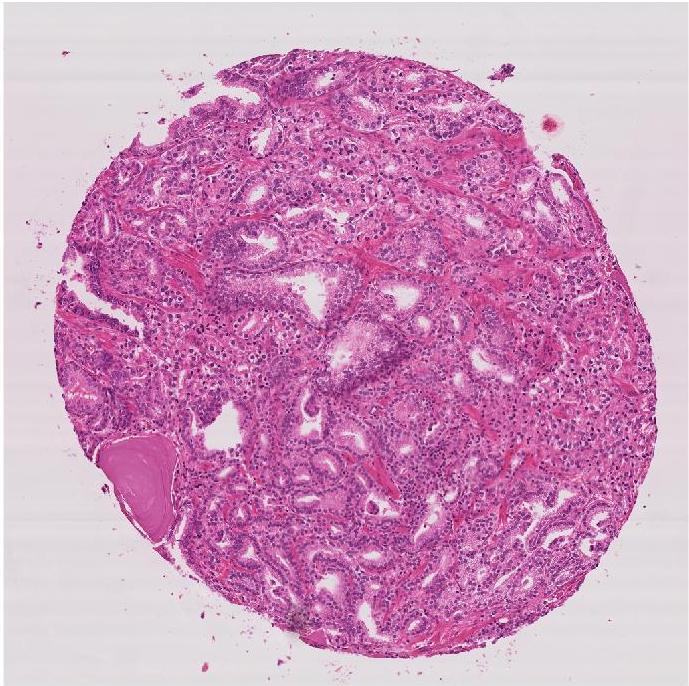} & 
\includegraphics[height=2.5cm, width=2.5cm]{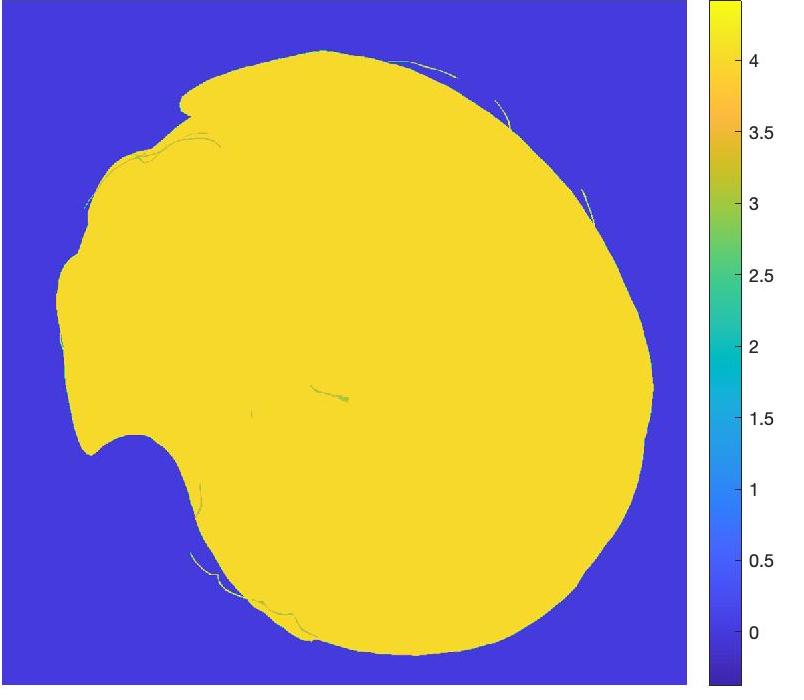} & 
\includegraphics[height=2.5cm, width=2.5cm]{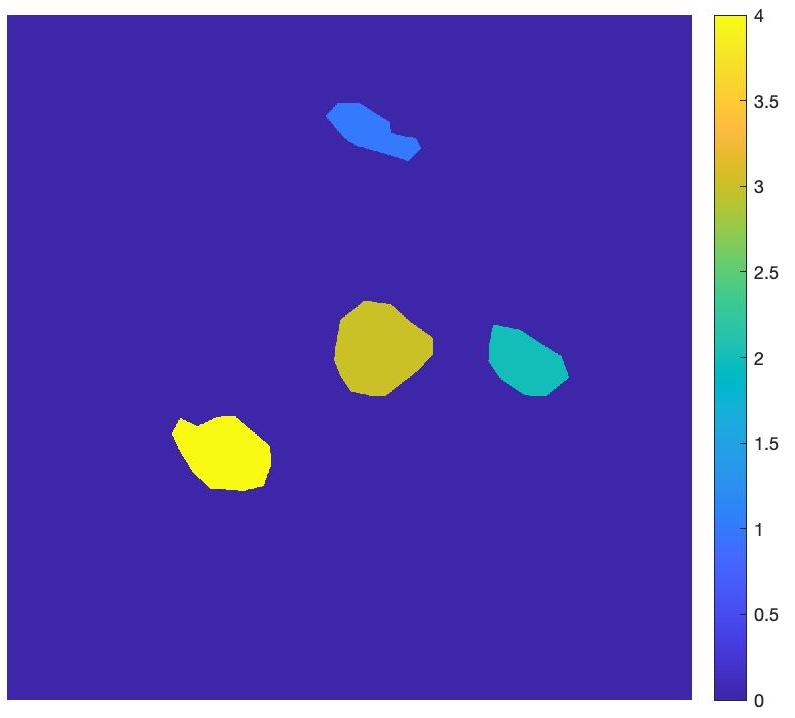} \\
 & & DM$=0.37$ \\
%--------
\includegraphics[height=2.5cm, width=2.5cm]{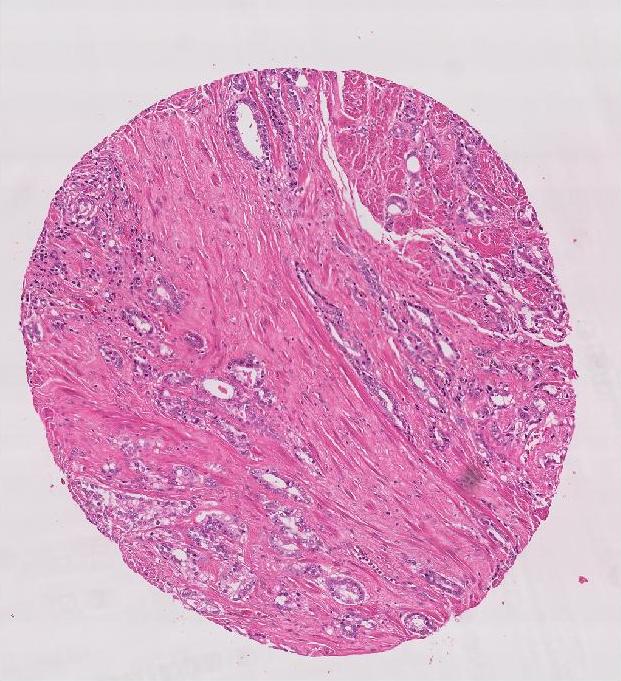} & 
\includegraphics[height=2.5cm, width=2.5cm]{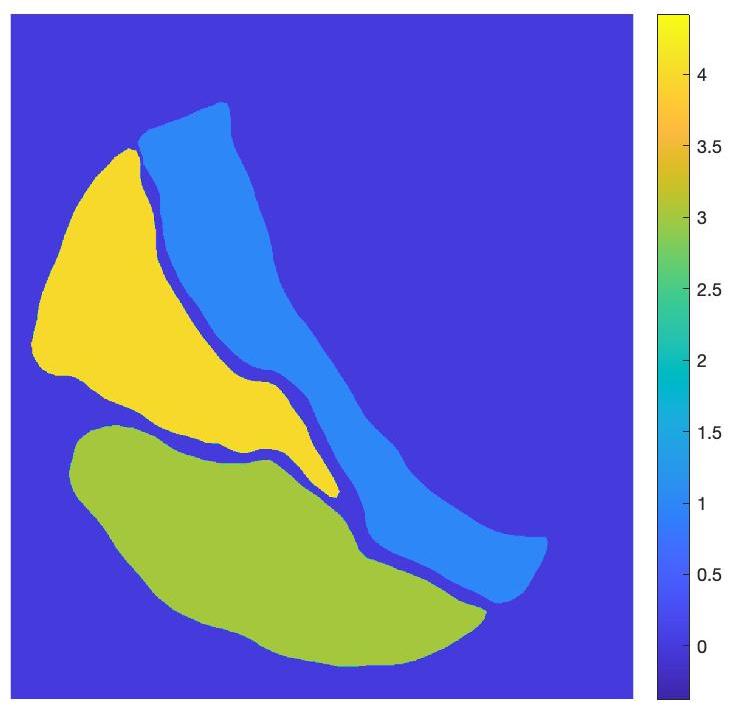} & 
\includegraphics[height=2.5cm, width=2.5cm]{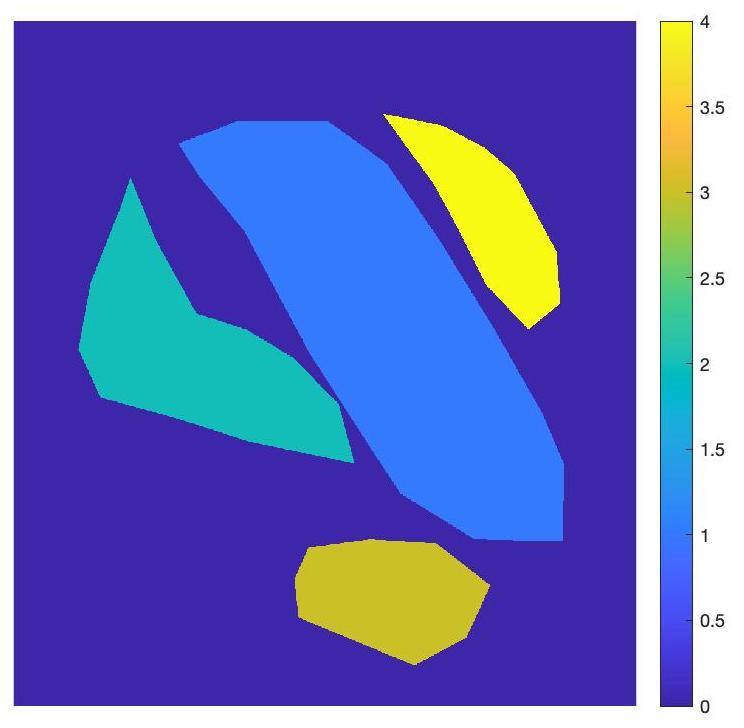} \\
 & & DM$=0.63$ \\
 (a) & (b) & (c) \\
\end{tabular}
\caption{Examples of failure cases: (a) original image; (b) manual segmentation map; (c) segmentation map obtained using  GeoGAN$_{WSS}$.}
\label{fig:failure}
\end{figure}

\snm{
 Figure~\ref{fig:failure} shows examples of some failure cases where our proposed algorithm's segmentation had very low dice score on the Gleason dataset. The underlying reason is due to various factors such as artifacts or similar-looking structures nearby. In future work on improving the algorithm, we aim to incorporate ways to overcome the challenges posed by dissimilar tissues being too close together. 
}

\section{Discussion and Conclusion}\label{sec:concl}

We have proposed a novel approach to generate high-quality synthetic histopathology images for segmenting Gleason graded PCa \snm{and other pathological conditions}. Our method exploits the inherent geometric relationship between different segmentation labels and uses it to guide the shape generation process. We propose a self-supervised learning approach where a shape restoration network (ShaRe-Net) learns to predict the original segmentation mask from its distorted version. The pre-trained ShaRe-Net is used to extract feature maps and integrate them in the training stage.
Considering the time and effort required to obtain manual segmentation maps, we propose a weakly supervised segmentation method that generates segmentation maps having high agreement with manual segmentations provided by clinicians. 

We also include an uncertainty sampling stage based on a Dense UNet architecture where we inject diversity in the image generation process. The inter-label geometric relationship learnt by ShaRe-Net is leveraged by the sampling stage to generate realistic  images % where the learnt inter-label relationship is preserved.
 which add diversity to the training data. %, and at the same time generated realistic images based on inter-label relationships. 
 Comparative results with other synthetic image generation methods show that the augmented dataset from our proposed GeoGAN$_{WSS}$ method outperforms standard data augmentation and other competing methods when applied to the segmentation of PCa (from the Gleason grading challenge) and other pathologies (from CAMELYON and GlaS dataset). The synergy between shape, classification, and sampling terms leads to improved and robust segmentation and each term is important in generating realistic images. %Furthermore, results on the Glas Segmentation challenge show the robustness of our method to different datasets. % e relative contributions of each term and conclude that the shape prior term makes a significant contribution to the output, while a .
%  Our approach can be used for other medical imaging modalities without major changes to the workflow.
 %
 Despite the good performance of our method we observe  failure cases when the base images are noisy due to inherent characteristics of the image acquisition procedure. In future work we aim to address this challenge. %Although the second scenario is not very common, it can be critical in the medical context. In future work we aim to evaluate our method's robustness on a wide range of medical imaging modalities such as MRI, Xray, etc. 

\snm{
Our method's effectiveness is not limited to the Gleason dataset but also exhibits good performance on the GlaS segmentation dataset and the CAMELYON16 dataset. %Results for the domain generalization scenario demonstrate that when presented with a new dataset, our network's performance is close to the supervised setting scenario. 
The high-performance gains are mainly due to learning the inter-label geometric relationships.  
%
%The supervised setting would give the most accurate results since the train, and test domains are similar. However, the use of our image synthesis method improves the baseline performance of the domain generalization method. 
 Despite its complexity, our approach is still relevant for clinical use since it doesn't require much effort from clinicians, shows superior performance than competing methods, and our weakly supervised segmentation step helps reduce clinician involvement. %Since the method performs better than competing approaches, it will be relevant for clinical use.
}

% \snm{
% Weakly supervised segmentation is an important component of our method. Many previous works have used weakly supervised segmentation approaches and achieved high segmentation accuracy. However most models may not generalize to all datasets. For example, our approach uses the domain knowledge of Gleason grades to assign labels to the two largest clusters, and this condition is not valid in many use cases. Going forward, in a general medical image analysis scenario the initial weak segmentation outputs may need to be refined based on fewer manual segmentations.
% }

\bibliographystyle{IEEEtran}
\bibliography{TMI_GeoGAN}

\end{document}